\documentclass[manuscript,acmsmall,nonacm]{acmart}
\usepackage{algorithmic}
\usepackage{amsmath}
\usepackage{graphicx}
\usepackage{caption}
\usepackage{subfigure}
\usepackage{multirow}
\usepackage{makecell}
\usepackage{booktabs}
\usepackage{color}

\usepackage{pifont}
\usepackage[ruled,vlined,linesnumbered]{algorithm2e}
\usepackage{caption}
\usepackage{tcolorbox}
\usepackage{rotating}
\usepackage{diagbox}
\usepackage[edges]{forest}
\definecolor{hidden-draw}{RGB}{20,68,106}
\definecolor{hidden-pink}{RGB}{255,245,247}

\renewcommand\footnotetextcopyrightpermission[1]{} % removes footnote with conference information in first column
\AtBeginDocument{%
  }

\begin{document}

%%
%% The "title" command has an optional parameter,
%% allowing the author to define a "short title" to be used in page headers.
\title{A Survey on Deep Text Hashing: Efficient Semantic Text Retrieval with Binary Representation}

%%
%% The "author" command and its associated commands are used to define
%% the authors and their affiliations.
%% Of note is the shared affiliation of the first two authors, and the
%% "authornote" and "authornotemark" commands
%% used to denote shared contribution to the research.
\author{Liyang He}
\email{heliyang@mail.ustc.edu.cn}
\affiliation{%
  \department{State Key Laboratory of Cognitive Intelligence}
  \institution{University of Science and Technology of China}
  \city{Hefei}
  \country{China}
}

\author{Zhenya Huang}
\email{huangzhy@ustc.edu.cn}
\affiliation{%
  \department{State Key Laboratory of Cognitive Intelligence}
  \institution{University of Science and Technology of China}
  \city{Hefei}
  \country{China}
}
\affiliation{%
  \department{Institute of Artificial Intelligence}
  \institution{Hefei Comprehensive National Science Center}
  \city{Hefei}
  \country{China}
}

\author{Cheng Yang}
\email{yangchengyc@mail.ustc.edu.cn}
\author{Rui Li}
\email{ruili2000@mail.ustc.edu.cn}
\author{Zheng Zhang}
\email{zhangzheng@mail.ustc.edu.cn}
\author{Kai Zhang}
\email{kkzhang08@ustc.edu.cn}
\affiliation{%
  \department{State Key Laboratory of Cognitive Intelligence}
  \institution{University of Science and Technology of China}
  \city{Hefei}
  \country{China}
}

\author{Zhi Li}
\email{zhilizl@sz.tsinghua.edu.cn}
\affiliation{%
  \department{Shenzhen International Graduate School}
  \institution{Tsinghua University}
  \city{Shenzhen}
  \country{China}
}

\author{Qi Liu}
\email{qiliuql@ustc.edu.cn}
\author{Enhong Chen}
\email{cheneh@ustc.edu.cn}
\affiliation{%
  \department{State Key Laboratory of Cognitive Intelligence}
  \institution{University of Science and Technology of China}
  \city{Hefei}
  \country{China}
}

%%
%% By default, the full list of authors will be used in the page
%% headers. Often, this list is too long, and will overlap
%% other information printed in the page headers. This command allows
%% the author to define a more concise list
%% of authors' names for this purpose.
\renewcommand{\shortauthors}{Liyang He et al.}

%%
%% The abstract is a short summary of the work to be presented in the
%% article.
\begin{abstract}
With the rapid growth of textual content on the Internet, efficient large-scale semantic text retrieval has garnered increasing attention from both academia and industry. Text hashing, which projects original texts into compact binary hash codes, is a crucial method for this task. By using binary codes, the semantic similarity computation for text pairs is significantly accelerated via fast Hamming distance calculations, and storage costs are greatly reduced. With the advancement of deep learning, deep text hashing has demonstrated significant advantages over traditional, data-independent hashing techniques. By leveraging deep neural networks, these methods can learn compact and semantically rich binary representations directly from data, overcoming the performance limitations of earlier approaches. This survey investigates current deep text hashing methods by categorizing them based on their core components: semantic extraction, hash code quality preservation, and other key technologies. We then present a detailed evaluation schema with results on several popular datasets, followed by a discussion of practical applications and open-source tools for implementation. Finally, we conclude by discussing key challenges and future research directions, including the integration of deep text hashing with large language models to further advance the field. The project for this survey can be accessed at https://github.com/hly1998/DeepTextHashing.
\end{abstract}

%%
%% The code below is generated by the tool at http://dl.acm.org/ccs.cfm.
%% Please copy and paste the code instead of the example below.
%%


%%
%% Keywords. The author(s) should pick words that accurately describe
%% the work being presented. Separate the keywords with commas.

% \received{20 February 2007}
% \received[revised]{12 March 2009}
% \received[accepted]{5 June 2009}

%%
%% This command processes the author and affiliation and title
%% information and builds the first part of the formatted document.
\maketitle

\section{Introduction}

Online content is growing exponentially. Estimates suggest that zettabytes of new text data emerge annually from diverse platforms such as web pages, social media, scientific literature, and enterprise repositories \cite{loubnaissn}. Effectively retrieving relevant information from this massive data deluge is critical. It powers essential applications, including search engines \cite{cordeiro2023lesse,zhuang2024setwise}, question answering systems \cite{zhu2021retrieving,kratzwald2018adaptive,wang2019multi,nie2020dc}, and digital libraries \cite{li2019problems,malakhov2023developing}. Traditional keyword methods often fail to capture nuanced meaning within this vast text volume. Therefore, semantic text retrieval, which focuses on understanding the underlying intent and relationships, has become essential \cite{guo2022semantic,zhao2024dense}. Addressing the twin needs for semantic understanding and large-scale efficiency involves mainly combining two techniques: representation learning \cite{bengio2013representation,zhong2016overview,ericsson2022self} to encode semantics in vectors \cite{mikolov2013efficient,le2014distributed}, and approximate nearest neighbor (ANN) search \cite{abbasifard2014survey,li2019approximate,peng2023efficient} for rapid retrieval. Within the ANN field, hashing techniques \cite{salakhutdinov2009semantic,ge2013optimized,dolatshah2015ball,malkov2018efficient} are notably prominent. Their significant advantages in computational speed and storage efficiency make them particularly well-suited for handling large-scale datasets.

Among the earliest and most influential techniques for text data hashing is Locality Sensitive Hashing (LSH) \cite{charikar2002similarity,dasgupta2011fast,kulis2011kernelized}. The purpose of LSH is to map the original data into multiple hash buckets, ensuring that the closer the original distance between objects, the higher the likelihood of them falling into the same hash bucket. However, LSH is data-independent, which means that in complex textual data, it often leads to a significant decline in recall performance due to information loss. To address this issue, researchers proposed data-dependent learning-to-hash methods \cite{weiss2008spectral,salakhutdinov2009semantic,zhang2010self,liu2011hashing,wang2013semantic,liu2012supervised,kulis2009learning,wang2013learning,wang2017survey}. These approaches are designed to preserve the structural similarity of the original data features, drawing inspiration from techniques such as spectral clustering \cite{ng2001spectral}, latent semantic indexing \cite{deerwester1990indexing}, and other related technologies. 

With the success of deep learning across various fields \cite{lecun2015deep,zhuang2023efficiently,cheng2024towards}, researchers began considering deep models as hash functions to achieve more accurate data representations. At this point, differences in data types led to variations in deep learning models and methods, resulting in various branches, such as deep image hashing for image retrieval \cite{yuan2020central,fan2020deep,wang2023deep,he2024bit,he2024one}, deep text hashing for text retrieval \cite{chaidaroon2017variational,shen2018nash,hansen2019unsupervised,hansen2020unsupervised,ye2020unsupervised}, deep cross-modal hashing for text-image retrieval \cite{luo2018asymmetric,tan2022teacher,zhang2017semi,yang2020nonlinear,wang2021high}, and deep multi-modal hashing for multi-model retrieval \cite{lu2019efficient,zheng2019fast,zhu2020deep,tan2022bit}, etc. Among them, deep text hashing faces distinct challenges due to differences in data characteristics. Generally, the semantic information of categories in image data is relatively easier to extract, whereas capturing the semantics of textual data proves more challenging, necessitating a more intricate representation space. However, the limited capacity of the Hamming space exacerbates the difficulty of representing textual data within it. Although deep cross-modal hashing and deep multi-modal hashing incorporate textual elements, the text in these cases often serves as a simple supplement or description of the image data. In contrast, deep text hashing focuses on more complex and intricate textual content.

This survey focuses on deep text hashing due to its significant challenges and the lack of comprehensive reviews in this area. While pioneering surveys have explored hashing, their focus has largely been on general learning to hash~\cite{wang2015learning,wang2017survey,chi2017hashing} image-based hashing~\cite{rodrigues2020deep,luo2023survey}, or cross-modal hashing~\cite{zhu2023multi}. To the best of our knowledge, this is the first survey dedicated specifically to deep text hashing. Besides, a key distinction of our work lies in the novel, component-based taxonomy we propose. Our framework moves beyond the traditional supervised/unsupervised dichotomy and instead analyzes models by their core components. We first examine \textbf{Semantic Extraction}, which is how models capture meaning from text. This includes various strategies like reconstruction, pseudo-similarity, and learning from categories or relevance. Next, we analyze \textbf{Hash Code Quality Preservation}. This part focuses on the properties of the binary output, such as code compactness, its distribution balance, and low quantization error. Finally, our framework covers \textbf{Other Technologies} that provide further enhancements, including robustness optimization, adaptation to indexing, and better gradient propagation. We believe this new taxonomy offers significant advantages. By focusing on the underlying mechanisms rather than just data supervision, it provides deeper technical insights, is more aligned with the primary challenges of the task, and offers a more modular framework to guide and inspire future research. Furthermore, this survey not only focuses on the hashing models but also considers the entire search process and practical application methods.

The organization of this work is structured as follows. We first give notations and definitions we will use later and then introduce the background of the deep text hashing in Section \ref{sec:back}. Following that, we introduce the representative deep text hashing approaches and detail them from different aspects in Section \ref{sec:semantic_extraction}, Section \ref{sec:code_quality}, and Section \ref{sec:other_tech}. Section \ref{sec:performance} introduces the evaluation schema, datasets, and performance results. Section \ref{sec:application} presents the applications of deep text hashing and some open-source supports. Finally, we conclude the paper and give some future directions in Section \ref{sec:future}.

\section{Background}
\label{sec:back}

\begin{table}[tbp]
\caption{NOTATION} \label{tab:used_notation} 
\begin{center}

\begin{tabular}{cl}
\toprule
Notation & Description \\
\hline
$\boldsymbol{X}$ & The dataset\\
$\boldsymbol{q}$ & A query of text\\
$\boldsymbol{x}, \boldsymbol{x_i}$ & The arbitrary text and the $i$-th text in dataset \\
$\boldsymbol{h}, \boldsymbol{h_i}$ & The hash code of 
$\boldsymbol{x}$ and $\boldsymbol{x_i}$\\
$\boldsymbol{z}, \boldsymbol{z_i}$ & The latent representation of 
$\boldsymbol{x}$ and $\boldsymbol{x_i}$\\
$\boldsymbol{y}, \boldsymbol{y_i}$ & The semantic label of 
$\boldsymbol{x}$ and $\boldsymbol{x_i}$\\
$\boldsymbol{\tilde{h}}$ & The binary-like representation of $\boldsymbol{x}$ \\
$\boldsymbol{w}_i$ & The $i$-th word in the corpora\\
$\boldsymbol{I}$ & The identity matrix\\
$\rm{dist}(\cdot,\cdot) $ & The Euclidean distance between two vectors. \\
$\rm{dist}_H(\cdot,\cdot)$ & The Hamming distance between two vectors. \\
$D_{KL}(\cdot \| \cdot)$ & The KL divergence between two variables. \\
$\rm sign(\cdot)$ & The signum function. \\
$N$ & The number of texts in $\boldsymbol{X}$\\
$V$ & The number of vocabulary words in corpora\\
$K$ & The number of returned texts in KNN search\\
$L$ & The number of labels/categories in dataset\\
$b$ & The hash code length\\
$r$ & Search radius in Hamming space\\
$\phi$, $\theta$, $\boldsymbol{W}_{*}$, $\boldsymbol{b}_{*}$ & The parameters of model\\
\bottomrule
\end{tabular}
\label{tab:notation}
\end{center}
\end{table}

For the notations, we use bold lowercase letters to represent vectors and bold uppercase letters to denote matrices. We provide formal notations and key concepts in Table \ref{tab:notation} for the sake of clarity.

\subsection{Nearest Neighbor Search}

Given a $d$-dimension query $\boldsymbol{q} \in \mathbb{R}^d$, the exact nearest neighbor search aims to find the item $\rm{NN}(\boldsymbol{q})$ from a set of $N$ items $\boldsymbol{X}=\{\boldsymbol{x_1}, \boldsymbol{x_2},...,\boldsymbol{x}_N\}, \boldsymbol{x_i} \in \mathbb{R}^d$ so that:
\begin{align}
\rm{NN}(\boldsymbol{q}) = \rm{argmin}_{\boldsymbol{x} \in \boldsymbol{X}} \rm{dist}(\boldsymbol{q},\boldsymbol{x}),
\end{align}
where $\rm{dist}(\boldsymbol{q},\boldsymbol{x})$ is a distance computed between $\boldsymbol{q}$ and $\boldsymbol{x}$. The $K$-nearest Neighbor Search (KNNs) can be accordingly defined, where we need to find $K$ nearest neighbors. Another method frequently considered in deep hashing is the Point Location in Equal Balls (PLEB) search, which aims to find all texts within a fixed Hamming distance of a query. There exist efficient algorithms (e.g., KD trees \cite{zhou2008real}) for nearest neighbor search when the dimension $d$ is small. 

However, in large-scale high-dimensional cases, it turns out that the problem becomes hard and most algorithms even take higher computational costs than the naive solution, i.e., the linear scan. Instead of nearest neighbor search, a large number of practical techniques \cite{li2022deep} have been proposed for Approximate Nearest Neighbors (ANN) search, which relaxes the guarantee of accuracy for efficiency by evaluating a small subset of $\boldsymbol{X}$. To achieve this goal, the ANN methods typically first construct an index structure to organize data items and then execute a querying search algorithm based on this index to retrieve the nearest neighbor results for the given queries. ANN algorithms primarily encompass three categories: hashing-based methods \cite{salakhutdinov2009semantic,wang2012semi}, product quantization methods \cite{ge2013optimized,xu2018online,chen2020differentiable,liu2020online}, and graph-based methods \cite{malkov2018efficient,zhao2020song,zhang2022two,weng2024fast}. These algorithms have significantly enhanced searching efficiency while maintaining relatively high accuracy, making them widely utilized in the industry. Among these, researchers have studied hashing-based algorithms the longest and most extensively due to their substantial potential to improve computational efficiency and reduce memory costs.

\subsection{Fundamentals of deep text hashing}

Deep text hashing employs deep neural networks to efficiently and accurately retrieve textual data. Its goal is to learn a deep neural networks hashing function $f(\boldsymbol{x}): \boldsymbol{x} \rightarrow \boldsymbol{h} $, which projects high-dimensional data $\boldsymbol{x} \in \mathbb{R}^d$ into low-dimensional binary vector $\boldsymbol{h} \in \{-1, 1\}^b$ \footnote{Note that some works represent hash codes using $\{0,1\}$, while others use $\{-1,1\}$. In this paper, we do not differentiate between them, as they can be easily converted.}. We term the representation $\boldsymbol{h}$ as hash code, and $b$ is the code length. Besides, $\boldsymbol{x}$ can represent either the raw text data or a pre-computed feature vector, depending on whether a feature extractor (e.g., BERT \cite{kenton2019bert}) is applied beforehand. For the sake of simplicity in the subsequent discussion, we will refer to $\boldsymbol{x}$ as the text, but it is important to note that most deep text hashing models can also take features as direct input. Then, the Hamming distance is used as the distance metric to evaluate the similarities between two hash codes, which is defined as the number of different bits between two codes:
\begin{equation}
    \rm{dist}_{H}(\boldsymbol{h_1},\boldsymbol{h_2}) = POPCOUNT(\boldsymbol{h_1} \, XOR \, \boldsymbol{h_2}).
\end{equation}

This computation process is efficient by utilizing the $\rm XOR$ and $\rm POPCOUNT$ instructions. Meanwhile, deep text hashing models often take into account several optimization objectives, including: (1) Few-bit code: the hash code length \( b \) should be minimized; (2) Code balance: the hash codes should be evenly distributed across the Hamming space; and (3) Low quantization error: reducing information loss when converting real-valued representations into binary hash codes. These characteristics will be elaborated upon in detail in Section~\ref{sec:code_quality}.

\begin{figure}[t]
\centering
\includegraphics[width=1\textwidth]{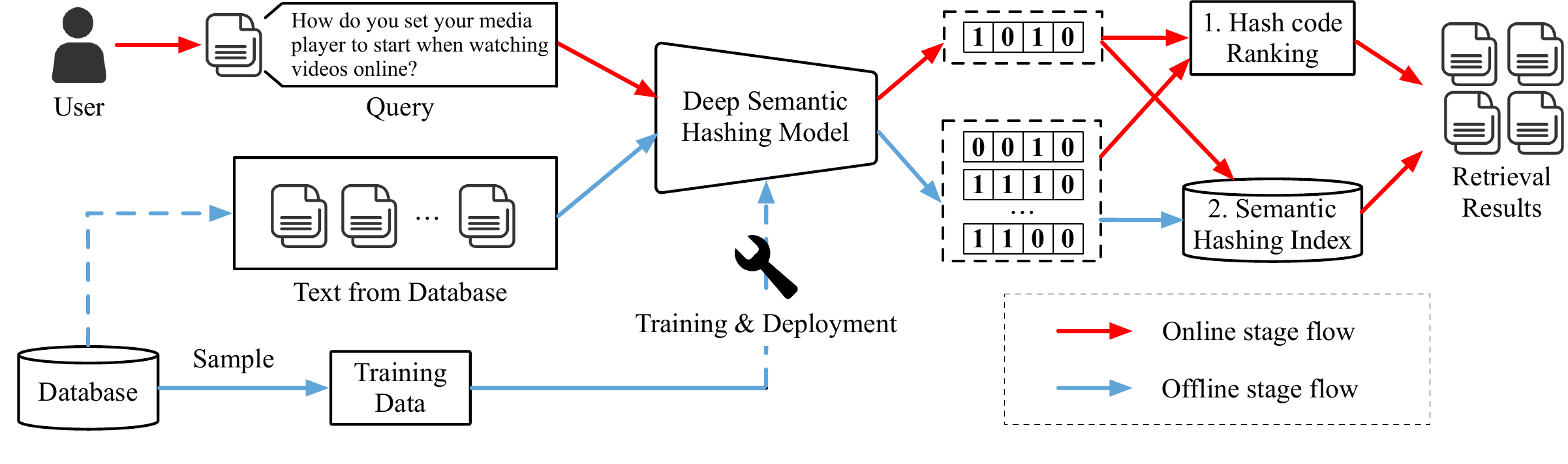}
\caption{The basic search framework for deep text hashing. }
\label{fig:semantic_hashing_search_pipline}
\end{figure}

\subsection{Search with Hash Code}
\label{sec:search_with_hash_code}

The basic search frameworks of deep text hashing are shown in Fig. \ref{fig:semantic_hashing_search_pipline}. At the offline stage, it samples text data from the database and trains a deep text hashing model. Subsequently, it deploys the trained deep text hashing model, maps all texts in the database to hash codes, and constructs an index \cite{norouzi2012fast}. At the online stage, given a query from a user or other systems, the deep text hashing model maps it to a binary vector using the trained deep text hashing model and then performs a rapid search using hashing code ranking or hash table lookup. 

Hash code ranking is suitable for use with relatively small datasets. It performs an exhaustive search: when a query comes, it computes the Hamming distance between the query and each text in the searching dataset, then it selects the points with relatively smaller Hamming distances as the candidates for nearest neighbor search. After that, a re-ranking process by the original features is usually followed to obtain the final nearest neighbor. Compared to computations in the original space, using hash codes for distance calculations is more efficient and requires less storage. For example, suppose the texts in the database are mapped to $d_f$ dimensional double-precision float vectors or $d_b$ bits hash codes for representations. Note that most modern CPUs can fuse the multiplication and addition as a single-cycle operation. Then, the total number of operations for the distance calculation is $d_f \times N \times C_{f}$. $C_{f}$ represents the computational cost of floating-point operations. The total number of operations for the hamming distance calculation is $d_b \times N \times 2 \times C_{b}$. $C_{b}$ denotes the computational cost of binary operations. Using the calculation method from \cite{rastegari2016xnor}, we assume the current generation of CPUs can perform $64$ binary operations in one clock cycle of the CPU. The speedup can be computed by:
\begin{equation}
    S = \frac{d_f \times N \times 64 \times C_{b}}{d_b \times N \times 2 \times C_{b}} = \frac{32d_f}{d_b},
\end{equation}
where \(d_f\) is much greater than \(d_b\). Besides, in terms of storage, a compression efficiency of \(\frac{64 d_f}{d_b}\) can be achieved.

\begin{figure}[t]
\centering
\includegraphics[width=0.8\textwidth]{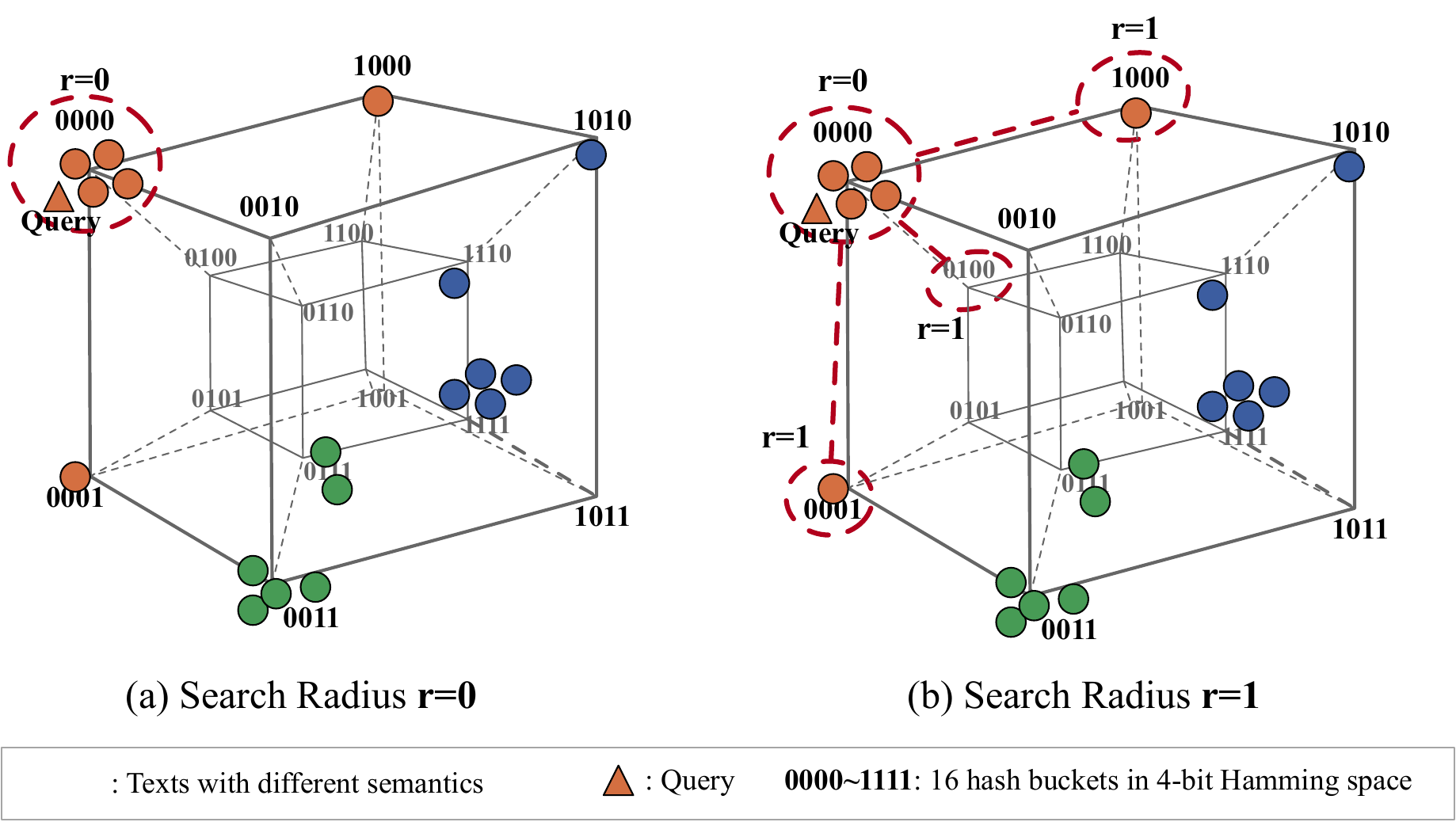}
\caption{An illustration of Hamming ball search during KNN search ($K=5$): (a) When the search distance \(r=0\), it accesses the same hash bucket as the query text, returning one text. Since the required number is not met, the search range is expanded. (b) Expanding the search distance to \(r=1\), it returns six texts, satisfying the required number, and the search stops.}
\label{fig:ball_search}
\end{figure}

However, in practical applications, $N$ typically exceeds tens of millions, making the computation of hash code ranking challenging. Hash table lookup aims to decrease the number of distance calculations to speed up searches. The structure of the hash table contains various buckets, each of which is indicated by one separate hash code. Each text is associated with a hash bucket that shares the same hash code. As suggested in \cite{salakhutdinov2009semantic,grauman2013learning}, hamming ball exploration is one way to find the related texts in deep hashing. As shown in Fig.~\ref{fig:ball_search}, it generally explores the Hamming space to acquire candidate texts by gradually increasing the search radius $r$ around the query. For hash codes of $b$ bits, the number of hash buckets to be examined can be formulated as:
\begin{equation}
    lookups(r, b) = \sum_{k=0}^{r}\tbinom{b}{k}.
\label{eq:look_up_bucket}
\end{equation}

When $b$ and $r$ are within a reasonable range, e.g., $b = 64$ and $r \leq 2$, this method can ensure the efficiency of the search. However, a major drawback of this method is that the search efficiency will break down for long code length and large search radius since the lookups will explode with the code length $b$ and $r$ increases. A promising way in deep text hashing is using multi-index hashing \cite{norouzi2012fast}. It splits the hash codes into $m$ substrings and builds hash tables for each substring. This way can significantly reduce the lookups and the total number of lookups is given by:
\begin{equation}
    lookups(r, b, m) \leq m2^{H(\frac{r}{b})\frac{b}{m}},
\label{eq:multi_bucket}
\end{equation}
where $H(a) = -alog_2a-(1-x)log_2(1-a)$ is the entropy of a Bernoulli distribution with probability $a$. Along this line, several improved multi-index hashing methods \cite{wang2015multi,miao2018approximate,lai2018improved} have been proposed.

Furthermore, since deep text hashing is essentially a method for obtaining binary representations, many other indexing methods can be employed \cite{malkov2018efficient,zhao2020song,zhang2022two,weng2024fast} as long as they can use Hamming distance for distance calculation. Regardless of the method employed, practical applications generally include a ranking step following the identification of the candidate set. The retrieved nearest neighbor candidates are ranked based on distances calculated using the original features to ascertain the nearest neighbors.

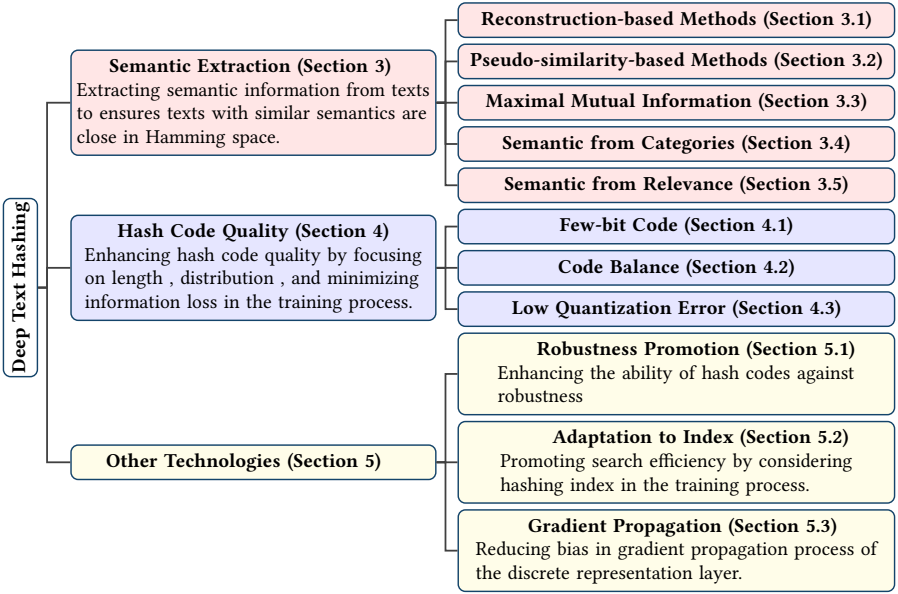
\begin{figure}[t!]
			\centering
			\resizebox{12cm}{!}{
				\begin{forest}
					forked edges,
					for tree={
						grow=east,
						reversed=true,
						anchor=base west,
						parent anchor=east,
						child anchor=west,
						base=left,
						font=\large,
						rectangle,
						draw=hidden-draw,
						rounded corners,
						align=left,
						text centered,
						minimum width=4em,
						edge+={darkgray, line width=1pt},
						s sep=3pt,
						inner xsep=2pt,
						inner ysep=3pt,
						line width=0.8pt,
						ver/.style={rotate=90, child anchor=north, parent anchor=south, anchor=center},
					},
					where level=1{text width=14em,font=\large,}{},
					where level=2{text width=15em,font=\large,}{},
					where level=3{text width=17em,font=\large,}{},
					[
					\textbf{Deep Text Hashing}, ver
					[
					{     \ \ \ } \textbf{Semantic Extraction (Section \ref{sec:semantic_extraction})} \\ Extracting semantic information from texts \\ to ensures texts with similar semantics are \\ close in Hamming space., fill=red!10, text width=20em
					[ 
					\textbf{Reconstruction-based Methods (Section \ref{sec:rec})}, fill=red!10, text width=24em    
					]
					[ 
					\textbf{Pseudo-similarity-based Methods (Section \ref{sec:pseu})}, fill=red!10, text width=24em
					]
					[
					\textbf{Maximal Mutual Information (Section \ref{sec:maximal})}, fill=red!10, text width=24em
					]
                    [
					\textbf{Semantic from Categories (Section \ref{sec:semantic_label})}, fill=red!10, text width=24em
					]
                    [
					\textbf{Semantic from Relevance (Section \ref{sec:semantic_relevance})}, fill=red!10, text width=24em
					]
					]
					[
					{     \ \ \ } \textbf{Hash Code Quality (Section \ref{sec:code_quality})} \\ Enhancing hash code quality by focusing \\ on length {,} distribution {,} and minimizing \\ information loss in the training process., fill=blue!10, text width=20em
					[
					\textbf{Few-bit Code (Section \ref{sec:few-bit})}, fill=blue!10, text width=24em
					]
					[
					\textbf{Code Balance (Section \ref{sec:code_balance})}, fill=blue!10, text width=24em
					]
					[
					\textbf{Low Quantization Error (Section \ref{sec:minimize_quan})}, fill=blue!10, text width=24em
					]
					]
					[
					\textbf{Other Technologies (Section \ref{sec:other_tech})} {     \ \ } , fill=yellow!10, text width=20em
					[
					{     \ \ \ \ } \textbf{Robustness Promotion (Section \ref{sec:robust})} \\ Enhancing the ability of hash codes against \\ robustness, fill=yellow!10, text width=24em
					]
					[
					{     \ \ \ \ \ \ } \textbf{Adaptation to Index (Section \ref{sec:adapt})} \\ Promoting search efficiency by considering \\ hashing index in the training process., fill=yellow!10, text width=24em
					]
                    [
					{     \ \ \ \ \ \ } \textbf{Gradient Propagation (Section \ref{sec:gradient})} \\ Reducing bias in gradient propagation process of \\ the discrete representation layer., fill=yellow!10, text width=24em
					]
					]
					]
				\end{forest}
			}
			\caption{The main content flow and summary of how to construct a deep text hashing model.}
			\label{summary}
		\end{figure}

Next, we delve into the intricate specifics of deep text hashing models. Each deep text hashing model incorporates various optimization and improvement methods. As shown in Fig.~\ref{summary}, constructing a deep text hashing model involves considering two primary aspects. One is how to extract semantic information, which includes multiple solutions, such as reconstruction-based methods, pseudo-similarity-based methods, maximal mutual information, learning semantic from labels, and learning semantic from relevance. The other is how to enhance the quality of the hash code, focusing on code length, hash code distribution, and minimizing information loss during the process. Additionally, other technologies aim to bolster the performance of deep text hashing models by addressing text noise, index, and bias in gradient propagation. Notably, a deep text hashing model may simultaneously employ multiple strategies, reflecting the flexibility of this field.

% \section{deep text hashing}
% \label{sec:deep_semamtic_hashing}

% In this section, we delve into the specific details of deep text hashing models. As illustrated in Fig~\ref{}, 
% As illustrated in Fig~\ref{}, the deep text hashing model primarily comprises three components: (1) a deep neural network architecture to map the raw features of the text to the hash space; (2) a mechanism to preserve the similarity information; (3) some method to promote the quality of hash code. We first introduce the related technology of the three components, followed by other techniques that enhance the performance of the deep text hashing model.

% In this section, we delve into the intricate specifics of deep text hashing models. As depicted in Figure~\ref{summary}, constructing a deep hashing model primarily entails consideration from three perspectives: (1) how to extract semantic information, (2) how to enhance the quality of the hash code, and (3) other technology to bolster the performance of deep text hashing models. The following sections will provide a detailed introduction to each part.

\tikzstyle{my-box}=[
		rectangle,
		draw=hidden-draw,
		rounded corners,
		text opacity=1,
		minimum height=1.5em,
		minimum width=5em,
		inner sep=2pt,
		align=center,
		fill opacity=.5,
		line width=0.8pt,
		]
		\tikzstyle{leaf}=[my-box, minimum height=1.5em,
		fill=hidden-pink!80, text=black, align=left,font=\large,
		inner xsep=2pt,
		inner ysep=4pt,
		line width=0.8pt,
		]

\section{Semantic Extraction}
\label{sec:semantic_extraction}

A fundamental requirement of deep text hashing is extracting semantic information from texts, which ensures that texts with similar semantics are close in Hamming space. In several surveys about image-based hashing \cite{luo2023survey} and cross-modal hashing \cite{zhu2023multi}, models are typically categorized into supervised and unsupervised frameworks. However, In terms of model design, deep text hashing models are more flexible compared to image-based hashing and cross-modal hashing. By adding modules or objectives, unsupervised models can be easily transformed into supervised ones. Thus, we do not distinguish significantly between supervised and unsupervised learning perspectives. We categorize these methods into five classes according to how semantic information is extracted, i.e., reconstruction-based, pseudo-similarity-based, maximal mutual information, semantics from categories, and semantics from relevance methods.

\subsection{Reconstruction-based Methods}
\label{sec:rec}

Reconstruction-based methods aim to harness reconstruction objective functions to learn the semantic information of text, thereby ensuring the retention of relevant information. In this process, some classical architectures such as auto-encoders (AE) \cite{hinton2006reducing} or variational auto-encoders (VAE) \cite{kingma2013auto} are employed, with specific enhancements made to accommodate the generation of hash codes.

AE is a simple yet flexible framework with a reconstruction objective function, thus various deep text hashing models have adopted it as the basic structure. The AE framework is defined as an encoding function that maps a text $\boldsymbol{x}_i$ into a latent representation $\boldsymbol{z}_i$, and a decoding function reconstructs the latent representation $\boldsymbol{z}_i$ to $\hat{\boldsymbol{x}}_i$. The reconstruction objective function can be written as the word-wise negative log-likelihood, averaged across all texts, as follows:
\begin{equation}
    \mathcal{L}_{AE} = \mathbb{E}_{\boldsymbol{x}_i \sim D_{\boldsymbol{X}}} [ \frac{1}{V} \sum_{j=1}^V - \log p(\hat{x}_{i,j} = x_{i,j}) ],
\label{eq:ae_loss}
\end{equation}
where \(\boldsymbol{x}_i\) is represented using a bag-of-words representation, \(x_{i,j}\) denotes the value of the \(j\)-th dimension of \(\boldsymbol{x}_i\), and $V$ is the number of words in the corpora. Some methods employ Eq.(\ref{eq:ae_loss}) to capture the semantic information of text and directly use the latent representation $\boldsymbol{z}_i$ as the feature representation of the hash code. For example, Deep Spectral Hashing (DSH) \cite{chen2016document} first employs Eq.(\ref{eq:ae_loss}) as a learning objective in an auto-encoder to obtain the hidden features of texts. These hidden features are then utilized as textual characteristics, and Spectral Hashing \cite{weiss2008spectral} is employed to derive the compact hash code. The Denoising Adversarial Binary Autoencoder (DABA) \cite{doan2020efficient} and denoiSing Memory-bAsed Semantic Hashing (SMASH) \cite{he2023efficient} applies the same idea to obtain textual features, and DABA proposes to use more sophisticated RNN \cite{kiros2015skip} and CNN \cite{zhang2017deconvolutional} structures as the encoder and decoder, aiming to learn more intricate text representations.

Based on the reconstruction learning objective, some deep text hashing models \cite{yu2015understanding,guo2022intra} have improved the basic auto-encoder framework to meet the requirements of deep text hashing. Stacked Auto-Encoders (SH-SAE) \cite{yu2015understanding} introduces a three-layer stacked auto-encoders for deep text hashing. It assigns specific tasks to each auto-encoder. The first auto-encoder takes the original text feature as input and obtains a fundamental hidden feature. The second auto-encoder uses the learned hidden feature as input and captures more abstract features, reducing the feature dimension as a transitional layer. The third auto-encoder adds Gaussian noise to render the learned abstract features as binary as possible. Besides, additional components can be integrated due to the auto-encoder framework's flexibility. Intra-category aware Hierarchical Supervised Document Hashing (IHDE) \cite{guo2022intra} incorporates an Intra-Category Component after the encoders, which aims to infuse supplementary information from a reference text representation into the query text representation.

\begin{figure}[t]
\centering
\includegraphics[width=0.8\textwidth]{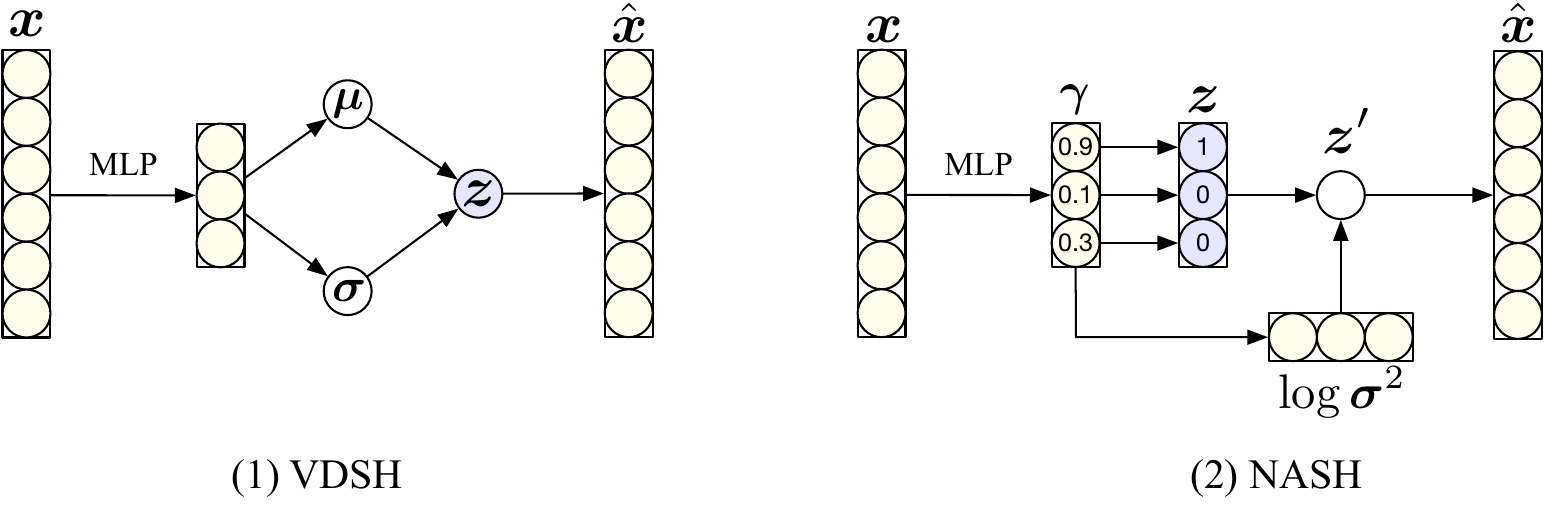}
\caption{An illustration of two mainstream deep text hashing frameworks employing the VAE architecture. (1) VDSH \cite{kingma2013auto} assumes the hash codes follow a Gaussian distribution, while (2) NASH \cite{shen2018nash} assumes the hash codes follow a Bernoulli distribution.}
\label{fig:VDSH_and_NASH}
\end{figure}

In addition to directly using the reconstruction learning objective to obtain latent representations of texts, many deep text hashing models incorporate prior distribution controls into the latent representations during this process like the VAE \cite{kingma2013auto} framework. This approach makes it more suitable for generating binary hash codes and ensures the preservation of semantic information. Variational deep text hashing (VDSH) \cite{chaidaroon2017variational} is the pioneer in employing such a concept to learn hash codes (Refer to Fig.~\ref{fig:VDSH_and_NASH} (1)). By applying the variational inference principle \cite{wainwright2008graphical}, it uses the following tractable lower bound of the text log-likelihood:
\begin{equation}
    \mathcal{L}=\mathbb{E}_{q_{\phi}(\boldsymbol{z}|\boldsymbol{x})}[\log q_{\theta}\left(\boldsymbol{x} \mid \boldsymbol{z}\right)]-D_{K L}(q_{\phi}(\boldsymbol{z} \mid \boldsymbol{x}) \| p(\boldsymbol{z})),
\label{vae_loss}
\end{equation}
where $D_{K L}( q_{\phi}(\boldsymbol{z}\mid\boldsymbol{x}) \| p(\boldsymbol{z}) )$ is the Kullback-Leibler (KL) divergence between the approximate
posterior distribution $q_{\phi}(\boldsymbol{z} \mid \boldsymbol{x})$ and the Gaussian prior $p(\boldsymbol{z})$. VDSH introduces a feedforward neural network encoder to condense text into compact continuous latent representation $\boldsymbol{z}$ as follows:
\begin{equation}
\begin{aligned}
    \boldsymbol{t} &= \rm{ReLU} ( \boldsymbol{W}_2  \rm{ReLU}(\boldsymbol{W}_1 \boldsymbol{x} + \boldsymbol{b}_1) + \boldsymbol{b}_2 ), \\
    \boldsymbol{\mu} &= \boldsymbol{W}_3 \boldsymbol{t}_2 + \boldsymbol{b}_3, \\
    \log \boldsymbol{\sigma} &= \boldsymbol{W}_4 \boldsymbol{t}_2 + \boldsymbol{b}_4, \\
    \boldsymbol{z} &\sim \mathcal{N}(\boldsymbol{\mu},\rm{diag}(\boldsymbol{\sigma}^2)),
\end{aligned}
\end{equation}
where $\mathcal{N}$ denotes Gaussian distribution. Similar to VAE, VSDH use reparameterization trick to turn the stochastic layer of $\boldsymbol{z}$ to be deterministic. Furthermore, VDSH employs a softmax decoder \( \sum_{i=1}^N p_{\theta}(\boldsymbol{w_i}|\boldsymbol{z}) \) to reconstruct orignal texts by independently generating words $\boldsymbol{w}_i$:
% \begin{equation}
% \begin{aligned}
%     p(\boldsymbol{w}_i|\boldsymbol{z}) &= \frac{exp(-\boldsymbol{z}^T\boldsymbol{W}_5\boldsymbol{w}_i+\boldsymbol{b}_5)}{\sum_{k=1}^Vexp(-\boldsymbol{z}^T\boldsymbol{W}_5\boldsymbol{w}_k+\boldsymbol{b}_5)}, \\
%     p(\boldsymbol{x}|\boldsymbol{z}) &= \prod_{i=1}^N P(\boldsymbol{w}_i|\boldsymbol{z}).
% \end{aligned}
% \end{equation}
\begin{equation}
\begin{aligned}
    p(\boldsymbol{w}_i|\boldsymbol{z}) &= \frac{exp(-\boldsymbol{z}^T\boldsymbol{W}_5\boldsymbol{w}_i+\boldsymbol{b}_5)}{\sum_{k=1}^Vexp(-\boldsymbol{z}^T\boldsymbol{W}_5\boldsymbol{w}_k+\boldsymbol{b}_5)}.
\end{aligned}
\end{equation}

In the training stage, \( \boldsymbol{z} \) is not directly converted into a binary representation due to the non-differentiability of the discrete representation layer. VDSH employs the median method in the inference stage to get the binary hash code. The median method first computes the median value of the latent representations $\{\boldsymbol{z}_i\}_{i=0}^N$ in the training set as the threshold. Then it sets the $h_{i,j}$ to $1$ if the $j$-th dimension of $h_i$ is larger than the threshold. Otherwise, it set the $\boldsymbol{h}_{i,j}$ to $-1$ or $0$. Several deep text hashing methods \cite{chaidaroon2018deep, xuan2019variational, xuan2020conditional, chaidaroon2020node2hash, zhang2024document} employ this foundational structure and have made some structural improvements.

An avenue of expansion involves considering alternative prior distributions to mitigate information loss's impact or incorporate more intricate relationship modeling. For example, Neural Architecture for Semantic Hashing (NASH) \cite{shen2018nash} adopts a similar variational autoencoder architecture as VDSH but models the hash codes as Bernoulli latent-variable (Refer to Fig.~\ref{fig:VDSH_and_NASH}
(2)). That means NASH casts the latent variable $\boldsymbol{z}$ as a binary vector and assumes a multivariate Bernoulli prior on $\boldsymbol{z}$: $p(\boldsymbol{z}) \sim {\rm Bernoulli}(\boldsymbol{\gamma})=\prod_{i=1}^b\gamma_i^{z_i}(1-\gamma_i)^{(1-z_i)}$. Here, $\gamma_i \in [0,1]$ is the i-th dimension of $\boldsymbol{\gamma}$. Based on the Bernoulli distribution prior $p(\boldsymbol{z})$, assume the encoding distribution $q_{\phi}(\boldsymbol{z}|\boldsymbol{x}) = {\rm Bernoulli} (\sigma(g_{\phi}(\boldsymbol{x})))$, where $\sigma(\cdot)$ is the sigmoid function. Then the Kullback-Leibler divergence $D_{KL}(q_{\phi}(\boldsymbol{z}|\boldsymbol{x})||p(\boldsymbol{z}))$ in Eq. \ref{vae_loss} can be calculated as follows:
\begin{equation}
D_{KL} = g_{\phi}(\boldsymbol{x}) \log \frac{g_{\phi}(\boldsymbol{x})}{\boldsymbol{\gamma}} + (1-g_{\phi}(\boldsymbol{x})) \log \frac{1-g_{\phi}(\boldsymbol{x})}{1-\boldsymbol{\gamma}}.
\end{equation}

The advantage of NASH lies in directly generating binary hash codes during the training process, thereby circumventing quantization loss and generating code-balanced hash codes. Through reparameterization techniques, there are two ways to obtain binary hash codes in NASH: deterministically or stochastically. They can be represented as follows:
\begin{equation}
    \boldsymbol{h} = \frac{{\rm sign}(\sigma(\boldsymbol{z}-\boldsymbol{a}))+1}{2},
\label{eq:sign_inference}
\end{equation}
where $\sigma(\cdot)$ is the sigmoid function. In the deterministic case, $\boldsymbol{a}$ is set to $0.5$, which can be understood as sampling from the Bernoulli distribution with the hyperparameter \(\gamma\) set to $0.5$ for each representation dimension, thus generating the hash code deterministically. In the stochastic case, \(\boldsymbol{a} \sim \text{Uniform}(0,1)\) is the result of sampling from a uniform distribution. During the training phase, both deterministic and stochastic methods can be employed. Then, the deterministic approach is typically used to ensure the consistency of the output hash codes during the inference phase. A lot of deep text hashing methods \cite{hansen2019unsupervised,mena2019binary,hansen2020unsupervised,ye2020unsupervised,dadaneh2020pairwise,nanculef2021self,hansen2021unsupervised,chen2023exploiting,huang2025confusing} have embraced the concept of sampling from the Bernoulli distribution. 

% Doc2Hash \cite{zhang2019doc2hash} argues that the Straight Through Estimator (STE) employed in NASH Bernoulli stochastic layer in NASH is non-differentiable. Although the Straight Through Estimator (STE) is employed to propagate gradients by bypassing the stochastic layer during backpropagation, it remains a biased estimator, introducing high-variance biased gradients of objectives during training. To solve this problem, Doc2Hash first assumes a multi-variate categorical prior on the latent variable $z$:

Some deep text hashing models consider additional priors to model more complex semantic information. GMSH/BMSH \cite{dong2019document} build upon the ideas of VDSH and NASH, but they tend to generate high-quality hashing codes by imposing mixture priors on generative models. This means replacing the original Gaussian or Bernoulli distribution priors with:
\begin{equation}
p(z) = \sum_{k=1}^K \pi_k \cdot \mathcal{N}(\mu_k, diag(\sigma_k^2)) \ \text{or} \
p(z) = \sum_{k=1}^K \pi_k \cdot {\rm Bernoulli}(\gamma_k),
\end{equation}
where $K$ is the number of mixture components, $\pi_k$ is the probability of choosing the $k$-th component. While Doc2Hash \cite{zhang2019doc2hash} assumes a multi-variate categorical prior ${\rm Cat}(\boldsymbol{\pi})$ on the latent variable $\boldsymbol{z}$:
\begin{equation}
p(\boldsymbol{z}) \sim {\rm Cat}(\boldsymbol{\pi}) = \prod_i^l I(\mathcal{C}(\boldsymbol{z})=l)\boldsymbol{\pi}_{ik},
\end{equation}
where \(\mathcal{C}(\boldsymbol{z})\) is used to obtain the category of \(\boldsymbol{z}\) and $\boldsymbol{\pi}_{ik}$ is the $l$-th class probability on $i$-th component of parameters $\boldsymbol{\pi}$. Then, the posterior distribution approximated by the encoding network is constrained in the form of $q_{\phi}(\boldsymbol{z}|\boldsymbol{x})= {\rm Cat} (\tilde{\boldsymbol{z}})$, where $\tilde{\boldsymbol{z}}$ is the output of the encoding network. The variational lower bound adopts the form of Eq.(\ref{vae_loss}), while the Kullback-Leibler divergence term $D_{KL}(q_{\phi}(\boldsymbol{z}|\boldsymbol{x}) \| p(\boldsymbol{z}))$ can be calculated as follows:
\begin{equation}
D_{KL}(q_{\phi}(\boldsymbol{z}|\boldsymbol{x})||p(\boldsymbol{z})) =\sum_i \sum_k g_\phi^{i k}(\boldsymbol{x}) \log \frac{g_\phi^{i k}(\boldsymbol{x})}{\boldsymbol{\pi}_{i k}}. 
\end{equation}

CorrSH \cite{zheng2020generative} adopts the Boltzmann-machine distribution \cite{ackley1985learning} as a variational posterior to capture various complex correlations among bits of hash codes. It restricts the posterior to the Boltzmann form and gets the following lower bound:
\begin{equation}
\mathcal{L}_{CorrSH}=\mathbb{E}_{q_\phi(\boldsymbol{z} \mid \boldsymbol{x})}\left[\log \frac{p_\theta(\boldsymbol{x} \mid \boldsymbol{z}) p(\boldsymbol{z})}{e^{-E_\phi(\boldsymbol{z})}}\right]+\log Z_\phi
\end{equation} 
where $E_\phi(\boldsymbol{z}) = -\frac{1}{2} \boldsymbol{z}^T \Sigma_\phi(\boldsymbol{x}) \boldsymbol{z}-\mu_\phi^T(\boldsymbol{x}) \boldsymbol{z}$ and $\Sigma_\phi(\boldsymbol{x})$ and $\mu_\phi(\boldsymbol{x})$ are functions parameterized by the encoder network with parameters $\phi$ and $\boldsymbol{x}$ as input. $Z_{\phi}$ is the normalization constant in Boltzmann-machine distribution.

These deep text hashing models consider solely the semantic information within the priors. However, it is widely observed that the neighborhood information among the texts is also beneficial in retrieval models \cite{huo2024deep,liu2024kat,frayling2024effective}. Thus, it is also possible to model complex relationships in the prior distribution.   Semantics-Neighborhood Unified Hashing (SNUH) \cite{ou2021integrating} proposes to encode the neighborhood information with a graph-induced Gaussian distribution. It assumes the representation $\boldsymbol{z}$ is drawn from a Gaussian
distribution with a neighborhood graph as follows:
\begin{equation}
\boldsymbol{z} \sim \mathcal{N}\left(\mathbf{0},\left(\boldsymbol{I}_N+\lambda \boldsymbol{A}\right) \otimes \boldsymbol{I}_d\right),
\end{equation}
where $\boldsymbol{A}$ is an affinity matrix and $\boldsymbol{I}_N+\lambda \boldsymbol{A}$ is a covariance matrix and denotes the neighborhood information of documents. This neighbor modeling strategy enables SNUH to learn better inter-text relationships. Multi-grained prototype-induced Hierarchical generative Hashing (HierHash) \cite{zhang2024document} models the intricate inter-textual relationships by considering hierarchical semantic information. Assume that the coarse-grained category and the fine-grained category of text \(\boldsymbol{x}\) are represented by \(l_c\) and \(l_f\), respectively. 
HierHash unifies the hierarchical semantic structure with the generative model by maximizing the evidence lower bound as follows:
\begin{equation}
\begin{aligned}
\mathcal{L}_{\text {HierHash}} &= \mathbb{E}_{q_\phi\left(l_c, l_f, z \mid x\right)}\left[\log \frac{p\left(x, l_c, l_f, z\right)}{q_\phi\left(l_c, l_f, z \mid x\right)}\right] \\ &= \mathbb{E}_{q_\phi(\boldsymbol{z}|\boldsymbol{x})}[\log q_{\theta}(x|z)] -D_{KL}(q_\phi(\boldsymbol{z} \mid \boldsymbol{x} ) \| p(\boldsymbol{z} \mid l_f )) \\ & -D_{KL}(q_\phi(l_f \mid \boldsymbol{z} ) \| p(l_f|l_c))-D_{KL}(q_\phi(l_c \mid l_f) \| p(l_c \mid l_f)) .
\end{aligned}
\end{equation}

Although HierHash utilizes category information instead of information from neighboring texts, the semantic representation of hierarchical categories in HierHash is derived from relational information between texts. Thus, HierHash can indirectly employ inter-textual relationships in the generative model modeling process.

In summary, the reconstruction-based approach serves as a flexible and foundational method for learning text representations without reliance on explicit semantic labels. Its key advantage is its ability to learn dense representations by capturing the global statistical structure of a corpus directly from the data itself. However, a significant limitation is that the reconstruction objective provides only an indirect signal for the ultimate goal of semantic retrieval. The model may learn to prioritize the reconstruction of frequent but semantically less important words.

\subsection{Pseudo-similarity-based Methods}
\label{sec:pseu}

Pseudo-similarity-based methods are designed to fabricate pseudo-similarity information, which is then used as a supervisory signal for deep semantic models to learn semantics or integrated implicitly into the model's training.

% In most image-based hashing methods \cite{hu2017pseudo,shen2019unsupervised,shi2020anchor}, there is an enthusiasm for generating pseudo-labels or pseudo-similar matrices through pre-trained modules or clustering approaches, such as K-means and spectral clustering. However, a more prevalent method in deep text hashing involves constructing mechanisms for pseudo-relevance feedback and utilizing generative models to learn relationships between texts.

Similar to most image-based hashing methods \cite{hu2017pseudo,shen2019unsupervised,shi2020anchor}, there is an enthusiasm for generating pseudo-labels or pseudo-similar matrices through pre-trained modules or clustering approaches, such as K-means \cite{ahmed2020k} and spectral clustering \cite{von2007tutorial,nascimento2011spectral}. Ranking-Based Semantic Hashing (RBSH) \cite{hansen2019unsupervised} first employing a weak supervision techniques to construct triplet feature inputs $(\boldsymbol{v}_1, \boldsymbol{v}_2, \boldsymbol{v}_3)$ as the pseudo-similar matrices. It utilizes Self-Taught Hashing \cite{zhang2010self} to generate hash codes in an unsupervised manner, then calculates the similarity between two texts as follows:
\begin{equation}
\begin{aligned}
    {\rm sim}_{1,2} = - || \boldsymbol{v}_1 - \boldsymbol{v}_2 ||_2.
\end{aligned}
\end{equation}
Subsequently, RBSH integrates inter-class similarity into the generation of hash codes by employing a modified version of the hinge loss to learn the correlations between texts:
\begin{align}
\mathcal{L}_{\text{rank}} = \begin{cases}
\max(0, \epsilon - \text{sign}_{1,2,3} D_{1,2,3}) & \text{if } \text{sim}_{1,2} \neq \text{sim}_{1,3} \\
|D_{1,2,3}| & \text{otherwise}
\end{cases}
\end{align}
Here, $\epsilon$ denotes the margin of the hinge loss, \({\rm sign}_{1,2,3}\) corresponds to the sign of the estimated pairwise text similarities, and \(D_{1,2,3}\) signifies the discrepancy between the squared Euclidean distances of the hash codes attributed to the text pairs, which can be articulated as follows:
\begin{equation}
\begin{aligned}
{\rm sign}_{1,2,3} &= {\rm sim}_{1,2}-{\rm sim}_{1,3}, \\
D_{1,2,3} &= ||\boldsymbol{z}_1-\boldsymbol{z}_3||_2^2 - ||\boldsymbol{z}_1-\boldsymbol{z}_2||_2^2.
\end{aligned}
\end{equation}
Similarly, Semantic-Alignment Promoting Multiple Features Hashing (SAMFH) \cite{chen2023exploiting} leverages a connection graph constructed with the KNN algorithm on the raw features of texts to construct the training set. Then, it performs contrastive learning on the hash
codes of connected texts in the connection graph to align the semantics between them by regarding the connected texts in the graph as positive sample pairs. While HierHash \cite{zhang2024document} gets the positive samples through the dropout of the BERT model as has been proposed in SimCSE \cite{gao2021simcse} and uses the in-batch sample to get the negative samples. Then a contrastive loss is used on the representation of the original $\boldsymbol{x}$ and its constructed texts.

\begin{figure}[t]
\centering
\includegraphics[width=0.8\textwidth]{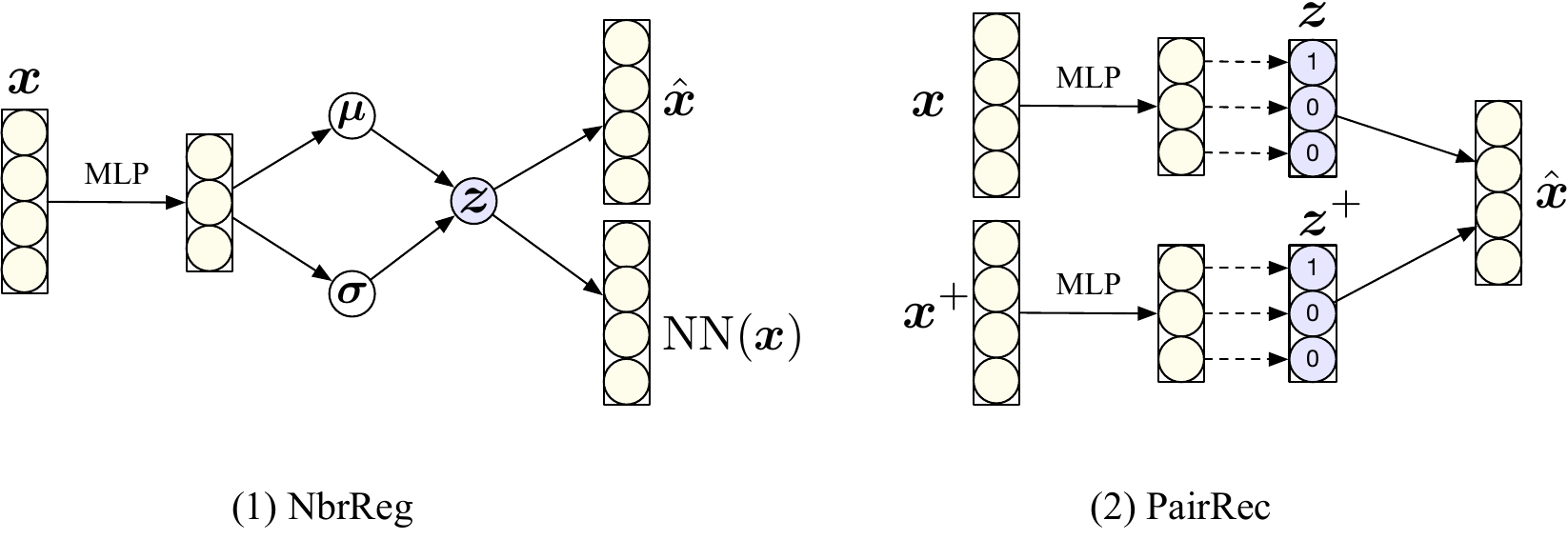}
\caption{An illustration of NbrReg \cite{chaidaroon2018deep} and PairRec \cite{hansen2020unsupervised}: (1) NbrReg uses the original text to reconstruct the text of its neighbors, while (2) PairRec uses the text of neighbors to reconstruct the original text.}
\label{fig:NbrReg_and_PairRec}
\end{figure}

Some deep text hashing methods implicitly integrate the semantic relationship between texts into the generative models during the decoder process. Neighborhood Recognition Model (NbrReg) \cite{chaidaroon2018deep} first uses the BM25 algorithm \cite{robertson2009probabilistic} to retrieve the K-nearest texts ${\rm NN}(\boldsymbol{x})$ for each text $\boldsymbol{x}$ in the training dataset. It assumes that the texts in the nearest neighbor set ${\rm NN}(\boldsymbol{x})$ bear the same label as the text $\boldsymbol{x}$. Consequently, a binary code of texts within the same vicinity in the vector space model should exhibit greater similarity (Refer to Fig.~\ref{fig:NbrReg_and_PairRec} (1)). Based on the generative framework, NbrReg assumes a semantic vector $\boldsymbol{z}$ drawn from a standard normal Gaussian distribution. For a word $\boldsymbol{w}_i$ in text $\boldsymbol{x}$, draw $\boldsymbol{w}_i$ from $p(\boldsymbol{w}_i|\boldsymbol{z})$. For a word $\tilde{\boldsymbol{w}}_i$ in the neighborhood set ${\rm NN}(\boldsymbol{x})$, draw $\tilde{\boldsymbol{w}}_i$ from $p(\tilde{\boldsymbol{w}}_i|\boldsymbol{z})$. By setting text likelihood $p(\boldsymbol{x})=\prod_{i}p(\boldsymbol{w}_i|\boldsymbol{z})$ and neighborhood likelihood $p({\rm NN}(\boldsymbol{x}))=\prod_{i}p(\tilde{\boldsymbol{w}}_i|\boldsymbol{z})$ as a product of word probabilities, NbrReg proposes to maximizing the variational lower bound of the log-likelihood of $p(\boldsymbol{x},{\rm NN}(\boldsymbol{x}))$ as follows:
\begin{equation}
\begin{aligned}
    \mathcal{L}_{\text{NbrReg}} &= \mathbb{E}_{q_{\phi}(\boldsymbol{z} | \boldsymbol{x})}\left[\log q_{\theta}\left(\boldsymbol{x} \mid \boldsymbol{z} \right)\right] + \mathbb{E}_{q_{\phi}(\boldsymbol{z}| \boldsymbol{x})}\left[\log q_{\theta}\left({\rm NN}(\boldsymbol{x}) \mid \boldsymbol{z} \right)\right] \\  &- D_{K L}(q_{\phi}(\boldsymbol{z} | \boldsymbol{x}) \| p(\boldsymbol{z})).
\label{nbr_loss}
\end{aligned}
\end{equation}
This target enables $\boldsymbol{z}$ to concurrently represent text \(\boldsymbol{x}\) and texts from its neighboring set ${\rm NN}(\boldsymbol{x})$. Besides, considering the neighbor documents also provide useful signals to enhance representation learning \cite{xie2024improving,inje2024document,yu2024tprf,wang2023colbert}, NbrReg uses a centroid representation obtained from the neighbor texts ${\rm NN}(\boldsymbol{x})$ of the input text $\boldsymbol{x}$. Node2hash \cite{chaidaroon2020node2hash} also adopts a similar architecture. 

Semantic Hashing with Pairwise Reconstruction (PairRec) \cite{hansen2020unsupervised} and Multi-Index Semantic Hashing (MISH) \cite{hansen2021unsupervised} use a pairwise reconstruction approach to encapsulate the local neighborhood structure within the hash code. Similar to NbrReg and RBSH, a set of some most similar texts ${\rm NN}(\boldsymbol{x})$ can be obtained by a weak supervised method. Then, assume a training pair $(\boldsymbol{x},\boldsymbol{x}^+)$ is constructed from the text $\boldsymbol{x}$ and a text $\boldsymbol{x}^+$ sampled from set ${\rm NN}(\boldsymbol{x})$, they use the following variational lower bound:
\begin{equation}
\begin{aligned}
 \mathcal{L}_{PairRec} = &-\mathbb{E}_{q_{\phi}(\boldsymbol{z}|\boldsymbol{x})}\left[\log q_{\theta}\left(\boldsymbol{x} \mid \boldsymbol{z} \right)\right] + \beta D_{K L}(q_{\phi}(\boldsymbol{z} | \boldsymbol{x}) \| p(\boldsymbol{z})) \\ &-\mathbb{E}_{q_{\phi}(\boldsymbol{z}^+|\boldsymbol{x}^+)}\left[\log q_{\theta}\left(\boldsymbol{x} \mid \boldsymbol{z}^+ \right)\right] \\ &+ \beta D_{K L}(q_{\phi}(\boldsymbol{z^+} | \boldsymbol{x}^+) \| p(\boldsymbol{z^+})).
\label{eq:pairrec_loss}
\end{aligned}
\end{equation}
This loss comprises two components: an ordinary variational lower bound for text $\boldsymbol{x}$, and the other involves using text $\boldsymbol{x}^+$ in the encoding process. At the same time, the decoding pertains to document $\boldsymbol{x}$. This method transfers the local neighborhood structure from the text space into the Hamming space, as $\boldsymbol{z}^+$ must be capable of reconstructing the original $\boldsymbol{x}$. 

This approach offers a notable advantage over the purely reconstruction-based methods discussed previously. By generating and utilizing pseudo-similarity information, it provides a more direct learning signal that is better aligned with the retrieval task, bridging the gap between learning from raw text structure and learning from explicit labels. The main drawback, however, is that the model's performance is fundamentally capped by the quality of the generated pseudo-signals. If the underlying similarity source (e.g., K-means) is noisy or semantically weak, these imperfections can be propagated or even amplified. Therefore, this approach is most suitable for scenarios where a reasonably reliable, albeit weak, similarity signal can be extracted from the data's intrinsic features.

% \subsubsection{Contrastive Learning-based Methods}
\subsection{Maximal Mutual Information Methods}
\label{sec:maximal}

Maximal mutual information has shown significant promise in the field of deep learning, as evidenced by several studies \cite{ma2024volta, yang2022learning, qian2022learning, chen2024learning}. As a result, some deep hashing methods are exploring ways to incorporate maximal mutual information to better capture and understand complex semantic structures.

AMMI \cite{stratos2020learning} estimates a distribution over latent variables without modeling raw signals by maximizing the mutual information between the latent variables and a label variable. In contrast to maximizing the evidence lower bound like some deep text hashing model, AMMI proposes to learn a text encoder $p^{\phi}_{Z|Y}$ by the following adversarial formulation of the mutual information between a random variable $Y$ corresponding to a text and binary hash code $Z$:
\begin{equation}
\max _\phi \min _\theta H_{\phi, \theta}^{+}(Z)-H_\phi(Z \mid Y),
\label{eq:amm}
\end{equation}
where $H_{\phi, \theta}^{+}(Z)$ is the cross entropy between ${p_{Z}^{\phi}}$ and ${p_{Z}^{\phi}}$, and $H_\phi(Z \mid Y)$ denotes the conditional entropy. 
Based on this motivation, AMMI proposes an algorithm to estimate these two terms in Eq. (\ref{eq:amm}). Maximizing the mutual information between the target and latent variables can also serve as a partial objective for deep text hashing. HierHash \cite{zhang2024document} seeks to maximize the mutual information between latent variables and fine-grained pseudo labels, thus averting model collapse where all samples might be allocated to a single fine-grained category.

Deep Hash InfoMax (DHIM) \cite{ou2021refining} and USH-SER \cite{tong2024efficient} propose to maximize both local and global mutual information. Specifically, they first employ BERT to obtain the embedding of a text, then pass the embedding through a textual CNN to derive $T$ local features \(\{\boldsymbol{z}^{l}_1, \boldsymbol{z}^{l}_2, ..., \boldsymbol{z}^{l}_T\}\). Subsequently, a global feature \(\boldsymbol{z}^{g}\) is obtained by applying a READOUT function to these local features. Then they apply the Jensen-Shannon divergence estimator (JSDE) \cite{nowozin2016f} to estimate the maximum information and optimize the model as follows:
\begin{equation}
\begin{aligned}
\tilde{I}_\phi(\boldsymbol{z}^{l}_i;\boldsymbol{z}^{g}) = &- softplus(-D_{\phi}(\boldsymbol{z}^{l}_i,\boldsymbol{z}^{g})) -E_{\mathbb{P}}[softplus(D_{\phi}(\tilde{\boldsymbol{z}}^{l}_i,\boldsymbol{z}^{g}))],
\end{aligned}
\end{equation}
where $\tilde{\boldsymbol{z}}^{l}_i$ is the $i$-th local representation of negative samples generated from the empirical distribution $\mathbb{P}$. $D_{\phi}(\cdot,\cdot)$ is a discriminator realized by a neural network with parameter $\phi$. Besides, in addition to maximizing mutual information between local hash codes and global hash codes, DHIM also maximizes the mutual information between the CLS token embedding from the BERT output and the global hash codes. 

From a theoretical standpoint, maximizing mutual information provides a more principled learning objective compared to the heuristic nature of reconstruction or the potential noise in pseudo-similarity. This approach encourages the model to learn representations that are maximally informative and can enhance robustness. However, its primary challenge is practical: the estimation and optimization of mutual information in high-dimensional spaces are notoriously difficult. These methods often require complex estimators or intricate adversarial training schemes.

\subsection{Semantic from Categories}
\label{sec:semantic_label}

The semantics of categories are crucial for distinguishing texts. Therefore, in scenarios where text categories are accessible, some deep text hashing methods extract semantic information from these categories. These methods usually add a classification layer to map the latent feature into label distributions, and then the hash codes are enhanced with the standard classification loss in label space.

\begin{figure}[t]
\centering
\includegraphics[width=0.8\textwidth]{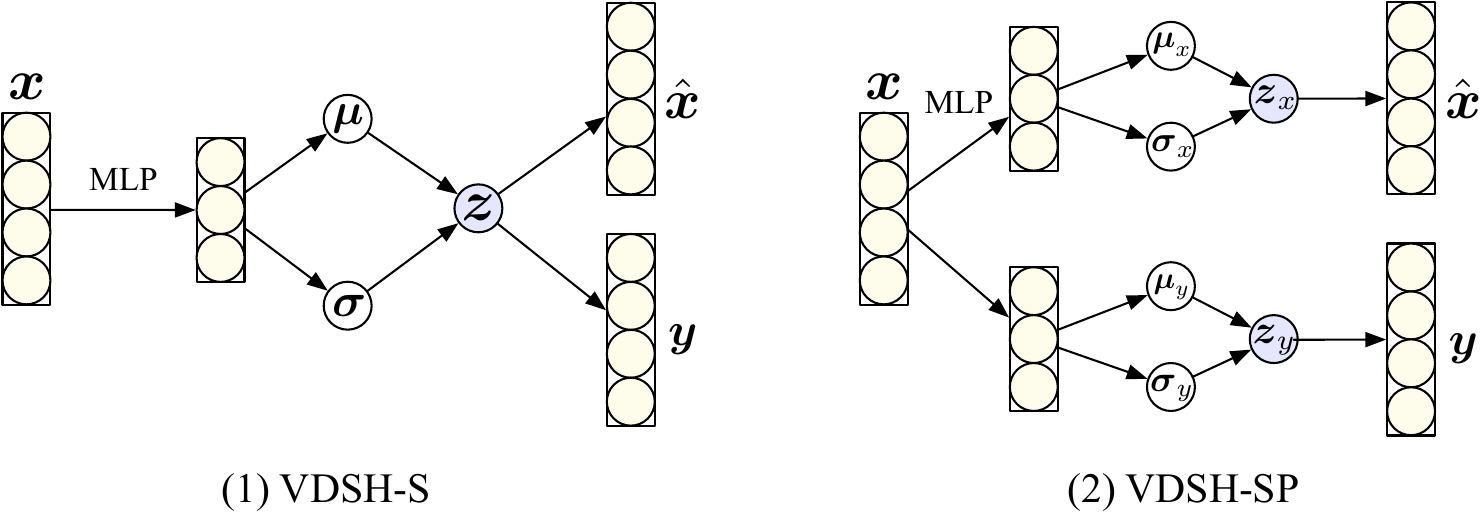}
\caption{An illustration of VDSH-S and VDSH-SP: (1) VDSH-S assumes that texts and labels are generated by the same latent semantic vector, while (2) VDSH-SP introduces a private text variable that is not shared by the labels.}
\label{fig:VDSH_S_and_P}
\end{figure}

In the previous sections, we mentioned that VDSH \cite{chaidaroon2017variational} is a highly representative work utilizing VAE to generate hash codes. In its original paper, the authors introduced its supervised version, named VDSH-S (refer to Fig.~\ref{fig:VDSH_S_and_P} (1)). Let $\boldsymbol{y}_j \in \{0,1\}^L$ denote the one-hot representation of the label $j$ in the label set, and $L$ is the total number of labels. VDSH-S assumes that both words and labels are generated based on the same latent semantic vector. It uses the following logistic function to model the generative probabilistic:
\begin{equation}
\begin{aligned}
p(\boldsymbol{y}_j|f(\boldsymbol{z};\tau))=\frac{1}{1+exp(-\boldsymbol{y}_i^Tf(\boldsymbol{z};\tau))},
\end{aligned}
\end{equation}
where $\tau$ is the parameter of a classification network \( f(s; \tau) \). VDSH-S assumes texts and labels are generated by the same latent semantic vector. Still, it could sometimes be difficult to find a common representation for both documents and labels. VDSH-SP introduces a text private variable, which is not shared by the labels $\boldsymbol{y}$ (refer to Fig.~\ref{fig:VDSH_S_and_P} (2)). In other words, the variational lower bound additionally considers a component for the label. These two methods have been flexibly applied to other reconstruction-based deep hashing models \cite{shen2018nash, dong2019document, zhang2019doc2hash, zhang2020discrete} to address model training problems in scenarios with categories and are employed in analogous forms by various other structures \cite{xu2015convolutional,yu2015understanding,cui2019short,nanculef2021self,huang2025confusing}.

Using explicit category labels gives a strong and clean signal for learning, which is more direct than signals from reconstruction (Section~\ref{sec:rec}) or pseudo-similarity (Section~\ref{sec:pseu}) methods. This makes these models very good at separating texts into different topics. However, a key limitation is that these methods rely on broad category labels. This forces the model to treat all texts within the same category as identical, ignoring their finer semantic differences. This results in a model that can separate broad topics well but cannot distinguish between similar texts inside the same topic. Therefore, this approach is best for topic-level sorting, not for tasks that need nuanced ranking.

\subsection{Semantic from Relevance}
\label{sec:semantic_relevance}

Learning semantic information from data relevance is also important in deep text hashing. Variational Pairwise Supervised text Hashing (VPSH) \cite{xuan2019variational} and CVAE/MVAE \cite{xuan2020conditional} use a pairwise loss based on the VAE framework. They use a two-tower structure VAE and draw the latent variable $\boldsymbol{z}_1$ and $\boldsymbol{z}_2$ from different towers. Then, a pairwise loss is introduced as follows:
\begin{equation}
\mathcal{L}_{label} = s_{1,2}-sim(\boldsymbol{z}_1,\boldsymbol{z}_2),
\end{equation}
where \(s_{1,2}\) represents the similarity information between \(x_1\) and \(x_2\) and $sim(\cdot,\cdot)$ is the similarity function of the two inputs. Similar objective functions are also used in LASH \cite{guo2021lash}, CoSHC \cite{gu2022accelerating}, and HDR-BERT \cite{lan2023towards} with partial modifications. The fundamental idea is to preserve similarity in the Hamming space using pre-defined or constructed similarity information between two texts. Intra-category aware Hierarchical Supervised Document Hashing (IHDH) \cite{guo2022intra} employs a similar pair-wise loss, but it utilizes the data within a batch for mutual similarity comparison, thereby learning relationships more efficiently. Building upon a similar variational inference framework, Pairwise Supervised Hashing (PSH) \cite{dadaneh2020pairwise} introduces a pairwise loss to minimize the distance between latent codes of similar documents while simultaneously maximizing the distance between latent codes of documents belonging to different categories as follows:
\begin{equation}
\begin{aligned}
\mathcal{L}_{PSH} 
= & \mathbf{1}_{y^{(1)}=y^{(2)}} d\left(\boldsymbol{z}^{(1)}, \boldsymbol{z}^{(2)}\right) - \mathbf{1}_{y^{(1)} \neq y^{(2)}} d\left(\boldsymbol{z}^{(1)}, \boldsymbol{z}^{(2)}\right),
\label{eq:psh}
\end{aligned}
\end{equation}
where $d(\cdot, \cdot)$ is a distance metric and $\boldsymbol{1}_S$ is the indicator function being equal to one when $S$ is true. However, SSB-VAE \cite{nanculef2021self} argues that the pairwise loss Eq.(\ref{eq:psh}) can deteriorate faster than a method using only point-wise supervision in label scarcity scenarios, because if the labeled subset is reduced to a fraction $\rho$ of the training set, the fraction of pairs that can be generated reduces to $\rho^2$. This issue makes the method more prone to over-fitting. To address this problem, SSB-VAE proposes a new  learning as follows:
\begin{equation}
\begin{aligned}
\mathcal{L}_{SSB} 
= \tilde{y}_1^T\tilde{y}_2d(\boldsymbol{z}_1,\boldsymbol{z}_2) - (1-\tilde{y}_1^T\tilde{y}_2)d'(\boldsymbol{z}_1,\boldsymbol{z}_2),
\label{eq:ssb}
\end{aligned}
\end{equation}
where $\tilde{y}$ is a prediction label from a fully connected layer. Eq.(\ref{eq:ssb}) uses the prediction label $\tilde{y}$ to approximate the true label $y$ for many unlabelled observations.  BPR \cite{yamada2021efficient} uses the hinge-loss in a candidate generation phase to enhance candidate generation as follows:
\begin{equation}
\begin{aligned}
\mathcal{L}_{\text {cand }}=\sum_{j=1}^n \max (0,-(\tilde{\boldsymbol{h}}_{q_i} \cdot \tilde{\boldsymbol{h}}_{p_i^{+}}+\tilde{\boldsymbol{h}}_{q_i} \cdot \tilde{\boldsymbol{h}}_{p_{i, j}^{-}})+\alpha).
\end{aligned}
\end{equation}
where the question $q_i$ and the positive candidate $p_i^+$ and negative candidate $p_i^-$ are darw from a pre-constructed set $\mathcal{D}=\left\{\left\langle q_i, p_i^{+}, p_{i, 1}^{-},..., p_{i, n}^{-}\right\rangle\right\}_{i=1}^m$. $\cdot$ denotes the inner product. A similar loss is also employed in DHSH \cite{huang2025confusing} to facilitate hierarchical semantic learning at both the parent and child levels. Besides, BPR introduces a contrastive learning loss in the ranking phase:
\begin{equation}
\begin{aligned}
\mathcal{L}_{\text {re }}=-\log \frac{\exp (\boldsymbol{z}_{q_i} \cdot \tilde{\boldsymbol{h}}_{p_i^{+}})}{\exp (\boldsymbol{z}_{q_i} \cdot \tilde{\boldsymbol{h}}_{p_i^{+}})+\sum_{j=1}^n \exp (\boldsymbol{z}_{q_i} \cdot \tilde{\boldsymbol{h}}_{p_{i, j}^{-}})}.
\end{aligned}
\end{equation}

Through a two-step training process, BPR can effectively learn the high-order semantic relationships between texts.

This approach directly addresses the low intra-class resolution of category-based methods (Section~\ref{sec:semantic_label}). By employing pairwise or triplet objectives, it captures fine-grained, instance-level relationships instead of treating all documents within a class as equally similar. This makes relevance-based methods inherently better suited for ranking-oriented tasks, such as question answering, where relative similarity is paramount. Their main challenge is computational complexity, especially the efficient sampling of informative training pairs from a polynomially growing space. Consequently, they are most effective when fine-grained relevance judgments are available for ranking-focused applications.

\section{Hash Code Quality Preservation}
\label{sec:code_quality}

The generation of binary representations is the key distinction between deep text hashing and traditional retrieval models. Therefore, a primary challenge is to ensure the quality of these compact codes. In this section, we provide a detailed analysis of existing efforts to preserve three desirable properties of hash codes: the compactness of the code (few-bit code), the uniformity of its distribution (code balance), and the fidelity of the binarization process (low quantization error).

\subsection{Few-bit Code} 
\label{sec:few-bit}

\begin{figure}[t]
\centering
\subfigure[]{
\label{fig_first_case}
\includegraphics[width=0.4\textwidth]{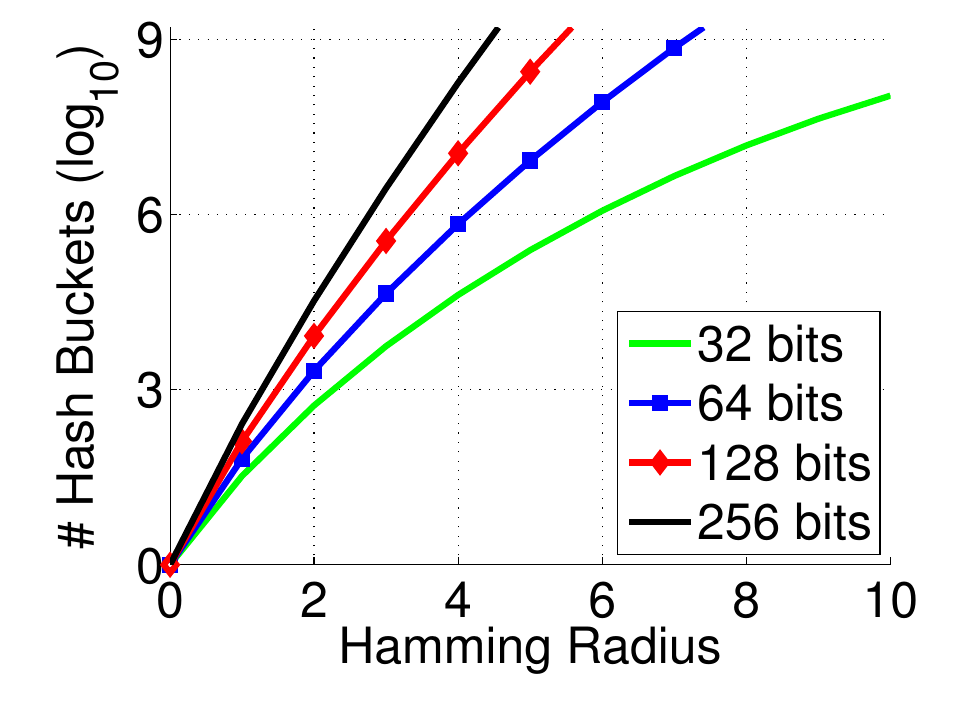}}
\subfigure[]{
\label{fig_second_case}
\includegraphics[width=0.4\textwidth]{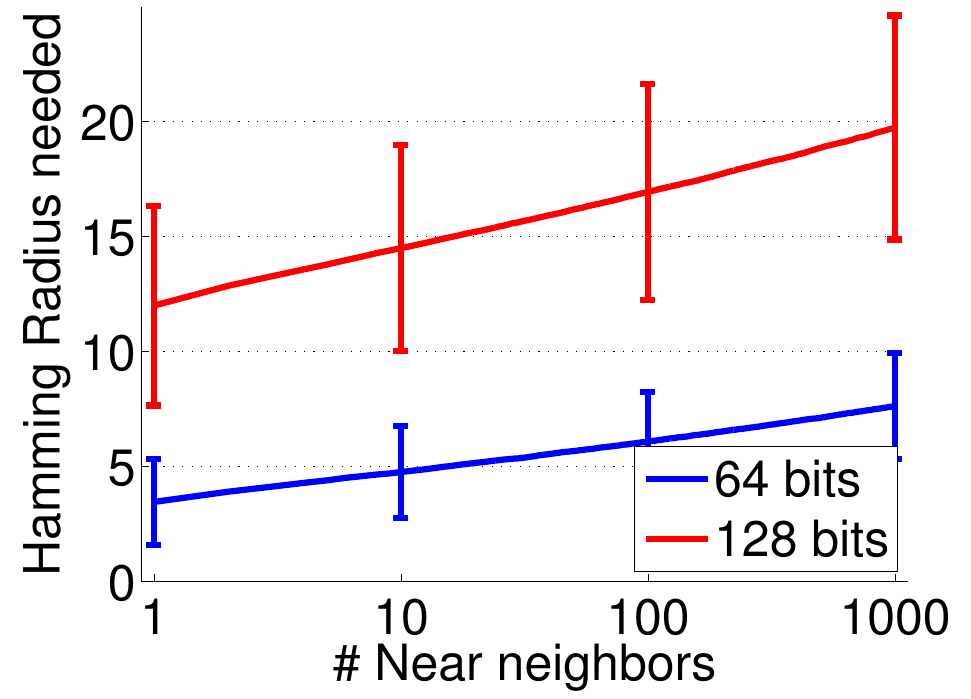}}
\caption{Norouzi et al. \cite{norouzi2012fast} shows (a) the number of distinct hash table indices (buckets) within a Hamming ball of radius \( r \) and (b) the expected search radius required for K-NN search as a function of \( K \), based on a dataset of 1B SIFT descriptors. It is evident that the length of the hash codes significantly impacts speed, especially when the \( K \) value or radius \( r \) is large.}
\label{fig:code_length_analysis}
\end{figure}

In deep text hashing, an essential characteristic of hash codes is compactness, which means there is a desire to generate the shortest possible hash codes, also known as few-bit codes. Shorter code lengths imply faster search speeds, as shown in Fig.~\ref{fig:code_length_analysis}. Meanwhile, Eq. \ref{eq:look_up_bucket} and Eq. \ref{eq:multi_bucket} formally demonstrate this effect. However, directly setting a short code length can easily result in a loss of precision due to information loss. Thus, in this regard, the primary consideration is ensuring the quality of retrieval results while keeping the code length \( b \) relatively small, also called the few-bits hashing problem \cite{ye2020unsupervised}. Besides, the code length is related to the number of semantics (in deep text hashing work, a category usually represents a semantic). The Hamming space must ensure that different semantics have unique codes. Thus, we have:
\begin{equation}
    b \leq \log_2 L,
\end{equation}
where $L$ is the dataset's category number. The key to obtaining a few-bit code is to retain as much effective information as possible while reducing the information representation space.

FeW-bIts Semantic Hashing (WISH) \cite{ye2020unsupervised} introduces a set of auxiliary implicit topic vectors to address the information loss caused by the few-bit problem. It employs a variational inference framework similar to NASH \cite{shen2018nash}, but assuming each word \( \boldsymbol{w}_i \) in \( \boldsymbol{x} \) is drawn from some selected topics $\mathcal{T}_{\boldsymbol{z}}$. Then, the variational lower bound of the text log-likelihood can be expressed as follows:
\begin{equation}
\begin{aligned}
\mathcal{L}_{Few}= & \mathbb{E}_{q_{\phi}(\boldsymbol{z}|\boldsymbol{x})}[\sum_{i=1}^N \log q_{\theta}\left(\boldsymbol{x} \mid f\left(g\left(\mathcal{T}_{\boldsymbol{z}}\right)\right)\right)] -D_{KL}(q_{\phi}(\boldsymbol{z} \mid \boldsymbol{x}) \| p(\boldsymbol{z})).
\end{aligned}
\end{equation}

This method successfully alleviates the few-bit problem because the implicit topic vectors play a crucial role in mitigating information loss in few-bit hashing. They are learned automatically according to the data distribution rather than being manually set up. These topic vectors serve as a repository for category information, thereby ensuring the effectiveness of the short hash codes. 

DenoiSing Memory-bAsed Semantic Hashing (SMASH) \cite{he2023efficient} chooses a long latent representation $\boldsymbol{z}^l$ and a short latent representation $\boldsymbol{z}^s$ within the autoencoder architecture. Define $|\boldsymbol{z}^l|$ and $|\boldsymbol{z}^s|$ are the dimension of $\boldsymbol{z}^l$ and $\boldsymbol{z}^s$ respectively, SMASH ensures $|\boldsymbol{z}^l| \gg |\boldsymbol{z}^s|$. Then, it introduces a relevance propagation objective to enable the short latent representation to learn the relational information of the long latent representation as follows:
\begin{equation}
    \mathcal{L}_{SMASH} = \frac{1}{B(B-1)} \sum_{i,j, i\neq j}^B \left| \frac{\boldsymbol{z}^l_i \boldsymbol{z}^l_j}{|\boldsymbol{z}^l|} - \frac{\boldsymbol{z}^s_i \boldsymbol{z}^s_j}{|\boldsymbol{z}^s|} \right|,
\end{equation}
where $B$ is the size of a mini-batch $\boldsymbol{B}=\{\boldsymbol{x_1}, \boldsymbol{x_2}, ..., \boldsymbol{x_B}\}$. Regarding the optimization procedure, SMASH ensures that texts proximate in the high-dimensional Hamming space remain proximate in the latent Hamming space.

\subsection{Code Balance}
\label{sec:code_balance}

\begin{figure}[t]
\centering
\subfigure[KNN search]{
\label{fig_first_case}
\includegraphics[width=0.45\textwidth]{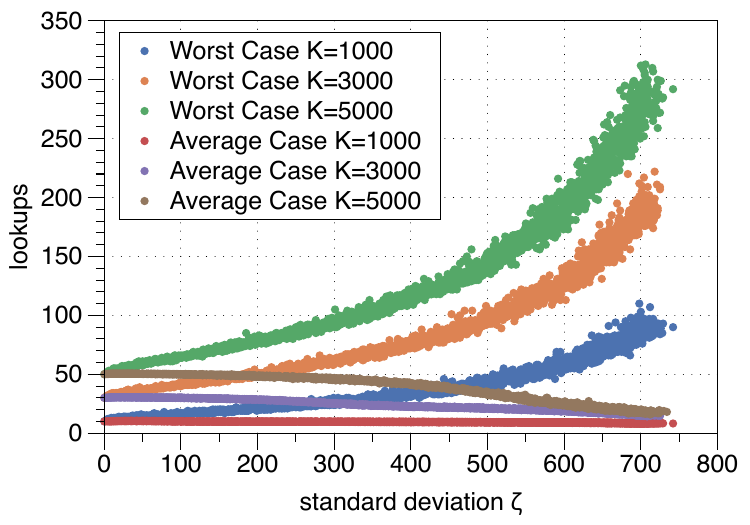}}
\subfigure[PLEB search]{
\label{fig_second_case}
\includegraphics[width=0.45\textwidth]{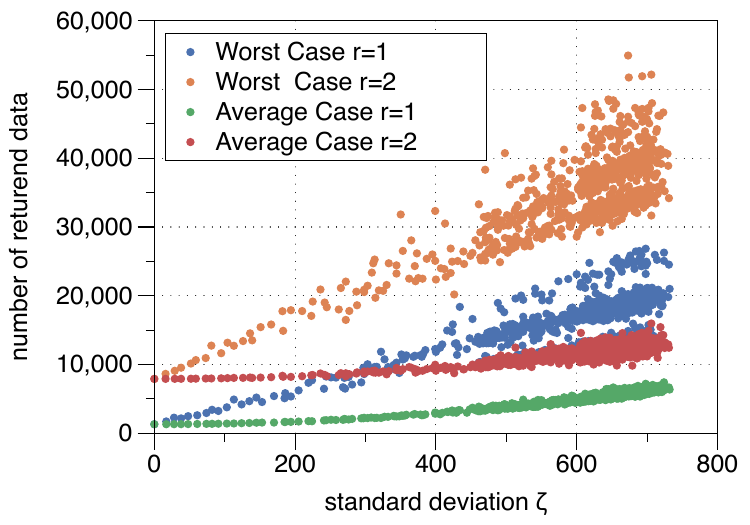}}
\caption{He et al. \cite{he2023efficient} empirically validated the impact of hash code distribution on both average and worst-case scenarios using synthetic data. They use the standard deviation $\zeta$ of the number of texts in all hash buckets to evaluate the degree of code balance. (a) plots the average and worst number of access hash buckets in the KNN search. (b) shows the average and worst number of returned texts in the PLEB search. This result shows that efficient hash codes should achieve code balance to mitigate bad search cases.}
\label{fig:code_balance_analysis}
\end{figure}

Code balance means hash codes are evenly spread throughout the hamming space. It can enhance retrieval performance in multiple aspects. First, code balance helps to reduce information redundancy in hash codes, then makes it better to preserve the original locality structure of the data \cite{doan2020efficient}. Second, as shown in Fig.~\ref{fig:code_balance_analysis}, code balance narrows the search latency gap between bad cases and average cases, therefore improving user experience \cite{he2011compact,he2023efficient}.

A direct method to obtain code-balanced hash codes is to sample from a Bernoulli distribution with a parameter \( \gamma = 0.5 \), akin to the methodologies employed by works such as NASH \cite{shen2018nash}. Another approach is to achieve code balance by incorporating certain constraints for the generated hash codes. Technically, code balance includes two optimization targets: bit balance and bit uncorrelation \cite{he2011compact}. The bit balance means each bit has the same probability of appearing. To achieve this goal, each bit in the hash code should have a 50\% chance of being 1 or –1. The bit uncorrelation means different bits should be as irrelevant as possible. In the early stage, Spectral Hashing \cite{weiss2008spectral} proposes two constraints to achieve these two targets:
\begin{equation}
    \sum_{i=0}^N \boldsymbol{h}_i =0, \ \ \  \frac{1}{n} \sum_{i=1}^N \boldsymbol{h}_i\boldsymbol{h}_i^T = \boldsymbol{I},
\label{eq:sp}
\end{equation}
where $\boldsymbol{I}$ represents the identity matrix. The first constraint requires each bit to have a $50\%$ chance to be $1$ or $-1$, and the second constraint requires the bits to be uncorrelated. This concept is continued by several methods. DSH \cite{chen2016document} employs an autoencoder to obtain the latent features. Then it applies Spectral Hashing to generate high-quality hash codes, where constraints in Eq.(\ref{eq:sp}) are employed. USH-SER \cite{tong2024efficient} uses these constraints as its learning objectives. SMASH \cite{he2023efficient} achieves code balance by introducing two similar constraints in their learning objective as follows:
\begin{equation}
    \mathcal{L}_{bb} = \frac{1}{b} \sum_{j=1}^b \alpha_i |\sum_{i=1}^{|\boldsymbol{B}|}| z_{ij}|, \ \ \mathcal{L}_{bd} = \frac{1}{b^2} ||\boldsymbol{A} \cdot (\frac{\boldsymbol{Z}^T\boldsymbol{Z}}{|\boldsymbol{B}|}-\boldsymbol{I}) ||_F^2,
\end{equation}
where $\alpha_i$ and $\boldsymbol{A}$ are two dynamic weight coefficients, which are learned from a global distribution. \(\boldsymbol{Z} = [\boldsymbol{z}_1, \boldsymbol{z}_2, \ldots, \boldsymbol{z}_B]\) represents the feature matrix composed of the current batch of data. Denoising Adversarial Binary Autoencoder (DABA) \cite{doan2020efficient} applies adversarial learning to encourage the latent variable \( \boldsymbol{z} \) to follow a univariate Bernoulli distribution ${\rm Bernoulli}(0.5)$ with the parameter $\gamma$ set to be $0.5$ for each dimension. Then each bit in \( \boldsymbol{z} \) will optimally partition the original space into two halves. Consequently, each bit has approximately a 50\% probability of being closer to $1$ or $-1$, ensuring bit balance. Moreover, the points within each half are more similar than those in the opposite half, ensuring the bit uncorrelation. DABA achieves this target through a game between the generator and the discriminator and transfers it to the optimal transport problem \cite{villani2021topics}. A similar operation that encourages hash codes to achieve code balance is also employed in WAE \cite{zhang2020discrete}.

% WISH \cite{ye2020unsupervised} uses a similar constraint in its topic vector. In the training process, it employs the following regularization function to :
% \begin{equation}
%     \mathcal{L} = ||T^TT-I||_F^2,
% \end{equation}
% where $||\cdot||_F$ denotes the Frobenius norm. 

\subsection{Low Quantization Error} 
\label{sec:minimize_quan}

Low quantization error aims to minimize information loss when encoding real-valued representations into binary hash codes. In most cases, we first generate a binary-like representation $\tilde{\boldsymbol{h}}$ and then use $\rm sign(\cdot)$ to get the binary hash code $\boldsymbol{h}$. If the values of $\tilde{\boldsymbol{h}}$ across its dimensions are close to zero, a significant amount of important information will be lost, leading to erroneous correlation assessments. For example, assume the binary-like representations of two relevant texts \( \boldsymbol{x}_1 \) and \( \boldsymbol{x}_2 \) are \(\tilde{\boldsymbol{h}}_1 = [0.8, 0.1, -0.1]\) and \(\tilde{\boldsymbol{h}}_2 = [0.7, -0.1, 0.1]\). Despite \(\tilde{\boldsymbol{h}}_1\) and \(\tilde{\boldsymbol{h}}_2\) being very close in Euclidean space, their resulting hash codes \( \boldsymbol{h}_1 = [1, -1, 1] \) and \( \boldsymbol{h}_2 = [1, 1, -1] \) are far apart in Hamming space. 

A typical approach involves incorporating a quantization loss during training to penalize the discrepancy between continuous codes and their binary counterparts. For example, HAS \cite{xu2020hashing}, LASH \cite{guo2021lash} and IHDH \cite{guo2022intra} apply following quantization loss:
\begin{equation}
\mathcal{L}_Q = || \boldsymbol{z} - {\rm sign}(\boldsymbol{z}) ||_F^2.
\end{equation}

Besides, some methods use activation functions to reduce quantization loss. For example, some methods \cite{cui2019short,doan2020efficient,ou2021integrating} use the sigmoid function to obtain the binary-like code within $0$ to $1$. While some methods \cite{yamada2021efficient,gu2022accelerating,he2023efficient,tong2024efficient} use the ${\rm tanh}(\cdot)$ function or its variants to encourage the model outputs to be close to $-1$ and $1$. In BPR \cite{yamada2021efficient} and CoSHC \cite{gu2022accelerating}, a scaled tanh function $\tilde{\boldsymbol{h}} = {\rm tanh}(\beta \boldsymbol{z})$ is employed. $\beta$ is increased by set $\beta = \sqrt{\gamma \cdot step + 1}$, where $step$ is the training step. It causes the function to become increasingly non-smooth. In USH-SER \cite{tong2024efficient}, a similar form of activation function $\tilde{\boldsymbol{h}} = {\rm tanh}(g(t)\cdot \boldsymbol{z}) $ is employed, where $g(t)$ is a function that varies over training time. A fundamental motivation here is that the model parameters undergo significant updates in the early stages of training. Therefore, a smoother activation function is used initially to minimize information loss. In the later stages of training, as the parameter updates become smaller, an activation function closer to the sign function can be employed. This approach aligns the training objectives more closely with the inference objectives, reducing quantization information loss during inference. These activation functions are typically used during training, while during inference, they usually apply $\rm{sign(\cdot)}$ or median method \cite{chaidaroon2017variational} to obtain binary hash codes.

\section{Other Technology}
\label{sec:other_tech}

Current deep text hashing primarily focuses on semantic extraction and hash code quality, yet other considerations are also necessary. For instance, enhancing the robustness of hash codes against noise, improving search efficiency by incorporating semantic hashing indices during training, and reducing bias in the gradient propagation process of non-differentiable functions.

% 1. 鲁棒性的问题
% 2. 梯度求导过程中bias的问题
% 3. 语义哈希索引的问题

% In addition to the primary methods, some deep semantic 
% hashing works also consider various other techniques. 

\subsection{Robustness Promotion}
\label{sec:robust}
In practical applications, text noise problems may arise, such as misspelling, non-standard abbreviations, and E-speak \& new words \cite{bazzo2020assessing,hanada2016effective,li2021dcspell}. Although we can clean the noisy text in the database, the noise can also be introduced by user queries. Therefore, generating robust hash codes is a significant research topic in deep text hashing.

% \begin{figure}[!t]
% \centering 
% \includegraphics[width=0.45\textwidth]{Figures/smash_noise_demo.pdf}
% \caption{The basic search framework for deep text hashing. }
% \label{fig:semantic_hashing_search_pipline}
% \end{figure}

DABA \cite{doan2020efficient} and SMASH \cite{he2023efficient} take into account the issue of textual Robustness Based on the autoencoder structure, they first construct a corrupted input \(\boldsymbol{x}'\) by removing or altering words in the original text \(\boldsymbol{x}\), and then the decoder endeavors to recover \(\boldsymbol{x}\). In this procedure, SMASH acknowledges the varying significance of individual words during the corruption process from \(\boldsymbol{x}\) to \(\boldsymbol{x}'\). For example, the original text is "What is the best exercise for losing weight in your upper thighs." By randomly deleting words, you might get ``What is the \#\#\# exercise for losing weight in \#\#\# upper thighs'', which doesn't significantly alter the original meaning and should be placed in the same hash bucket. However, the corrupted text could also be ``What is the best \#\#\# for losing \#\#\# in your upper thighs''. In this scenario, the semantics of this new text deviate significantly from the original. Thus, SMASH obtains a hidden representation from the structure of an auto-encoder and calculates the semantic similarity coefficient between them to derive a significant weight \(\epsilon_k\) for each text during reconstruction. Consequently, the learning objective becomes: 
% \begin{equation}
%     \mathcal{L}=E_{\boldsymbol{x} \sim D_x}[\frac{1}{V}(\sum_{j=1}^n \sum_{i=1}^{V}-\epsilon_j \log p(\hat{\boldsymbol{x}}_{(j)}^{\prime(i)}=\boldsymbol{x}^{(i)}))],
% \label{eq:dae_loss}
% \end{equation}
\begin{equation}
    \mathcal{L}=E_{\boldsymbol{x}_i \sim D_{\boldsymbol{X}}}[\frac{1}{V}(\sum_{k=1}^n \sum_{j=1}^{V}-\epsilon_k \log p(\hat{\boldsymbol{x}}_{i,j}^{\prime(k)}=\boldsymbol{x}_{i,j}))],
\label{eq:dae_loss}
\end{equation}
where \(\epsilon_k\) denote a latent weight variable for the \(k\)-th corrupted text, and \(\hat{\boldsymbol{x}}_{i,j}^{\prime(k)}\) is the \(k\)-th text reconstructed through the process of textual corruption. In addition to directly adding noise to the data, some methods \cite{shen2018nash,hansen2019unsupervised,hansen2020unsupervised,hansen2021unsupervised} introduce noise within the model's structure, thereby enhancing the robustness of the hash codes. They inject data-dependent noise into latent variable $\boldsymbol{z}$. It samples a \(\boldsymbol{z}'\) from \(\mathcal{N}(\boldsymbol{z},\sigma^2I)\) and utilizes \(\boldsymbol{z}'\) in the reconstruction process of the decoder.  NASH \cite{shen2018nash} employs a conventional rate-distortion trade-off problem to theoretically elucidate this process, where rate and distortion denote the effective code length, i.e., the number of bits used, and the distortion introduced by the encoding/decoding sequence, respectively. By controlling the variance \(\sigma\), the model can adaptively explore different trade-offs between the rate and distortion objectives.

% 感觉可能要从介绍multi-index hashing开始
\subsection{Adaptation to Index}
\label{sec:adapt}
Most deep text hashing models focus solely on the generation of hash codes from data, overlooking the performance of these hash codes in subsequent indexing. Nevertheless, the efficiency of the deep text hashing index is inherently linked to the distribution of the hash codes. Thus, some methods propose to process the generated hash codes further, ensuring that the processed results more closely align with the distribution characteristics required by the hashing index. In the early days, some approaches focused on post-processing the obtained hash codes to achieve an ideal hash code distribution suitable for multi-index hashing. Data Driven Multi-index Hashing (DDMIH) \cite{wan2013data} explores the statistic properties of the database and separates the highly correlated bits into different code segments, finally getting more uniform code distribution in each hash table. Data-oriented Multi-index Hashing (DOMIH) \cite{liu2015data} builds a training set to compute the correlations between bits of the codes and learn an adaptive projection vector for each substring, and then projects the substrings with corresponding projection vectors to generate new indices for the original multi-index hashing method \cite{norouzi2012fast}. However, DDMIH and DOMIH optimize exclusively for the generated hash codes without necessitating consideration of the underlying deep model. Multi-Index Semantic Hashing (MISH) \cite{hansen2021unsupervised} generates hash codes that are both effective and highly efficient by jointly optimizing deep hashing models and multi-index hashing. It introduces two learning objectives to ensure the learned hash codes are well-suited for multi-index hashing. Specifically, the candidate set estimated by multi-index hashing can be reduced by limiting the number of documents added by each hash table lookup, thus MISH applies a learning objective to limit the number of false positive matches as follows:
\begin{equation}
\mathcal{L}_{false-positive} = -d_H(\boldsymbol{z}_q^i,\boldsymbol{z}_s^i),
\end{equation}
here, $\boldsymbol{z}_q^i$ and $\boldsymbol{z}_s^i$ is the i-th substring of $\boldsymbol{z}_q$ and $\boldsymbol{z}_s$, $\boldsymbol{z}_s$ is sampled by given a $\boldsymbol{z}_q$ as follows:
\begin{equation}
\boldsymbol{z}_s={\rm argmax}_{\boldsymbol{z}_j}  D_H\left(\boldsymbol{z}_q, \boldsymbol{z}_j\right) \cdot \boldsymbol{1}_{\left[D_H\left(\boldsymbol{z}_q^i, \boldsymbol{z}_j^i\right) \leq r_i^*\right]} \cdot \boldsymbol{1}_{\left[D_H\left(\boldsymbol{z}_q, \boldsymbol{z}_j\right)>r\right]},
\end{equation}
where $r$ is the search radius and $r_i^*$ is the substring search radius for substring $i$. By sampling hash codes with the largest value of \(D_H(\boldsymbol{z}_q, \boldsymbol{z}_s)\), \(\boldsymbol{z}_s\) is unlikely to be within the top \(K\), but it would still appear in the candidate set due to the low substring Hamming distance. Another learning objective is to reduce the number of hash table lookups. Given a query hash code $\boldsymbol{z}_q$, MISH samples a hash code $\boldsymbol{z}_r$ with $d_H(\boldsymbol{z}_q,\boldsymbol{z}_r)=r$ and constructs the pair $(\boldsymbol{z}_q,\boldsymbol{z}_r)$. Then it defines a loss function that minimizes the Hamming distance of the pair:
\begin{equation}
\mathcal{L}_{radius} = d_H(\boldsymbol{z}_q,\boldsymbol{z}_r) \cdot \boldsymbol{1}_{[r>2m-1]},
\end{equation}
where \(m\) is the number of substrings, the indicator function ensures that the Hamming distance is minimized only when the search radius is excessively large. This controls the search radius \(r\), thereby limiting the exponential increase in the number of hash table lookups. 

\subsection{Gradient Propagation} 
\label{sec:gradient}

Gradient propagation in deep text hashing often encounters challenges, particularly when dealing with the discrete representation layer. For example, the $\rm sign(\cdot)$ function in Eq \ref{eq:sign_inference} leads to a gradient of zero when directly using backpropagation. To estimate the gradients for the $\rm{sign(\cdot)}$ function, the straight-through estimator (STE) \cite{yin2019understanding} is employed. STE simply backpropagates through the discrete representation layer by approximating the gradient $\frac{\partial \boldsymbol{z}}{\partial \phi}=1$, where $\phi$ is the parameter of the encoder. 

However, the STE method can lead to biased gradient estimation. Some works \cite{mena2019binary,zhang2019doc2hash} adopt the Gumbel-Softmax trick to overcome this problem. As we mentioned before, Doc2Vec \cite{zhang2019doc2hash} assumes that the hash codes follow a categorical distribution prior, and then it samples hash codes from the Gumbel distribution as follows:
\begin{equation}
    h_{i} = argmax_k(G_k + \log v_{i}^{(k)}), k \in \{0,1\},
\end{equation}
here, $G_1$ and $G_2$ is drawn i.i.d. from $Gumbel(0, 1) =
\log ( - \log \mu)$ where $\mu \sim {\rm Uniform}(0,1)$, following the uniform distribution. $v_{i}^{(k)}$ denotes the parameters of the k-th category. This process is used in forward propagation. During backpropagation, it employs the tempting softmax as a continuous relaxation of the $\rm{argmax}$ computation:
\begin{equation}
    h_{i} = \frac{exp((G_k + \log v_{i}^{(k)})/\gamma)}{\sum_k exp((G_k + \log v_{i}^{(k)})/\gamma)},k \in \{0,1\}.
\end{equation}
This relaxation approximates the discrete $\rm{argmax}$ computation as the temperature parameter $\gamma$ approaches $0$ yet keeps the relative order of $(G_k + \log v_i^{(k)})$. This method somewhat alleviates the issue of gradient estimation bias, but errors persist. To solve this problem, PSH \cite{dadaneh2020pairwise} further proposes to employ the unbiased ARM estimator \cite{dadaneh2020arsm}.

\begin{table}[tbp]
\caption{A Summary of deep text hashing Models w.r.t the Different Manner of Architecture Design, Similarity Preservation, Hash Code Quality Preservation, Other Technology, and Binarization method. Rec.= Reconstruction-based method, Robustness=Prompting robustness hash codes, Prior(X) = A prior X (X could be G: Gussain, B: Bernoulli, M: Mixture, C: Categorical, BM: Boltzmann, GA: Graph) control of the latent representation, Pse. = Pseudo-similarity-based method, MMI. = Maximal mutual information method, SFC. = Learning semantic from categories, SFR.=Learning semantic from relevance, FE.=Few-bit code, CB.=Code Balance, Quan(X) = Using quantization method (X could be Loss: quantization loss, Sgn: Signum function, Sigmoid: Sigmoid function, Tanh: Tanh function, Stanh: The scaled tanh function).}
\begin{center}
\scalebox{0.83}{
\begin{tabular}{c|c|c|c}
\toprule
Model & Semantic Extraction & Hash Code Quality & Other Technologies \\
\hline
THC \cite{xu2015convolutional} & SFC. &  &  \\
SH-SAE \cite{yu2015understanding} & Rec.+SFC.  & & Robustness \\
DSH \cite{chen2016document} & Rec. & CB. &   \\
VDSH \cite{chaidaroon2017variational} & Rec.+Prior(G) &  &    \\
NbrReg \cite{chaidaroon2018deep} & Rec.+Prior(G)+Pse. &  &  \\
NASH \cite{shen2018nash} & Rec.+Prior(B) & CB.+Quan(Sgn) & Robustness   \\
VPSH \cite{xuan2019variational} & Rec.+Prior(G)+SLR. &  &    \\
SCSE-DH \cite{cui2019short} & Rec.+SFC. & Quan(Sigmoid) & Robustness   \\
RBSH \cite{hansen2019unsupervised} & Rec.+Prior(B)+Pse. & CB.+Quan(Sgn) & Robustness  \\
B-VAE \cite{mena2019binary} & Rec.+Prior(B) & CB.+Quan(Sgn) & Gumbel-softmax   \\
Doc2Hash \cite{zhang2019doc2hash} & Rec.+Prior(C) &  & Gumbel-softmax  \\
GMSH/BMSH \cite{dong2019document} & Rec.+Prior(M) & BMSH:CB.+Quan(Sgn) &   \\
HAS \cite{xu2020hashing} & SFR. & Quan.+Quan(STanh) &    \\
% C\&D \cite{} & &  & &  &  \\
PairRec \cite{hansen2020unsupervised} & Rec.+Prior(B)+Pse. & CB.+Quan(Sgn) & Robustness  \\
Node2Hash \cite{chaidaroon2020node2hash} & Rec.+Prior(G)+Pse. & Drop. &    \\
AMMI \cite{stratos2020learning} & MMI. &  & Adversarial training   \\
WISH \cite{ye2020unsupervised} & Rec.+Prior(B) & CB.+FE.+Quan(Sgn) &    \\
CorrSH \cite{zheng2020generative} & Rec.+Prior(BM) & Quan(Sgn) &    \\
DABA \cite{doan2020efficient} & Rec. & CB.+Quan(Sigmoid) & Robustness  \\
PSH \cite{dadaneh2020pairwise} & Rec.+Prior(B).+SFR. & CB.+Quan(Sgn) & ARM gradient estimator   \\
WAE \cite{zhang2020discrete} & Rec. & CB. &   \\
CVAE/MVAE \cite{xuan2020conditional} & Rec.+Prior(G)+SFR. &  &    \\
SSB-VAE \cite{nanculef2021self} & Rec.+Prior(B)+SFC.+SFR. & CB.+Quan(Sgn) &    \\
MISH \cite{hansen2021unsupervised} & Rec.+Prior(B)+Pse. & CB.+Quan(Sgn) & Adaptation to index   \\
SNUH \cite{ou2021integrating} & Rec.+Prior(GA) & Quan(Sigmoid) &    \\
DHIM \cite{ou2021refining} & MMI. & CB.+Quan(Sgn) &    \\
BPR \cite{yamada2021efficient} & SFR. &  Quan(STanh) &    \\
LASH \cite{guo2021lash} & Rec.+SFR. & Quan(Loss) &    \\
CoSHC \cite{gu2022accelerating} & SFR. & Quan(STanh) &   \\
SAMFH \cite{chen2023exploiting} & Rec.+Prior(B)+Pse. & CB.+Quan(Sgn) &   \\
MASH/SMASH \cite{he2023efficient} & Rec. & CB.+FE.+Quan(Tanh) & Robustness   \\
HDR-BERT \cite{lan2023towards} & SFR. &  &   \\
IHDH \cite{guo2022intra} & Rec.+SFR. & Quan(Loss)+Quan(STanh) &   \\
USH-SER \cite{tong2024efficient} & MMI.+Pse. & CB.+Quan(STanh) &   \\
HierHash \cite{zhang2024document} & Rec.+Prior(G)+MMI.+Pse. &  &   \\
DHSH \cite{huang2025confusing} & Rec.+Prior(B)+SFC.+SFR. & CB.+Quan(Sgn) &   \\

\bottomrule
\end{tabular}}
\label{tab:method_summary}
\end{center}
\end{table}

\section{VALIDATION}
\label{sec:performance}

In this section, we review the commonly followed evaluation procedures on various datasets and the metrics adopted in the literature. We then showcase the dataset-specific improvements made by different methods over the past few years.

\subsection{Datasets for deep text hashing}

With the rapid progress in the deep text hashing domain, several datasets have been used for different tasks. This section primarily focuses on some commonly used evaluation datasets, which range in scale from small to large. We provide
a detailed overview of the datasets in Table~\ref{tab:datasets}. Deep text hashing distinguishes semantics through pre-assigned categories, meaning that texts with the same category are considered related. Moreover, if a dataset is multi-labeled, two texts are deemed related if they share at least one common category label. We describe some important datasets below:
% The scales of commonly used assessment datasets range from small to large. Here, we first introduce the three most commonly used datasets in most deep text hashing models, including 20Newsgroups \footnote{https://scikit-learn.org/0.19/datasets/twenty\_newsgroups.html}, Agnews \footnote{http://groups.di.unipi.it/gulli/AG\_corpus\_of\_news\_articles.html}, and Reuters \footnote{https://www.nltk.org/book/ch02.html}.

\begin{itemize}
\item \textit{20Newsgroups}\footnote{https://scikit-learn.org/0.19/datasets/twenty\_newsgroups.html} consists of 18,828 newsgroup articles from 20 different topics. These topics cover various domains, such as computer technology, sports, religion, etc. Each article is assigned to a specific newsgroup topic.
\item \textit{Agnews}\footnote{http://groups.di.unipi.it/gulli/AG\_corpus\_of\_news\_articles.html} is a commonly used text dataset that contains  127,600 news articles from Agence France-Presse (AFP). This dataset covers four main topics: world, national, business, and technology.  Each article is assigned to a topic.
\item \textit{Reuters}\footnote{https://www.nltk.org/book/ch02.html} is a collection of 10,788 news articles commonly used for text classification and information retrieval tasks. This multi-label dataset encompasses a diverse range of 90 topics, including business, finance, politics, sports, and more. In deep text hashing, the 20 most frequent categories are typically selected. 
\item \textit{RCV1}\footnote{https://www.csie.ntu.edu.tw/~cjlin/libsvmtools/datasets/multilabel.html} is an extensive collection of manually labeled 800,000 newswire stories provided by Reuters, covering a wide range of topics such as politics, economics, sports, and more. There are a total of 103 classes. Each news article can be associated with multiple categories or labels.
\item \textit{TMC}\footnote{https://catalog.data.gov/dataset/siam-2007-text-mining-competition-dataset} is a dataset that contains the air traffic reports provided by NASA and is used as part of the SIAM text-mining competition. It consists of 28,596  air traffic reports divided into 22 different categories.
\item \textit{DBpedia} \footnote{https://emilhvitfeldt.github.io/textdata/reference/dataset\_dbpedia.html} \cite{lehmann2015dbpedia} is a collection of 60,000 documents collected from DBpedia. It classifies these documents into 14 non-overlapping ontology classes.
\item \textit{Yahooanswer}\footnote{https://www.kaggle.com/soumikrakshit/yahoo-answers-dataset} is a large dataset includes  1,460,000 questions collected from the Yahoo Answers platform. These questions are split into 10 topics, including health, education, technology, entertainment, etc. 
\end{itemize}

Most deep text hashing works follow the approach of Chaidaroon et al. \cite{chaidaroon2017variational}, dividing the dataset in an 8:1:1 ratio into training, validation, and test sets. During validation and testing, the training data is used as the database, while the validation or test sets serve as queries. Some datasets have significant class imbalance issues, so they perform sampling to address this. For example, they select the 20 most populous categories from Reuters and the 4 most populous categories from RCV1.

\begin{table}
\begin{center}
\caption{Basic Statistic of Datasets}
\label{tab:datasets} 
\begin{tabular}{cccc}
\toprule
Datasets & Instance & Categories & Single-/Multi-Label \\
\hline
20Newsgroups & 18,846 & 20 & single-label\\
Agnews & 127,600 & 4 & single-label\\
Reuters & 10,788 & 90/20 & muti-label\\
DBpedia & 60,000 & 14 & single-label\\
RCV1 & 804,414 & 103/4 & muti-label\\
TMC & 28,596 & 22 & multi-label \\
NYT & 11,527 & 26 & single-label \\
Yahooanswer &  1,460,000 & 10 & single-label \\
\bottomrule
\end{tabular}
\end{center}
\end{table}

\subsection{Evaluation Metrics}

The effectiveness of the deep text hashing model is typically evaluated using search precision while maintaining a consistent hash code length \( b \). The most popular metrics include \( Precision@K \), \( Recall@K \), precision-recall curve.

$Precision@K$ is a metric used to evaluate the performance of an information retrieval system by measuring how many relevant items are included in the top $K$ retrieved results. The calculation of $Precision@K$ is as follows: 
\begin{equation}
    Precision@K = \frac{1}{N_c}\sum_{i=1}^{N_c}\frac{Top_K(x_i)}{K},
\end{equation}
where \(N_c\) represents the total number of query items in the test set, \(Top_K(x_i)\) denotes the number of relevant items in the top \(K\) of the returned set for the \(i\)-th query item. 
% $Precision@K$ ranges from $0$ to $1$, where a value closer to $1$ indicates a higher proportion of relevant texts in the top $K$ results, indicating better retrieval performance. 
$Recall@K$ measures how well the model covers relevant texts in the top $K$ returned results. The calculation of $Recall@K$ is as follows: 
\begin{equation}
    Recall@K = \frac{1}{N_c}\sum_{i=1}^{N_c}\frac{Top_K(x_i)}{Rel(x_i)},
\end{equation}
where \(Rel(x_i)\) represents the total number of relevant items for the \(i\)-th query item. 

$Precision@K$ and $Recall@K$ are used together to comprehensively evaluate the performance of deep text hashing models. While $Precision@K$ emphasizes accuracy, $Recall@K$ focuses on coverage. Thus, the precision-recall curve is used to show their relationship and trade-offs. The precision-recall curve is a graph plotted with precision on the y-axis and recall on the x-axis. The precision rate and recall rate in text retrieval are both influenced by the returned result number $K$. By adjusting the value of $K$, different combinations using precision rate and recall rate can be obtained. In the precision-recall curve, the ideal scenario is for the curve to be as close to the top-right corner as possible, indicating high precision and recall. Analyzing the Precision-Recall curve can help us choose appropriate thresholds to balance precision and recall, thereby better evaluating and optimizing the performance of deep text hashing models.

% The area under the curve (AUC) below the curve is also an important metric for evaluating classifier performance, where a larger AUC value indicates better classifier performance. Analyzing the Precision-Recall curve can help us choose appropriate thresholds to balance precision and recall, thereby better evaluating and optimizing the performance of models.
    
% $mAP@K$ first calculates the Average Precision (AP) for each query, and then the average of AP for all queries is computed to obtain $mAP@K$. Average Precision (AP) is the average precision of the model's returned results among all relevant texts. The calculation of AP is as follows:
%  \begin{equation}
%      AP = \frac{1}{F} \sum_{k=1}^{n} precision@k{\{T@k}\}
%  \end{equation}
% where ${T\@k}$ denotes the change in recall from item k − 1 to k. The sum of ${T\@k}$ is F and the core idea of AP is to evaluate a ranked list by averaging the precision at each position. Then MAP can be computed by taking the mean of the average precision of every query. $MAP\@K$ is calculated in terms of top K ranked retrieval results.

Additionally, a metric that can be explored in deep text hashing is \( Radius@R \). $Radius@R$ is calculated by setting the search radius to $K$ in the Hamming space retrieving the texts within this range. Then, it calculates their precision, which can be formulated as:
\begin{equation}
Radius@R=\frac{1}{N_c}\sum_{i=1}^{N_c}\frac{Rel(Radius_R(x_i))}{Radius_R(x_i)},
\end{equation}
where \( Radius_R(x_i) \) denotes the number of texts retrieved when the search radius for query \( x_i \) is \( R \), and \( Rel(Radius_R(x_i)) \) represents the number of relevant texts within this retrieval result.

% The calculation of $Radius@R$ is similar to the calculation of $Precision@K$ mentioned above, but there are some differences. $Radius@R$ is the ratio of the number of relevant texts retrieved within the search radius $R$ to the total number of texts retrieved within the search radius R, whereas Precision@K is the ratio of relevant documents to the total number of documents retrieved within the top K results. 

\subsection{Performance Analysis}

This section highlights the advancements accomplished by various deep text hashing methods on several commonly used datasets in recent years. We adopted the widely used metric $Precision@100$ to evaluate performance in three single-label datasets, including 20Newspaper, Agnews, and DBpedia, and two multilabel datasets, including Reuters and TMC. The lengths of the hash code are set to $b \in \{16, 32, 64\}$. The models selected for evaluation are the currently available open-source deep text hashing methods. The results are obtained according to the following rules: (a) If corresponding results are available in the original paper, we directly use them. (b) If the original paper lacks results for specific datasets or hash code lengths, we test them using the provided open-source code, and the results we obtain are marked with an asterisk ($*$). (c) In cases where the open-source code of certain deep text hashing models is missing necessary hyperparameters for specific datasets, making it impossible to run or yielding significantly lower results, we use a dash ($-$) as a placeholder. Furthermore, we attempt to extend the WISH and SMASH models into supervised scenarios by leveraging the VDSH-P method, resulting in two new models, WISH-S and SMASH-S, for further verification. We use a Linux server with a single NVIDIA GeForce RTX 4090 GPU and Intel(R) Xeon(R) Gold 6426Y CPU to conduct the experiments.

Table~\ref{tab:main_table_single} and Table~\ref{tab:main_table_multiple} comprehensively present the experimental results for both single-label and multi-label datasets, offering a clear comparison of performance across different settings and methods. Specifically, the upper sections of these tables summarize the outcomes achieved under unsupervised settings, while the lower sections focus on the results obtained in supervised settings. This division provides a structured view of how the availability of label information impacts the performance of deep text hashing models. In the unsupervised scenario, where no label information is used during training, we observe interesting trends across various datasets. Notably, on datasets such as 20Newspaper, Agnews, Reuters, and TMC, continuous advancements in deep text hashing techniques have led to significant performance improvements over time. These improvements underscore the rapid evolution of the field, as researchers introduce more sophisticated architectures and optimization strategies to capture the semantic structure of textual data better. For example, recent models have demonstrated their ability to generate compact and meaningful hash codes that effectively preserve the semantic relationships between documents. However, it is worth noting that on the DBpedia dataset, earlier methods such as VDSH continue to maintain a distinct advantage over newer approaches. This observation suggests that certain datasets, due to their unique characteristics or inherent structure, might still pose challenges for modern deep hashing models. The enduring success of VDSH on DBpedia highlights the importance of tailoring hashing techniques to specific dataset properties and encourages further exploration into why some methods excel in particular contexts. Overall, these findings emphasize the significant potential for additional innovation and refinement in the design of unsupervised deep text hashing models. In supervised settings, where label information is incorporated during training, the results demonstrate a substantial boost in performance for deep hashing models. The inclusion of labels enables these models to leverage explicit class information, which helps guide the learning process and results in more discriminative hash codes. As a result, supervised deep hashing models consistently outperform their unsupervised counterparts. These findings highlight the importance of exploiting label information when available in practice.

\begin{table}[t]
\centering
\scalebox{0.7}{
\begin{tabular}{c|c|ccc|ccc|ccc|c}
\toprule[1pt]
\multicolumn{1}{c|}{ \multirow{2}*{ Scenario } } & \multicolumn{1}{c|}{ \multirow{2}*{ Methods } } &  \multicolumn{3}{c|}{20Newspaper} & \multicolumn{3}{c|}{Agnews} & \multicolumn{3}{c|}{DBpedia} & \multicolumn{1}{c}{ \multirow{2}*{ Year } } \\

{} & {} & 16 bits & 32 bits & 64 bits & 16 bits & 32 bits & 64 bits & 16 bits & 32 bits & 64 bits & {}\\ 
\hline
\multicolumn{1}{c|}{ \multirow{12}*{ \makecell {Unsupervise}  } } 
& 
 VDSH \cite{chaidaroon2017variational} & 0.3904 & 0.4327 & 0.1731 & 0.7885* & 0.8110* & 0.8190* & 0.7645* & 0.8344* & 0.8592* & 2017\\ 
{} & NASH \cite{shen2018nash} & 0.5310 & 0.6225 & 0.5377 & 0.7097* & 0.7563* & 0.7748* & 0.7110 & 0.7319 & 0.7091 & 2018\\ 
{} & NbrReg \cite{chaidaroon2018deep} & 0.4470 & 0.4898 & 0.5118 & 0.7984 & 0.8149 & 0.8233 & 0.7437 & 0.7996 & 0.8238 & 2018\\ 
{} & Doc2Hash \cite{zhang2019doc2hash} & 0.3297* & 0.4467* & 0.4470* & 0.7597* & 0.7816* & 0.7817* & 0.8070 & 0.8376 & 0.8438 & 2019 \\
{} & RBSH \cite{hansen2019unsupervised} & 0.6087 & 0.6385 & 0.6655 & 0.8288 & 0.8363 & 0.8393 & 0.7145* & 0.7427* & 0.7327* & 2019 \\ 
{} & PairRec \cite{hansen2020unsupervised} & 0.3962* & 0.4226* & 0.4426* & 0.8354 & 0.8452 & 0.8492 & 0.7241* & 0.7436* & 0.7517* &  2020 \\ 
{} & AMMI \cite{stratos2020learning} & 0.3352* & 0.3638* & 0.3857* & 0.8173 & 0.8446 & 0.8506 & 0.7346* & 0.7423* & 0.7373* & 2020 \\
{} & WISH \cite{ye2020unsupervised} & 0.3597* & 0.3818* & 0.3581* & 0.7549* & 0.7583* & 0.7394* & 0.7409* & 0.7614* & 0.7438* & 2020 \\
{} & MISH \cite{hansen2021unsupervised} & 0.3401* & 0.3627* & 0.3923* & 0.7262* & 0.8375 & 0.8419 & 0.7215* & 0.7426* & 0.7638* & 2021 \\
{} & SNUH \cite{ou2021integrating} & 0.5775 & 0.6387 & 0.6646 & 0.7253* & 0.7448* & 0.7631* & 0.7326* & 0.7537* & 0.7825* & 2021 \\
{} & SMASH \cite{he2023efficient} & 0.3424* & 0.3595* & 0.4081* & 0.7496* & 0.7625* & 0.7976* & 0.7683* &  0.7862* & 0.8025* & 2023 \\
\hline
\multicolumn{1}{c|}{ \multirow{5}*{ \makecell {Supervise}  } } 
& 
VDSH-S \cite{chaidaroon2017variational} & 0.6791 & 0.7564 & 0.6850 & 0.8993* & 0.9020* & 0.8999* & 0.9711* & 0.9748* & 0.9758* & 2017\\
{} & VDSH-SP \cite{chaidaroon2017variational} & 0.6551 & 0.7125 & 0.7045 & 0.9008* & 0.9019* & 0.9006* & 0.9723* & 0.9745* & 0.9752* & 2017 \\
{} & NASH-S \cite{shen2018nash} & 0.6973 & 0.8069 & 0.8213 & 0.8883* & 0.8831* & 0.8705* & 0.8693* & 0.8637* & 0.8511* & 2018 \\
{} & Doc2Hash-S \cite{zhang2019doc2hash} & 0.5894* & 0.6789* & 0.6948* & 0.8209* & 0.8557* & 0.8805* & 0.9221* & 0.9436* & 0.9481* & 2019 \\
{} & WISH-S \cite{ye2020unsupervised} & 0.6904* & 0.6952* & 0.6785* & 0.8933* & 0.8862* & 0.8707* & 0.9695* & 0.9591* & 0.9501* & 2020 \\
{} & SSB-VAE \cite{nanculef2021self} & 0.8730 & 0.8800 &  0.8508* & 0.8507* & 0.8502* & 0.8498* & 0.8716* & 0.8715* & 0.8724* & 2021 \\
{} & SMASH-S \cite{he2023efficient} & 0.6765* & 0.6517* & 0.6457* & 0.8989* & 0.8742* & 0.8847* & 0.9688* & 0.9734* & 0.9753* & 2023 \\

\bottomrule[1pt]
\end{tabular}}
\caption{The Precision@100 results on six popular datasets for each year's state-of-the-art models, as reported in the original papers. *: The original paper lacks results for specific datasets or hash code lengths; we test them using the provided open-source code.}

\label{tab:main_table_single}
\end{table}

\begin{table}[t]
\centering

\scalebox{0.7}{
\begin{tabular}{c|c|ccc|ccc|c}
\toprule[1pt]
\multicolumn{1}{c|}{ \multirow{2}*{ Metric } } & \multicolumn{1}{c|}{ \multirow{2}*{ Methods } } &  \multicolumn{3}{c|}{Reuters} & \multicolumn{3}{c|}{TMC} & \multicolumn{1}{c}{ \multirow{2}*{ Year } } \\

{} & {} & 16 bits & 32 bits & 64 bits & 16 bits & 32 bits & 64 bits & {}\\ 
\hline
\multicolumn{1}{c|}{ \multirow{12}*{ \makecell {Unsupervise}  } } 
& 
 VDSH \cite{chaidaroon2017variational} & 0.7165 & 0.7753 & 0.7456 & 0.6853 & 0.7108 & 0.4410 & 2017 \\ 
{} & NASH \cite{shen2018nash} & 0.7624 & 0.7993 & 0.7812 & 0.6956 & 0.7327 & 0.7010 & 2018 \\
{} & NbrReg \cite{chaidaroon2018deep} & 0.2902* & 0.3015* & 0.3311* & 0.4833* & 0.4862* & 0.4855* & 2018\\ 
{} & Doc2Hash \cite{zhang2019doc2hash} & 0.7224 & 0.7473 & 0.7532 & 0.7224 & 0.7473 & 0.7532 & 2019 \\ 
{} & RBSH \cite{hansen2019unsupervised} & 0.7740 & 0.8149 & 0.8120 & 0.7253* & 0.7436* & 0.7551* & 2019 \\
{} & PairRec \cite{hansen2020unsupervised} & 0.8028 & 0.8268 & 0.8329 & 0.7991 & 0.8239 & 0.8280 & 2020 \\
{} & AMMI \cite{stratos2020learning} & 0.8173 & 0.8446 & 0.8506 & 0.7096 & 0.7416 & 0.7522 & 2020 \\
{} & WISH \cite{ye2020unsupervised} & 0.7129* & 0.7268* & 0.7044* & 0.7484* & 0.7573* & 0.7404* & 2020 \\
{} & MISH \cite{hansen2021unsupervised} & 0.7025* & 0.8286 & 0.8377 & 0.7527* & 0.8156 & 0.8261 & 2021 \\
{} & SNUH \cite{ou2021integrating} & 0.8320 & 0.8466 & 0.8560 & 0.7251 & 0.7543 & 0.7658 & 2021 \\
{} & SMASH \cite{he2023efficient} & 0.7006* & 0.7061* & 0.7385* & 0.5214* & 0.5282* & 0.7791* & 2023 \\
\hline
\multicolumn{1}{c|}{ \multirow{5}*{ \makecell {Supervise}  } } 
& 
VDSH-S \cite{chaidaroon2017variational} & 0.9121 & 0.9337 & 0.9407  & 0.7883 & 0.7967 & 0.8018& 2017\\
{} & VDSH-SP \cite{chaidaroon2017variational} & 0.9326 & 0.9283 & 0.9286 & 0.7891 & 0.7888 & 0.7970 & 2017 \\
{} & NASH-S \cite{shen2018nash} & 0.8693* & 0.8637* & 0.8511* & 0.7946 & 0.7987 & 0.8014& 2018 \\
{} & Doc2Hash-S \cite{zhang2019doc2hash} & 0.9338 & 0.9557 & 0.9602 & 0.8472 & 0.8490 & 0.8492  & 2019 \\
{} & WISH-S \cite{ye2020unsupervised} & 0.8636* & 0.8520* & 0.8248* & 0.8576* & 0.8637* & 0.8479* & 2020 \\
{} & SSB-VAE \cite{nanculef2021self} & 0.8425* & 0.8639* & 0.9196* & 0.8080 & 0.8180 & 0.8191* & 2021 \\
{} & SMASH-S \cite{he2023efficient} & 0.8462* & 0.8338* & 0.8251* & 0.8273* & 0.8647* & 0.8561* & 2023 \\
\bottomrule[1pt]
\end{tabular}}

\caption{The Precision@100 results on six popular datasets for each year's state-of-the-art models, as reported in the original papers. *: The original paper lacks results for specific datasets or hash code lengths; we test them using the provided open-source code.}

\label{tab:main_table_multiple}
\end{table}

\section{Application and Open-source Tools}
\label{sec:application}

\subsection{Application}
Deep text hashing techniques have been successfully applied to various text-related downstream tasks, enabling efficient large-scale retrieval and matching. The key advantages of deep text hashing in these applications include reduced storage requirements, faster retrieval times, and the ability to preserve semantic similarities for better retrieval results. We can broadly categorize these applications into: (1) General Information Retrieval, (2) Specialized Domain Information Retrieval, and (3) Intelligent Systems and Task-Specific Support. We summarize these major application areas in this section.

\subsubsection{General Information Retrieval}
General Information Retrieval represents the most fundamental and widely adopted application area for deep text hashing. The core task is to identify documents within a large collection that exhibit similarity to a given query document \cite{park2002web,muangprathub2021document,arabi2022improving}. This capability is crucial for various common tasks such as retrieving similar news articles, detecting analogous web content, and conducting plagiarism analysis. The process typically involves comparing the content, structure, or semantic meaning of documents to uncover those most relevant or closely related. Many benchmark datasets frequently employed in deep text hashing research, including 20Newsgroups, Agnews, Reuters, RCV1, TMC, and DBpedia, are directly associated with this application domain. Furthermore, the methods developed for general similar document retrieval often serve as a foundation upon which researchers build specialized deep text hashing approaches through adaptive improvements tailored to specific downstream applications.

\subsubsection{Specialized Domain Information Retrieval}
Beyond general documents, deep text hashing is increasingly applied to retrieve information within specific, structured domains, often leveraging domain-specific characteristics.

\paragraph{Academic Literature Retrieval} The literature retrieval task involves searching and extracting relevant academic papers, articles, or documents from large databases or repositories \cite{liu2014literature,ding2021attention}. This is vital in research and academia where efficient access to the right literature informs and supports new studies. As literature databases grow exponentially, deep text hashing methods offer a scalable solution. A pioneering work \cite{yaxue2020convolutional} applied deep text hashing using convolutional neural networks to extract text features and generate compact binary hash codes for rapid retrieval. Recognizing the importance of citation networks, Large-Scale Academic deep text hashing (LASH) \cite{guo2021lash} combines word embeddings and citation network information to generate hash codes, allowing efficient retrieval while preserving both semantic and structural similarities between papers.

\paragraph{Code Retrieval} Code retrieval aims to find relevant pieces of code from large repositories or databases in response to a user's query \cite{feng2020codebert,li2024consider,arakelyan2022ns3,wang2023codet5}. This task is essential in software development and maintenance, enabling developers to quickly locate specific code snippets, functions, or libraries. Code Search with Deep Hashing and Code Classification (CoSHC) \cite{gu2022accelerating} demonstrated the feasibility of this approach by generating binary hash codes for both source code and queries based on representations from existing models. Evaluation results showed that CoSHC could preserve over 99\% of the performance of baseline models while significantly accelerating the search process, highlighting the potential of deep text hashing in the software engineering domain.

\subsubsection{Intelligent Systems and Task-Specific Support}
Deep text hashing also serves as a crucial component within larger intelligent systems, optimizing specific tasks that require efficient matching or retrieval over textual data.

\paragraph{Question Answering Systems} Retrieval for question answering involves finding relevant information from a large corpus of documents to answer a specific question \cite{zhu2021retrieving,kratzwald2018adaptive,wang2019multi,nie2020dc}. Deep hashing techniques enhance the efficiency of these systems. For instance, Hash-based Answer Selection (HAS) \cite{xu2020hashing} uses hashing to learn binary matrix representations for answers, significantly reducing the memory cost compared to previous techniques in answer selection tasks. Binary Paragraph Retriever (BPR) \cite{yamada2021efficient} integrates hashing into the dense paragraph retriever (DPR) \cite{karpukhin2020dense}, using binary codes for efficient candidate generation and continuous vectors for accurate re-ranking. Weighted Binary Passage Retriever (WBPR) \cite{xuan2024fast} further refines the distance computation in Hamming space by incorporating dimension weights, building upon the foundation of BPR. HDR-BERT \cite{lan2023towards} employs deep text hashing for an efficient coarse-grained response selection subsystem, reducing index storage while maintaining high recall accuracy.

\paragraph{Intelligent Education} In the field of intelligent education, deep text hashing facilitates applications like personalized learning and resource recommendation. A key task is similar exercise retrieval, which involves finding exercises or problems similar to a given one \cite{hage2006ice,liu2018finding,huang2021disenqnet,tong2020exploiting}. This is valuable for students seeking practice variations or educators creating assessments. USH-SER \cite{tong2024efficient} is an example of an efficient similar question retrieval technology based on deep text hashing. It utilizes time-varying activation functions to reduce encoding information loss and incorporates bit balance and independence objectives during optimization to maximize the effective use of the Hamming space for representing and retrieving educational exercises.

\subsection{Open-source Tools}
Deep text hashing reduces the computational complexity of the retrieval process by performing hashing operations on representations. However, commonly used deep learning libraries like PyTorch and TensorFlow do not support hashing operation acceleration. Thus, this section introduces some open-source tools that support hashing operations during retrieval.
 
\begin{figure}[t]
\centering
\subfigure[]{
\label{fig_first_case}
\includegraphics[width=0.46\textwidth]{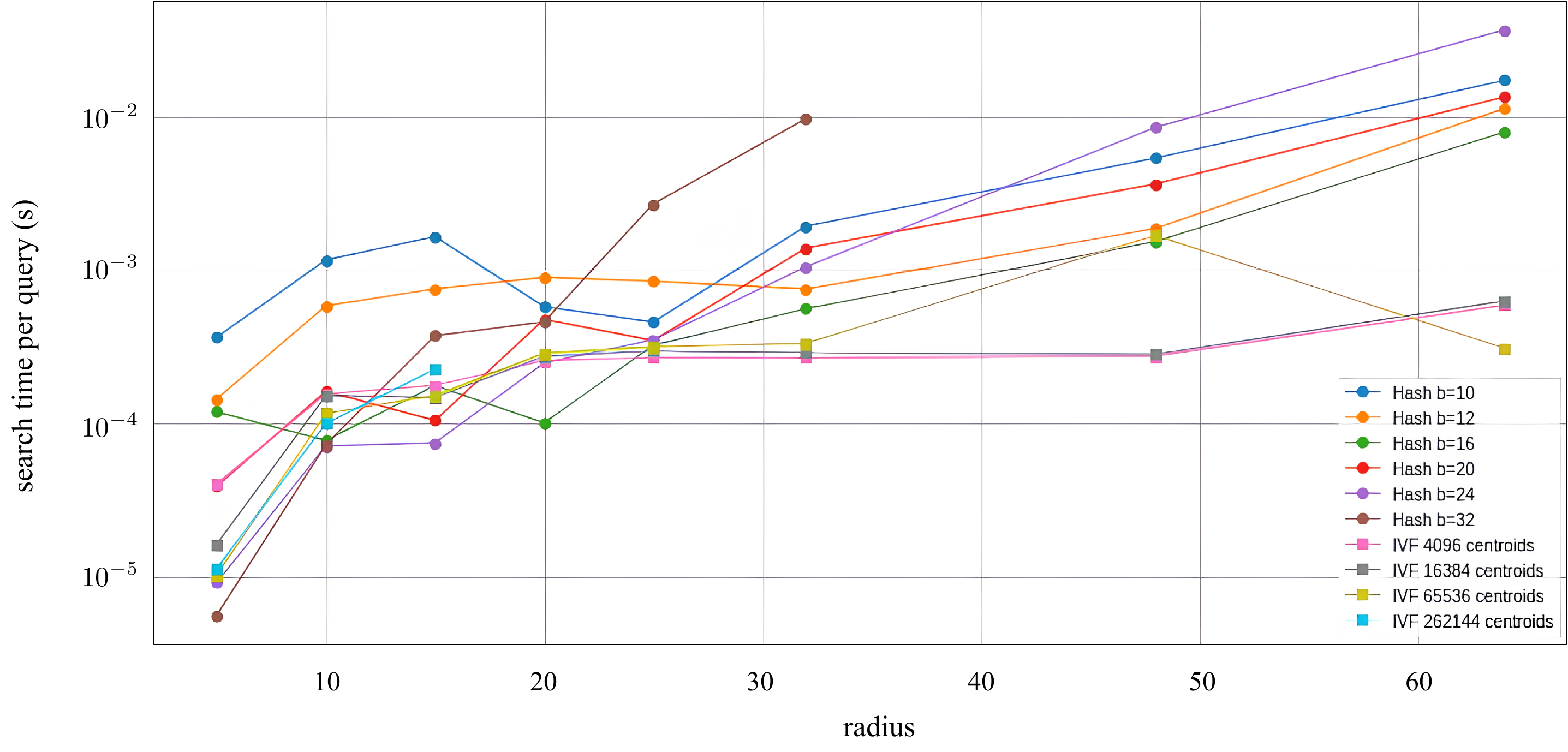}}
\subfigure[]{
\label{fig_second_case}
\includegraphics[width=0.46\textwidth]{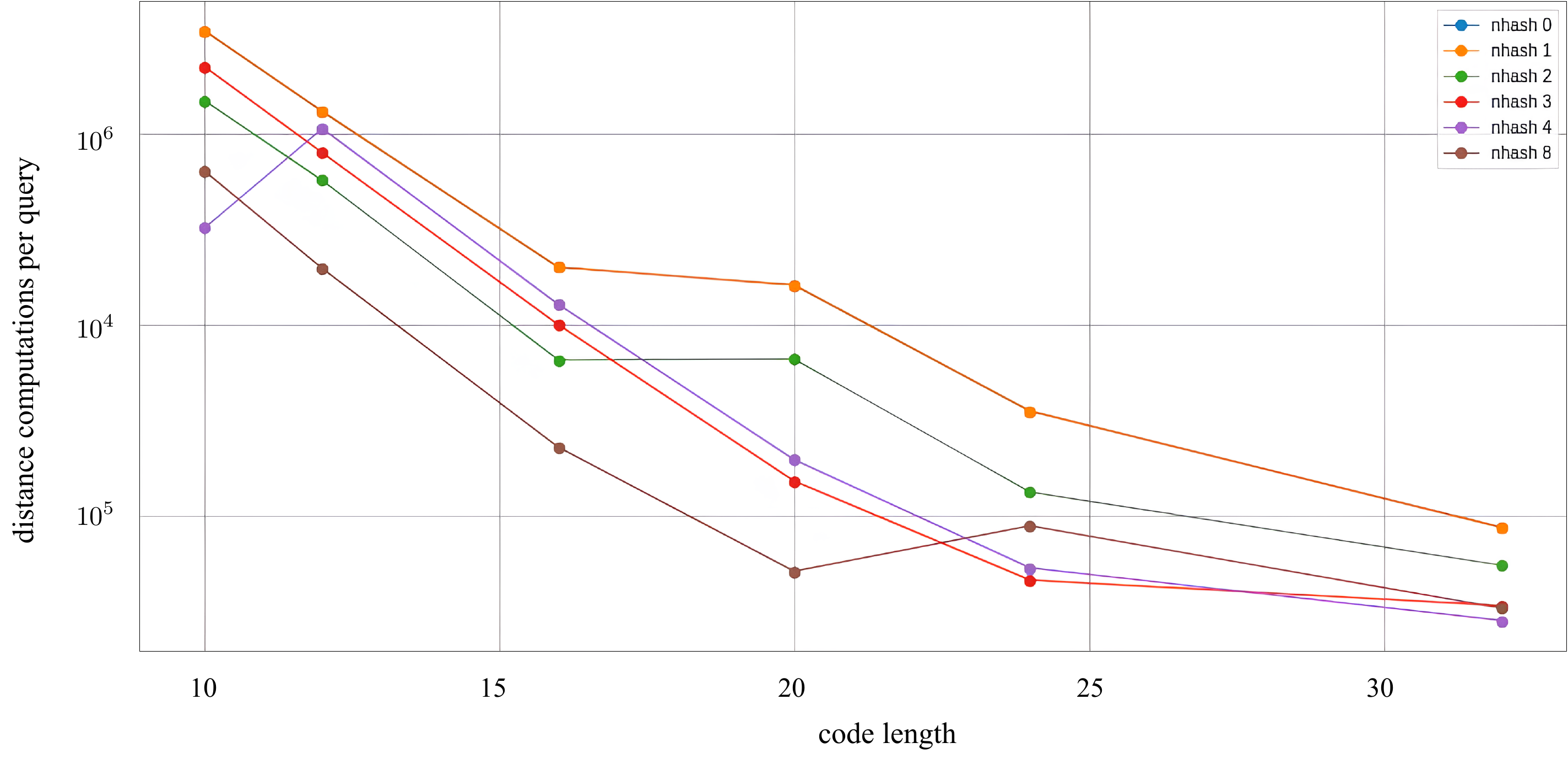}}
\caption{The Faiss group \cite{douze2024faiss} employs a database of 50 million 256-bit binary hash vectors and a query set of 10,000 256-bit binary hash vectors to conduct experiments (a) the comparison between hashing index (IndexBinaryHash) and the inverted file index (IndexBinaryIVF). They fix the target recall at 99\% and select the most cost-efficient operating point that achieves this recall for each radius. (b) The comparison between Multi-index hashing (\( nhash > 1 \)) and the hashing index (\( nhash = 0, 1 \)). }
\label{fig:search_time}
\end{figure}

Facebook AI Similarity Search (FAISS) \cite{douze2024faiss} is an open-source library developed by Facebook AI Research for efficient similarity search and dense vector clustering, particularly suitable for handling large-scale datasets with billions of entries. For deep text hashing, FAISS proposes two specific indexes, IndexBinaryHash and IndexBinaryMultiHash. IndexBinaryHash is the classical method to access hash buckets within a  Hamming radius $r$ from the query vector's hash code, and IndexBinaryMultiHash is the implementation of multi-index hashing \cite{norouzi2012fast}. Additionally, it supports other indexing binary representation methods, which can also be utilized by deep text hashing, including IndexBinaryFlat, IndexBinaryIVF, and IndexBinaryHNSW. IndexBinaryFlat performs an exhaustive search and optimizes it using popcount CPU instructions. This process is equivalent to the previously mentioned hash code ranking method. IndexBinaryIVF speeds up the search by clustering the vectors using an inverse vector file. IndexBinaryHNSW uses binary vectors to represent data and constructs an HNSW index \cite{malkov2018efficient}. Figure~\ref{fig:search_time} illustrates an analysis of efficiency across IndexBinaryFlat, IndexBinaryHash, and IndexBinaryMultiHash. The findings reveal that: (a) for smaller radius, IndexBinaryHash excels in minimizing the number of distance computations. However, as the radius increases beyond 32, IndexBinaryIVF demonstrates superior performance. (b) Increasing the number of hash tables—effectively transforming the Hashing Index into Multi-Index Hashing—slightly reduces the number of distance computations. Nevertheless, this increase in hash tables proportionally amplifies the number of random accesses required. Since each distance computation necessitates a corresponding random access, this ultimately results in a decline in overall performance.

In addition to FAISS, which specifically develops indexing methods for deep text hashing searches, there are numerous open-source tools available that support searches with binary representations. For example, Qdrant \footnote{\url{https://qdrant.tech}} is a high-performance vector database for providing vector similarity search and storage services, which can be easily integrated with modern deep learning models. It provides binary quantization to reduce memory consumption and improve retrieval speeds up to 40 times. Non-Metric Space Library (Nmslib) \footnote{\url{https://github.com/nmslib/nmslib}} is an efficient similarity search library designed for data analysis and large-scale approximate nearest neighbor search. The library constructs efficient approximate graph structures based on the HNSW algorithm to find nearby data points in high-dimensional spaces quickly. It also supports search in the Hamming space, which is particularly useful for binary data representations. Approximate Nearest Neighbors Oh Yeah (Annoy) \footnote{\url{https://github.com/spotify/annoy}} is a fast nearest neighbor search library, which can create large read-only file-based data structures that are mapped into memory so that many processes may share the same data. It supports the Hamming distance computation between binary vectors and uses built-in bit count primitives for fast search. In addition to the aforementioned open-source tools, USearch \footnote{https://github.com/unum-cloud/usearch}, Milvus \footnote{https://github.com/milvus-io/milvus}, Vespa AI \footnote{https://github.com/vespa-engine/vespa}, and Weaviate \footnote{https://github.com/weaviate/weaviate} also support similar operations.

% Sentence Transformers \footnote{\url{https://github.com/UKPLab/sentence-transformers}} is an open-source library for embedding sentences, paragraphs, and documents into high-dimensional dense vector spaces. Its purpose is to simplify the conversion of pre-trained language models like BERT into sentence-level semantic representations. 

% Initially developed by UKP Lab, its purpose is to simplify the conversion of pre-trained language models like BERT into sentence-level semantic representations. By fine-tuning the original BERT or RoBERTa models, the library generates vector embeddings suitable for similarity computation, clustering, and information retrieval, while allowing users to customize training strategies for fine-tuning. 

% However, it should be noted that although Sentence Transformers can achieve high-precision model binarization, it cannot directly utilize binary vectors to accelerate retrieval and often needs to be combined with other libraries, such as FAISS, for this purpose.

\section{Conclusion and Future Directions}
\label{sec:future}

In this survey, we offer a comprehensive review of the literature on deep text hashing. We begin by systematically categorizing various approaches based on two key aspects emphasized by current deep text hashing models: semantic extraction and hash code quality. Subsequently, we present performance evaluation results on several widely used benchmark datasets and summarize the prevailing directions of application. While significant progress has been made in recent years, the field still faces several open challenges that hinder its deployment in more realistic and diverse scenarios. In this section, we summarize the key limitations of current approaches and outline four major directions for future research: (1) constructing more realistic and fine-grained evaluation benchmarks, (2) designing adaptive and robust models for dynamic environments, (3) scaling deep text hashing with large language models, and (4) expanding its applicability beyond retrieval.

\textbf{Constructing More Realistic and Fine-Grained Evaluation Benchmarks}. Most existing deep text hashing methods are evaluated on datasets with a limited number of coarse-grained categories, such as 20Newsgroups and RCV1. However, real-world retrieval systems often operate in fine-grained semantic spaces with thousands of categories, where such datasets fall short in reflecting practical challenges. Moreover, widely used text retrieval benchmarks like MS MARCO~\cite{Bajaj} and BEIR~\cite{Thakur} have not yet been systematically explored in the context of deep text hashing. Thus, future research could focus on: (1) Developing or adapting large-scale, fine-grained datasets specifically tailored for hashing-based retrieval. (2) Establishing standardized evaluation protocols to better assess the performance of deep text hashing models in realistic settings.

\textbf{Designing Adaptive and Robust Models for Open-world Environments}. Current deep text hashing models are primarily designed for static, offline environments, with limited attention paid to online or dynamic settings. In many real-world applications, data is continuously generated and updated, necessitating models that can adapt incrementally without retraining from scratch. Thus, future research could focus on: (1) Developing mechanisms for updating hash functions or representations incrementally as new data arrives while maintaining consistency in the hash space. (2) Addressing the robustness of hash codes to noisy or adversarial inputs, especially in user-generated queries where spelling errors, ambiguity, or incomplete information are common. Although some existing works~\cite{doan2020efficient,he2023efficient} propose basic denoising techniques, a deeper understanding of how noise affects binary representations is still lacking. (3) Enabling models to handle unseen or evolving semantic concepts during inference, which is crucial for deployment in open-domain or long-tail scenarios.

\textbf{Scaling Deep Text Hashing with Large Language Models}. The advent of large language models (LLMs) such as BGE-m3~\cite{chen2024bge}, NV-Embed~\cite{lee2024nv}, and SFR-Embedding-2~\cite{mengsfr} has significantly advanced the state of semantic representation in natural language processing. These models, which often contain billions of parameters, exhibit strong few-shot generalization and semantic understanding capabilities. A recent study~\cite{shakir2024quantization} shows that simple binarization (e.g., sign function) can retain over 90\% of the performance of some large embedding models. This opens up several promising research avenues: (1) Investigating how to effectively binarize embeddings produced by LLMs while minimizing information loss. (2) Exploring parameter-efficient fine-tuning strategies (e.g., adapter layers, LoRA, prompt tuning) to adapt LLMs for hashing tasks without incurring high computational overhead. (3) Addressing the trade-off between the accuracy gains from LLMs and the computational cost during training and inference, especially in resource-constrained environments.

\textbf{Expanding the Applicability of Deep Text Hashing Beyond Retrieval}. Although most of the existing work focuses on retrieval-oriented tasks, the core objective of deep text hashing, learning compact, semantically meaningful binary representations, makes it a valuable tool for a broader range of applications in representation learning and efficient computation. Future research could explore: (1) Applying hash codes as intermediate or final representations in tasks such as text classification, clustering, or recommendation, where storage and computation efficiency are critical.
(2) Investigating whether binary representation learning can serve as a pretraining objective that benefits downstream tasks, similar to masked language modeling in traditional NLP pipelines. (3) Leveraging the compact and less interpretable nature of hash codes to enhance privacy and communication efficiency in decentralized learning settings.

\begin{acks}
To Robert, for the bagels and explaining CMYK and color spaces.
\end{acks}

%%
%% The next two lines define the bibliography style to be used, and
%% the bibliography file.
\bibliographystyle{ACM-Reference-Format}
\bibliography{reference}

%%% -*-BibTeX-*-
%%% Do NOT edit. File created by BibTeX with style
%%% ACM-Reference-Format-Journals [18-Jan-2012].

\begin{thebibliography}{169}

%%% ====================================================================
%%% NOTE TO THE USER: you can override these defaults by providing
%%% customized versions of any of these macros before the \bibliography
%%% command.  Each of them MUST provide its own final punctuation,
%%% except for \shownote{}, \showDOI{}, and \showURL{}.  The latter two
%%% do not use final punctuation, in order to avoid confusing it with
%%% the Web address.
%%%
%%% To suppress output of a particular field, define its macro to expand
%%% to an empty string, or better, \unskip, like this:
%%%
%%% \newcommand{\showDOI}[1]{\unskip}   % LaTeX syntax
%%%
%%% \def \showDOI #1{\unskip}           % plain TeX syntax
%%%
%%% ====================================================================

\ifx \showCODEN    \undefined \def \showCODEN     #1{\unskip}     \fi
\ifx \showDOI      \undefined \def \showDOI       #1{#1}\fi
\ifx \showISBNx    \undefined \def \showISBNx     #1{\unskip}     \fi
\ifx \showISBNxiii \undefined \def \showISBNxiii  #1{\unskip}     \fi
\ifx \showISSN     \undefined \def \showISSN      #1{\unskip}     \fi
\ifx \showLCCN     \undefined \def \showLCCN      #1{\unskip}     \fi
\ifx \shownote     \undefined \def \shownote      #1{#1}          \fi
\ifx \showarticletitle \undefined \def \showarticletitle #1{#1}   \fi
\ifx \showURL      \undefined \def \showURL       {\relax}        \fi
% The following commands are used for tagged output and should be
% invisible to TeX
\providecommand\bibfield[2]{#2}
\providecommand\bibinfo[2]{#2}
\providecommand\natexlab[1]{#1}
\providecommand\showeprint[2][]{arXiv:#2}

\bibitem[Abbasifard et~al\mbox{.}(2014)]%
        {abbasifard2014survey}
\bibfield{author}{\bibinfo{person}{Mohammad~Reza Abbasifard},
  \bibinfo{person}{Bijan Ghahremani}, {and} \bibinfo{person}{Hassan Naderi}.}
  \bibinfo{year}{2014}\natexlab{}.
\newblock \showarticletitle{A survey on nearest neighbor search methods}.
\newblock \bibinfo{journal}{\emph{International Journal of Computer
  Applications}} \bibinfo{volume}{95}, \bibinfo{number}{25}
  (\bibinfo{year}{2014}).
\newblock


\bibitem[Ackley et~al\mbox{.}(1985)]%
        {ackley1985learning}
\bibfield{author}{\bibinfo{person}{David~H Ackley}, \bibinfo{person}{Geoffrey~E
  Hinton}, {and} \bibinfo{person}{Terrence~J Sejnowski}.}
  \bibinfo{year}{1985}\natexlab{}.
\newblock \showarticletitle{A learning algorithm for Boltzmann machines}.
\newblock \bibinfo{journal}{\emph{Cognitive science}} \bibinfo{volume}{9},
  \bibinfo{number}{1} (\bibinfo{year}{1985}), \bibinfo{pages}{147--169}.
\newblock


\bibitem[Ahmed et~al\mbox{.}(2020)]%
        {ahmed2020k}
\bibfield{author}{\bibinfo{person}{Mohiuddin Ahmed}, \bibinfo{person}{Raihan
  Seraj}, {and} \bibinfo{person}{Syed Mohammed~Shamsul Islam}.}
  \bibinfo{year}{2020}\natexlab{}.
\newblock \showarticletitle{The k-means algorithm: A comprehensive survey and
  performance evaluation}.
\newblock \bibinfo{journal}{\emph{Electronics}} \bibinfo{volume}{9},
  \bibinfo{number}{8} (\bibinfo{year}{2020}), \bibinfo{pages}{1295}.
\newblock


\bibitem[Arabi and Akbari(2022)]%
        {arabi2022improving}
\bibfield{author}{\bibinfo{person}{Hamed Arabi} {and} \bibinfo{person}{Mehdi
  Akbari}.} \bibinfo{year}{2022}\natexlab{}.
\newblock \showarticletitle{Improving plagiarism detection in text document
  using hybrid weighted similarity}.
\newblock \bibinfo{journal}{\emph{Expert Systems with Applications}}
  \bibinfo{volume}{207} (\bibinfo{year}{2022}), \bibinfo{pages}{118034}.
\newblock


\bibitem[Arakelyan et~al\mbox{.}(2022)]%
        {arakelyan2022ns3}
\bibfield{author}{\bibinfo{person}{Shushan Arakelyan}, \bibinfo{person}{Anna
  Hakhverdyan}, \bibinfo{person}{Miltiadis Allamanis}, \bibinfo{person}{Luis
  Garcia}, \bibinfo{person}{Christophe Hauser}, {and} \bibinfo{person}{Xiang
  Ren}.} \bibinfo{year}{2022}\natexlab{}.
\newblock \showarticletitle{NS3: Neuro-symbolic semantic code search}.
\newblock \bibinfo{journal}{\emph{Advances in Neural Information Processing
  Systems}}  \bibinfo{volume}{35} (\bibinfo{year}{2022}),
  \bibinfo{pages}{10476--10491}.
\newblock


\bibitem[Bajaj et~al\mbox{.}(2016)]%
        {Bajaj}
\bibfield{author}{\bibinfo{person}{Payal Bajaj}, \bibinfo{person}{Daniel
  Campos}, \bibinfo{person}{Nick Craswell}, \bibinfo{person}{Li Deng},
  \bibinfo{person}{Jianfeng Gao}, \bibinfo{person}{Xiaodong Liu},
  \bibinfo{person}{Rangan Majumder}, \bibinfo{person}{Andrew McNamara},
  \bibinfo{person}{Bhaskar Mitra}, \bibinfo{person}{Tri Nguyen},
  {et~al\mbox{.}}} \bibinfo{year}{2016}\natexlab{}.
\newblock \showarticletitle{Ms marco: A human generated machine reading
  comprehension dataset}.
\newblock \bibinfo{journal}{\emph{arXiv preprint arXiv:1611.09268}}
  (\bibinfo{year}{2016}).
\newblock


\bibitem[Bazzo et~al\mbox{.}(2020)]%
        {bazzo2020assessing}
\bibfield{author}{\bibinfo{person}{Guilherme~Torresan Bazzo},
  \bibinfo{person}{Gustavo~Acauan Lorentz}, \bibinfo{person}{Danny
  Suarez~Vargas}, {and} \bibinfo{person}{Viviane~P Moreira}.}
  \bibinfo{year}{2020}\natexlab{}.
\newblock \showarticletitle{Assessing the impact of OCR errors in information
  retrieval}. In \bibinfo{booktitle}{\emph{Advances in Information Retrieval:
  42nd European Conference on IR Research, ECIR 2020, Lisbon, Portugal, April
  14--17, 2020, Proceedings, Part II 42}}. Springer, \bibinfo{pages}{102--109}.
\newblock


\bibitem[Bengio et~al\mbox{.}(2013)]%
        {bengio2013representation}
\bibfield{author}{\bibinfo{person}{Yoshua Bengio}, \bibinfo{person}{Aaron
  Courville}, {and} \bibinfo{person}{Pascal Vincent}.}
  \bibinfo{year}{2013}\natexlab{}.
\newblock \showarticletitle{Representation learning: A review and new
  perspectives}.
\newblock \bibinfo{journal}{\emph{IEEE transactions on pattern analysis and
  machine intelligence}} \bibinfo{volume}{35}, \bibinfo{number}{8}
  (\bibinfo{year}{2013}), \bibinfo{pages}{1798--1828}.
\newblock


\bibitem[Chaidaroon et~al\mbox{.}(2018)]%
        {chaidaroon2018deep}
\bibfield{author}{\bibinfo{person}{Suthee Chaidaroon}, \bibinfo{person}{Travis
  Ebesu}, {and} \bibinfo{person}{Yi Fang}.} \bibinfo{year}{2018}\natexlab{}.
\newblock \showarticletitle{Deep semantic text hashing with weak supervision}.
  In \bibinfo{booktitle}{\emph{The 41st international acm sigir conference on
  research \& development in information retrieval}}.
  \bibinfo{pages}{1109--1112}.
\newblock


\bibitem[Chaidaroon and Fang(2017)]%
        {chaidaroon2017variational}
\bibfield{author}{\bibinfo{person}{Suthee Chaidaroon} {and} \bibinfo{person}{Yi
  Fang}.} \bibinfo{year}{2017}\natexlab{}.
\newblock \showarticletitle{Variational deep semantic hashing for text
  documents}. In \bibinfo{booktitle}{\emph{Proceedings of the 40th
  International ACM SIGIR Conference on Research and Development in Information
  Retrieval}}. \bibinfo{pages}{75--84}.
\newblock


\bibitem[Chaidaroon et~al\mbox{.}(2020)]%
        {chaidaroon2020node2hash}
\bibfield{author}{\bibinfo{person}{Suthee Chaidaroon},
  \bibinfo{person}{Dae~Hoon Park}, \bibinfo{person}{Yi Chang}, {and}
  \bibinfo{person}{Yi Fang}.} \bibinfo{year}{2020}\natexlab{}.
\newblock \showarticletitle{node2hash: Graph aware deep semantic text hashing}.
\newblock \bibinfo{journal}{\emph{Information Processing \& Management}}
  \bibinfo{volume}{57}, \bibinfo{number}{6} (\bibinfo{year}{2020}),
  \bibinfo{pages}{102143}.
\newblock


\bibitem[Charikar(2002)]%
        {charikar2002similarity}
\bibfield{author}{\bibinfo{person}{Moses~S Charikar}.}
  \bibinfo{year}{2002}\natexlab{}.
\newblock \showarticletitle{Similarity estimation techniques from rounding
  algorithms}. In \bibinfo{booktitle}{\emph{Proceedings of the thiry-fourth
  annual ACM symposium on Theory of computing}}. \bibinfo{pages}{380--388}.
\newblock


\bibitem[Chen et~al\mbox{.}(2016)]%
        {chen2016document}
\bibfield{author}{\bibinfo{person}{Hong Chen}, \bibinfo{person}{Jungang Xu},
  \bibinfo{person}{Qi Wang}, {and} \bibinfo{person}{Ben He}.}
  \bibinfo{year}{2016}\natexlab{}.
\newblock \showarticletitle{A Document Modeling Method Based on Deep Generative
  Model and Spectral Hashing}. In \bibinfo{booktitle}{\emph{Knowledge Science,
  Engineering and Management: 9th International Conference, KSEM 2016, Passau,
  Germany, October 5-7, 2016, Proceedings 9}}. Springer,
  \bibinfo{pages}{402--413}.
\newblock


\bibitem[Chen and Su(2023)]%
        {chen2023exploiting}
\bibfield{author}{\bibinfo{person}{Jiayang Chen} {and}
  \bibinfo{person}{Qinliang Su}.} \bibinfo{year}{2023}\natexlab{}.
\newblock \showarticletitle{Exploiting Multiple Features for Hash Codes
  Learning with Semantic-Alignment-Promoting Variational Auto-encoder}. In
  \bibinfo{booktitle}{\emph{CCF International Conference on Natural Language
  Processing and Chinese Computing}}. Springer, \bibinfo{pages}{563--575}.
\newblock


\bibitem[Chen et~al\mbox{.}(2024b)]%
        {chen2024bge}
\bibfield{author}{\bibinfo{person}{Jianlv Chen}, \bibinfo{person}{Shitao Xiao},
  \bibinfo{person}{Peitian Zhang}, \bibinfo{person}{Kun Luo},
  \bibinfo{person}{Defu Lian}, {and} \bibinfo{person}{Zheng Liu}.}
  \bibinfo{year}{2024}\natexlab{b}.
\newblock \showarticletitle{Bge m3-embedding: Multi-lingual,
  multi-functionality, multi-granularity text embeddings through self-knowledge
  distillation}.
\newblock \bibinfo{journal}{\emph{arXiv preprint arXiv:2402.03216}}
  (\bibinfo{year}{2024}).
\newblock


\bibitem[Chen et~al\mbox{.}(2020)]%
        {chen2020differentiable}
\bibfield{author}{\bibinfo{person}{Ting Chen}, \bibinfo{person}{Lala Li}, {and}
  \bibinfo{person}{Yizhou Sun}.} \bibinfo{year}{2020}\natexlab{}.
\newblock \showarticletitle{Differentiable product quantization for end-to-end
  embedding compression}. In \bibinfo{booktitle}{\emph{International Conference
  on Machine Learning}}. PMLR, \bibinfo{pages}{1617--1626}.
\newblock


\bibitem[Chen et~al\mbox{.}(2024a)]%
        {chen2024learning}
\bibfield{author}{\bibinfo{person}{Xin Chen}, \bibinfo{person}{Hanxian Huang},
  \bibinfo{person}{Yanjun Gao}, \bibinfo{person}{Yi Wang},
  \bibinfo{person}{Jishen Zhao}, {and} \bibinfo{person}{Ke Ding}.}
  \bibinfo{year}{2024}\natexlab{a}.
\newblock \showarticletitle{Learning to Maximize Mutual Information for
  Chain-of-Thought Distillation}.
\newblock \bibinfo{journal}{\emph{arXiv preprint arXiv:2403.03348}}
  (\bibinfo{year}{2024}).
\newblock


\bibitem[Cheng et~al\mbox{.}(2024)]%
        {cheng2024towards}
\bibfield{author}{\bibinfo{person}{Cheng Cheng}, \bibinfo{person}{GuanHao
  Zhao}, \bibinfo{person}{Zhenya Huang}, \bibinfo{person}{Yan Zhuang},
  \bibinfo{person}{Zhaoyuan Pan}, \bibinfo{person}{Qi Liu},
  \bibinfo{person}{Xin Li}, {and} \bibinfo{person}{Enhong Chen}.}
  \bibinfo{year}{2024}\natexlab{}.
\newblock \showarticletitle{Towards Explainable Computerized Adaptive Testing
  with Large Language Model}. In \bibinfo{booktitle}{\emph{Findings of the
  Association for Computational Linguistics: EMNLP 2024}}.
  \bibinfo{pages}{2655--2672}.
\newblock


\bibitem[Chi and Zhu(2017)]%
        {chi2017hashing}
\bibfield{author}{\bibinfo{person}{Lianhua Chi} {and} \bibinfo{person}{Xingquan
  Zhu}.} \bibinfo{year}{2017}\natexlab{}.
\newblock \showarticletitle{Hashing techniques: A survey and taxonomy}.
\newblock \bibinfo{journal}{\emph{ACM Computing Surveys (Csur)}}
  \bibinfo{volume}{50}, \bibinfo{number}{1} (\bibinfo{year}{2017}),
  \bibinfo{pages}{1--36}.
\newblock


\bibitem[Cordeiro et~al\mbox{.}(2023)]%
        {cordeiro2023lesse}
\bibfield{author}{\bibinfo{person}{Nuno~Pablo Cordeiro},
  \bibinfo{person}{Jo{\~a}o Dias}, {and} \bibinfo{person}{Pedro~A Santos}.}
  \bibinfo{year}{2023}\natexlab{}.
\newblock \showarticletitle{LeSSE—A Semantic Search Engine Applied to
  Portuguese Consumer Law}. In \bibinfo{booktitle}{\emph{EPIA Conference on
  Artificial Intelligence}}. Springer, \bibinfo{pages}{118--130}.
\newblock


\bibitem[Cui et~al\mbox{.}(2019)]%
        {cui2019short}
\bibfield{author}{\bibinfo{person}{Wanqiu Cui}, \bibinfo{person}{Junping Du},
  \bibinfo{person}{Dawei Wang}, \bibinfo{person}{Xunpu Yuan},
  \bibinfo{person}{Feifei Kou}, \bibinfo{person}{Liyan Zhou}, {and}
  \bibinfo{person}{Nan Zhou}.} \bibinfo{year}{2019}\natexlab{}.
\newblock \showarticletitle{Short text analysis based on dual semantic
  extension and deep hashing in microblog}.
\newblock \bibinfo{journal}{\emph{ACM Transactions on Intelligent Systems and
  Technology (TIST)}} \bibinfo{volume}{10}, \bibinfo{number}{4}
  (\bibinfo{year}{2019}), \bibinfo{pages}{1--24}.
\newblock


\bibitem[Dadaneh et~al\mbox{.}(2020a)]%
        {dadaneh2020pairwise}
\bibfield{author}{\bibinfo{person}{Siamak~Zamani Dadaneh},
  \bibinfo{person}{Shahin Boluki}, \bibinfo{person}{Mingzhang Yin},
  \bibinfo{person}{Mingyuan Zhou}, {and} \bibinfo{person}{Xiaoning Qian}.}
  \bibinfo{year}{2020}\natexlab{a}.
\newblock \showarticletitle{Pairwise supervised hashing with Bernoulli
  variational auto-encoder and self-control gradient estimator}. In
  \bibinfo{booktitle}{\emph{Conference on Uncertainty in Artificial
  Intelligence}}. PMLR, \bibinfo{pages}{540--549}.
\newblock


\bibitem[Dadaneh et~al\mbox{.}(2020b)]%
        {dadaneh2020arsm}
\bibfield{author}{\bibinfo{person}{Siamak~Zamani Dadaneh},
  \bibinfo{person}{Shahin Boluki}, \bibinfo{person}{Mingyuan Zhou}, {and}
  \bibinfo{person}{Xiaoning Qian}.} \bibinfo{year}{2020}\natexlab{b}.
\newblock \showarticletitle{Arsm gradient estimator for supervised learning to
  rank}. In \bibinfo{booktitle}{\emph{ICASSP 2020-2020 IEEE International
  Conference on Acoustics, Speech and Signal Processing (ICASSP)}}. IEEE,
  \bibinfo{pages}{3157--3161}.
\newblock


\bibitem[Dasgupta et~al\mbox{.}(2011)]%
        {dasgupta2011fast}
\bibfield{author}{\bibinfo{person}{Anirban Dasgupta}, \bibinfo{person}{Ravi
  Kumar}, {and} \bibinfo{person}{Tam{\'a}s Sarl{\'o}s}.}
  \bibinfo{year}{2011}\natexlab{}.
\newblock \showarticletitle{Fast locality-sensitive hashing}. In
  \bibinfo{booktitle}{\emph{Proceedings of the 17th ACM SIGKDD international
  conference on Knowledge discovery and data mining}}.
  \bibinfo{pages}{1073--1081}.
\newblock


\bibitem[Deerwester et~al\mbox{.}(1990)]%
        {deerwester1990indexing}
\bibfield{author}{\bibinfo{person}{Scott Deerwester}, \bibinfo{person}{Susan~T
  Dumais}, \bibinfo{person}{George~W Furnas}, \bibinfo{person}{Thomas~K
  Landauer}, {and} \bibinfo{person}{Richard Harshman}.}
  \bibinfo{year}{1990}\natexlab{}.
\newblock \showarticletitle{Indexing by latent semantic analysis}.
\newblock \bibinfo{journal}{\emph{Journal of the American society for
  information science}} \bibinfo{volume}{41}, \bibinfo{number}{6}
  (\bibinfo{year}{1990}), \bibinfo{pages}{391--407}.
\newblock


\bibitem[Ding and Luo(2021)]%
        {ding2021attention}
\bibfield{author}{\bibinfo{person}{Haoran Ding} {and} \bibinfo{person}{Xiao
  Luo}.} \bibinfo{year}{2021}\natexlab{}.
\newblock \showarticletitle{Attention-based unsupervised keyphrase extraction
  and phrase graph for COVID-19 medical literature retrieval}.
\newblock \bibinfo{journal}{\emph{ACM Transactions on Computing for Healthcare
  (HEALTH)}} \bibinfo{volume}{3}, \bibinfo{number}{1} (\bibinfo{year}{2021}),
  \bibinfo{pages}{1--16}.
\newblock


\bibitem[Doan and Reddy(2020)]%
        {doan2020efficient}
\bibfield{author}{\bibinfo{person}{Khoa~D Doan} {and}
  \bibinfo{person}{Chandan~K Reddy}.} \bibinfo{year}{2020}\natexlab{}.
\newblock \showarticletitle{Efficient implicit unsupervised text hashing using
  adversarial autoencoder}. In \bibinfo{booktitle}{\emph{Proceedings of the Web
  Conference 2020}}. \bibinfo{pages}{684--694}.
\newblock


\bibitem[Dolatshah et~al\mbox{.}(2015)]%
        {dolatshah2015ball}
\bibfield{author}{\bibinfo{person}{Mohamad Dolatshah}, \bibinfo{person}{Ali
  Hadian}, {and} \bibinfo{person}{Behrouz Minaei-Bidgoli}.}
  \bibinfo{year}{2015}\natexlab{}.
\newblock \showarticletitle{Ball*-tree: Efficient spatial indexing for
  constrained nearest-neighbor search in metric spaces}.
\newblock \bibinfo{journal}{\emph{arXiv preprint arXiv:1511.00628}}
  (\bibinfo{year}{2015}).
\newblock


\bibitem[Dong et~al\mbox{.}(2019)]%
        {dong2019document}
\bibfield{author}{\bibinfo{person}{Wei Dong}, \bibinfo{person}{Qinliang Su},
  \bibinfo{person}{Dinghan Shen}, {and} \bibinfo{person}{Changyou Chen}.}
  \bibinfo{year}{2019}\natexlab{}.
\newblock \showarticletitle{Document Hashing with Mixture-Prior Generative
  Models}. In \bibinfo{booktitle}{\emph{Proceedings of the 2019 Conference on
  Empirical Methods in Natural Language Processing and the 9th International
  Joint Conference on Natural Language Processing (EMNLP-IJCNLP)}}.
  \bibinfo{pages}{5226--5235}.
\newblock


\bibitem[Douze et~al\mbox{.}(2024)]%
        {douze2024faiss}
\bibfield{author}{\bibinfo{person}{Matthijs Douze}, \bibinfo{person}{Alexandr
  Guzhva}, \bibinfo{person}{Chengqi Deng}, \bibinfo{person}{Jeff Johnson},
  \bibinfo{person}{Gergely Szilvasy}, \bibinfo{person}{Pierre-Emmanuel
  Mazar{\'e}}, \bibinfo{person}{Maria Lomeli}, \bibinfo{person}{Lucas
  Hosseini}, {and} \bibinfo{person}{Herv{\'e} J{\'e}gou}.}
  \bibinfo{year}{2024}\natexlab{}.
\newblock \showarticletitle{The faiss library}.
\newblock \bibinfo{journal}{\emph{arXiv preprint arXiv:2401.08281}}
  (\bibinfo{year}{2024}).
\newblock


\bibitem[Ericsson et~al\mbox{.}(2022)]%
        {ericsson2022self}
\bibfield{author}{\bibinfo{person}{Linus Ericsson}, \bibinfo{person}{Henry
  Gouk}, \bibinfo{person}{Chen~Change Loy}, {and} \bibinfo{person}{Timothy~M
  Hospedales}.} \bibinfo{year}{2022}\natexlab{}.
\newblock \showarticletitle{Self-supervised representation learning:
  Introduction, advances, and challenges}.
\newblock \bibinfo{journal}{\emph{IEEE Signal Processing Magazine}}
  \bibinfo{volume}{39}, \bibinfo{number}{3} (\bibinfo{year}{2022}),
  \bibinfo{pages}{42--62}.
\newblock


\bibitem[Fan et~al\mbox{.}(2020)]%
        {fan2020deep}
\bibfield{author}{\bibinfo{person}{Lixin Fan}, \bibinfo{person}{Kam~Woh Ng},
  \bibinfo{person}{Ce Ju}, \bibinfo{person}{Tianyu Zhang}, {and}
  \bibinfo{person}{Chee~Seng Chan}.} \bibinfo{year}{2020}\natexlab{}.
\newblock \showarticletitle{Deep Polarized Network for Supervised Learning of
  Accurate Binary Hashing Codes.}. In \bibinfo{booktitle}{\emph{IJCAI}}.
  \bibinfo{pages}{825--831}.
\newblock


\bibitem[Feng et~al\mbox{.}(2020)]%
        {feng2020codebert}
\bibfield{author}{\bibinfo{person}{Zhangyin Feng}, \bibinfo{person}{Daya Guo},
  \bibinfo{person}{Duyu Tang}, \bibinfo{person}{Nan Duan},
  \bibinfo{person}{Xiaocheng Feng}, \bibinfo{person}{Ming Gong},
  \bibinfo{person}{Linjun Shou}, \bibinfo{person}{Bing Qin},
  \bibinfo{person}{Ting Liu}, \bibinfo{person}{Daxin Jiang}, {et~al\mbox{.}}}
  \bibinfo{year}{2020}\natexlab{}.
\newblock \showarticletitle{CodeBERT: A Pre-Trained Model for Programming and
  Natural Languages}. In \bibinfo{booktitle}{\emph{Findings of the Association
  for Computational Linguistics: EMNLP 2020}}. \bibinfo{pages}{1536--1547}.
\newblock


\bibitem[Frayling et~al\mbox{.}(2024)]%
        {frayling2024effective}
\bibfield{author}{\bibinfo{person}{Erlend Frayling}, \bibinfo{person}{Sean
  MacAvaney}, \bibinfo{person}{Craig Macdonald}, {and} \bibinfo{person}{Iadh
  Ounis}.} \bibinfo{year}{2024}\natexlab{}.
\newblock \showarticletitle{Effective Adhoc Retrieval Through Traversal of a
  Query-Document Graph}. In \bibinfo{booktitle}{\emph{European Conference on
  Information Retrieval}}. Springer, \bibinfo{pages}{89--104}.
\newblock


\bibitem[Gao et~al\mbox{.}(2021)]%
        {gao2021simcse}
\bibfield{author}{\bibinfo{person}{Tianyu Gao}, \bibinfo{person}{Xingcheng
  Yao}, {and} \bibinfo{person}{Danqi Chen}.} \bibinfo{year}{2021}\natexlab{}.
\newblock \showarticletitle{Simcse: Simple contrastive learning of sentence
  embeddings}.
\newblock \bibinfo{journal}{\emph{arXiv preprint arXiv:2104.08821}}
  (\bibinfo{year}{2021}).
\newblock


\bibitem[Ge et~al\mbox{.}(2013)]%
        {ge2013optimized}
\bibfield{author}{\bibinfo{person}{Tiezheng Ge}, \bibinfo{person}{Kaiming He},
  \bibinfo{person}{Qifa Ke}, {and} \bibinfo{person}{Jian Sun}.}
  \bibinfo{year}{2013}\natexlab{}.
\newblock \showarticletitle{Optimized product quantization}.
\newblock \bibinfo{journal}{\emph{IEEE transactions on pattern analysis and
  machine intelligence}} \bibinfo{volume}{36}, \bibinfo{number}{4}
  (\bibinfo{year}{2013}), \bibinfo{pages}{744--755}.
\newblock


\bibitem[Grauman and Fergus(2013)]%
        {grauman2013learning}
\bibfield{author}{\bibinfo{person}{Kristen Grauman} {and} \bibinfo{person}{Rob
  Fergus}.} \bibinfo{year}{2013}\natexlab{}.
\newblock \showarticletitle{Learning binary hash codes for large-scale image
  search}.
\newblock \bibinfo{journal}{\emph{Machine learning for computer vision}}
  (\bibinfo{year}{2013}), \bibinfo{pages}{49--87}.
\newblock


\bibitem[Gu et~al\mbox{.}(2022)]%
        {gu2022accelerating}
\bibfield{author}{\bibinfo{person}{Wenchao Gu}, \bibinfo{person}{Yanlin Wang},
  \bibinfo{person}{Lun Du}, \bibinfo{person}{Hongyu Zhang},
  \bibinfo{person}{Shi Han}, \bibinfo{person}{Dongmei Zhang}, {and}
  \bibinfo{person}{Michael~R Lyu}.} \bibinfo{year}{2022}\natexlab{}.
\newblock \showarticletitle{Accelerating code search with deep hashing and code
  classification}.
\newblock \bibinfo{journal}{\emph{arXiv preprint arXiv:2203.15287}}
  (\bibinfo{year}{2022}).
\newblock


\bibitem[Guo et~al\mbox{.}(2022a)]%
        {guo2022semantic}
\bibfield{author}{\bibinfo{person}{Jiafeng Guo}, \bibinfo{person}{Yinqiong
  Cai}, \bibinfo{person}{Yixing Fan}, \bibinfo{person}{Fei Sun},
  \bibinfo{person}{Ruqing Zhang}, {and} \bibinfo{person}{Xueqi Cheng}.}
  \bibinfo{year}{2022}\natexlab{a}.
\newblock \showarticletitle{Semantic models for the first-stage retrieval: A
  comprehensive review}.
\newblock \bibinfo{journal}{\emph{ACM Transactions on Information Systems
  (TOIS)}} \bibinfo{volume}{40}, \bibinfo{number}{4} (\bibinfo{year}{2022}),
  \bibinfo{pages}{1--42}.
\newblock


\bibitem[Guo et~al\mbox{.}(2021)]%
        {guo2021lash}
\bibfield{author}{\bibinfo{person}{Jia-Nan Guo}, \bibinfo{person}{Xian-Ling
  Mao}, \bibinfo{person}{Tian Lan}, \bibinfo{person}{Rong-Xin Tu},
  \bibinfo{person}{Wei Wei}, {and} \bibinfo{person}{Heyan Huang}.}
  \bibinfo{year}{2021}\natexlab{}.
\newblock \showarticletitle{LASH: Large-scale academic deep semantic hashing}.
\newblock \bibinfo{journal}{\emph{IEEE Transactions on Knowledge and Data
  Engineering}} \bibinfo{volume}{35}, \bibinfo{number}{2}
  (\bibinfo{year}{2021}), \bibinfo{pages}{1734--1746}.
\newblock


\bibitem[Guo et~al\mbox{.}(2022b)]%
        {guo2022intra}
\bibfield{author}{\bibinfo{person}{Jia-Nan Guo}, \bibinfo{person}{Xian-Ling
  Mao}, \bibinfo{person}{Wei Wei}, {and} \bibinfo{person}{Heyan Huang}.}
  \bibinfo{year}{2022}\natexlab{b}.
\newblock \showarticletitle{Intra-category aware hierarchical supervised
  document hashing}.
\newblock \bibinfo{journal}{\emph{IEEE Transactions on Knowledge and Data
  Engineering}} \bibinfo{volume}{35}, \bibinfo{number}{6}
  (\bibinfo{year}{2022}), \bibinfo{pages}{6003--6013}.
\newblock


\bibitem[Hage and Aimeru(2006)]%
        {hage2006ice}
\bibfield{author}{\bibinfo{person}{Hicham Hage} {and} \bibinfo{person}{E
  Aimeru}.} \bibinfo{year}{2006}\natexlab{}.
\newblock \showarticletitle{ICE: A system for identification of conflicts in
  exams}. In \bibinfo{booktitle}{\emph{IEEE International Conference on
  Computer Systems and Applications, 2006.}} IEEE, \bibinfo{pages}{980--987}.
\newblock


\bibitem[Hanada et~al\mbox{.}(2016)]%
        {hanada2016effective}
\bibfield{author}{\bibinfo{person}{Raiza Hanada}, \bibinfo{person}{Maria da
  Gra{\c{c}}a~C Pimentel}, \bibinfo{person}{Marco Cristo}, {and}
  \bibinfo{person}{Fernando~Anglada Lores}.} \bibinfo{year}{2016}\natexlab{}.
\newblock \showarticletitle{Effective spelling correction for eye-based typing
  using domain-specific information about error distribution}. In
  \bibinfo{booktitle}{\emph{Proceedings of the 25th ACM International on
  Conference on Information and Knowledge Management}}.
  \bibinfo{pages}{1723--1732}.
\newblock


\bibitem[Hansen et~al\mbox{.}(2019)]%
        {hansen2019unsupervised}
\bibfield{author}{\bibinfo{person}{Casper Hansen}, \bibinfo{person}{Christian
  Hansen}, \bibinfo{person}{Jakob~Grue Simonsen}, \bibinfo{person}{Stephen
  Alstrup}, {and} \bibinfo{person}{Christina Lioma}.}
  \bibinfo{year}{2019}\natexlab{}.
\newblock \showarticletitle{Unsupervised neural generative semantic hashing}.
  In \bibinfo{booktitle}{\emph{Proceedings of the 42nd International ACM SIGIR
  Conference on Research and Development in Information Retrieval}}.
  \bibinfo{pages}{735--744}.
\newblock


\bibitem[Hansen et~al\mbox{.}(2020)]%
        {hansen2020unsupervised}
\bibfield{author}{\bibinfo{person}{Casper Hansen}, \bibinfo{person}{Christian
  Hansen}, \bibinfo{person}{Jakob~Grue Simonsen}, \bibinfo{person}{Stephen
  Alstrup}, {and} \bibinfo{person}{Christina Lioma}.}
  \bibinfo{year}{2020}\natexlab{}.
\newblock \showarticletitle{Unsupervised semantic hashing with pairwise
  reconstruction}. In \bibinfo{booktitle}{\emph{Proceedings of the 43rd
  International ACM SIGIR Conference on Research and Development in Information
  Retrieval}}. \bibinfo{pages}{2009--2012}.
\newblock


\bibitem[Hansen et~al\mbox{.}(2021)]%
        {hansen2021unsupervised}
\bibfield{author}{\bibinfo{person}{Christian Hansen}, \bibinfo{person}{Casper
  Hansen}, \bibinfo{person}{Jakob~Grue Simonsen}, \bibinfo{person}{Stephen
  Alstrup}, {and} \bibinfo{person}{Christina Lioma}.}
  \bibinfo{year}{2021}\natexlab{}.
\newblock \showarticletitle{Unsupervised multi-index semantic hashing}. In
  \bibinfo{booktitle}{\emph{Proceedings of the Web Conference 2021}}.
  \bibinfo{pages}{2879--2889}.
\newblock


\bibitem[He et~al\mbox{.}(2011)]%
        {he2011compact}
\bibfield{author}{\bibinfo{person}{Junfeng He}, \bibinfo{person}{Shih-Fu
  Chang}, \bibinfo{person}{Regunathan Radhakrishnan}, {and}
  \bibinfo{person}{Claus Bauer}.} \bibinfo{year}{2011}\natexlab{}.
\newblock \showarticletitle{Compact hashing with joint optimization of search
  accuracy and time}. In \bibinfo{booktitle}{\emph{CVPR 2011}}. IEEE,
  \bibinfo{pages}{753--760}.
\newblock


\bibitem[He et~al\mbox{.}(2023)]%
        {he2023efficient}
\bibfield{author}{\bibinfo{person}{Liyang He}, \bibinfo{person}{Zhenya Huang},
  \bibinfo{person}{Enhong Chen}, \bibinfo{person}{Qi Liu},
  \bibinfo{person}{Shiwei Tong}, \bibinfo{person}{Hao Wang},
  \bibinfo{person}{Defu Lian}, {and} \bibinfo{person}{Shijin Wang}.}
  \bibinfo{year}{2023}\natexlab{}.
\newblock \showarticletitle{An efficient and robust semantic hashing framework
  for similar text search}.
\newblock \bibinfo{journal}{\emph{ACM Transactions on Information Systems}}
  \bibinfo{volume}{41}, \bibinfo{number}{4} (\bibinfo{year}{2023}),
  \bibinfo{pages}{1--31}.
\newblock


\bibitem[He et~al\mbox{.}({[n.\,d.]})]%
        {he2024one}
\bibfield{author}{\bibinfo{person}{Liyang He}, \bibinfo{person}{Zhenya Huang},
  \bibinfo{person}{Chenglong Liu}, \bibinfo{person}{Rui Li},
  \bibinfo{person}{Runze Wu}, \bibinfo{person}{Qi Liu}, {and}
  \bibinfo{person}{Enhong Chen}.} \bibinfo{year}{[n.\,d.]}\natexlab{}.
\newblock \showarticletitle{One-bit Semantic Hashing: Towards
  Resource-Efficient Hashing Model with Binary Neural Network}. In
  \bibinfo{booktitle}{\emph{ACM Multimedia 2024}}.
\newblock


\bibitem[He et~al\mbox{.}(2024)]%
        {he2024bit}
\bibfield{author}{\bibinfo{person}{Liyang He}, \bibinfo{person}{Zhenya Huang},
  \bibinfo{person}{Jiayu Liu}, \bibinfo{person}{Enhong Chen},
  \bibinfo{person}{Fei Wang}, \bibinfo{person}{Jing Sha}, {and}
  \bibinfo{person}{Shijin Wang}.} \bibinfo{year}{2024}\natexlab{}.
\newblock \showarticletitle{Bit-mask Robust Contrastive Knowledge Distillation
  for Unsupervised Semantic Hashing}. In \bibinfo{booktitle}{\emph{Proceedings
  of the ACM on Web Conference 2024}}. \bibinfo{pages}{1395--1406}.
\newblock


\bibitem[Hinton and Salakhutdinov(2006)]%
        {hinton2006reducing}
\bibfield{author}{\bibinfo{person}{Geoffrey~E Hinton} {and}
  \bibinfo{person}{Ruslan~R Salakhutdinov}.} \bibinfo{year}{2006}\natexlab{}.
\newblock \showarticletitle{Reducing the dimensionality of data with neural
  networks}.
\newblock \bibinfo{journal}{\emph{science}} \bibinfo{volume}{313},
  \bibinfo{number}{5786} (\bibinfo{year}{2006}), \bibinfo{pages}{504--507}.
\newblock


\bibitem[Hu et~al\mbox{.}(2017)]%
        {hu2017pseudo}
\bibfield{author}{\bibinfo{person}{Qinghao Hu}, \bibinfo{person}{Jiaxiang Wu},
  \bibinfo{person}{Jian Cheng}, \bibinfo{person}{Lifang Wu}, {and}
  \bibinfo{person}{Hanqing Lu}.} \bibinfo{year}{2017}\natexlab{}.
\newblock \showarticletitle{Pseudo label based unsupervised deep discriminative
  hashing for image retrieval}. In \bibinfo{booktitle}{\emph{Proceedings of the
  25th ACM international conference on Multimedia}}.
  \bibinfo{pages}{1584--1590}.
\newblock


\bibitem[Huang et~al\mbox{.}(2025)]%
        {huang2025confusing}
\bibfield{author}{\bibinfo{person}{Tian Huang}, \bibinfo{person}{Jian Wang},
  {and} \bibinfo{person}{Yuqing Sun}.} \bibinfo{year}{2025}\natexlab{}.
\newblock \showarticletitle{De-confusing Hard Samples for Text Semantic
  Hashing}. In \bibinfo{booktitle}{\emph{ICASSP 2025-2025 IEEE International
  Conference on Acoustics, Speech and Signal Processing (ICASSP)}}. IEEE,
  \bibinfo{pages}{1--5}.
\newblock


\bibitem[Huang et~al\mbox{.}(2021)]%
        {huang2021disenqnet}
\bibfield{author}{\bibinfo{person}{Zhenya Huang}, \bibinfo{person}{Xin Lin},
  \bibinfo{person}{Hao Wang}, \bibinfo{person}{Qi Liu}, \bibinfo{person}{Enhong
  Chen}, \bibinfo{person}{Jianhui Ma}, \bibinfo{person}{Yu Su}, {and}
  \bibinfo{person}{Wei Tong}.} \bibinfo{year}{2021}\natexlab{}.
\newblock \showarticletitle{Disenqnet: Disentangled representation learning for
  educational questions}. In \bibinfo{booktitle}{\emph{Proceedings of the 27th
  ACM SIGKDD Conference on Knowledge Discovery \& Data Mining}}.
  \bibinfo{pages}{696--704}.
\newblock


\bibitem[Huo et~al\mbox{.}(2024)]%
        {huo2024deep}
\bibfield{author}{\bibinfo{person}{Yadong Huo}, \bibinfo{person}{Qin Qibing},
  \bibinfo{person}{Jiangyan Dai}, \bibinfo{person}{Wenfeng Zhang},
  \bibinfo{person}{Lei Huang}, {and} \bibinfo{person}{Chengduan Wang}.}
  \bibinfo{year}{2024}\natexlab{}.
\newblock \showarticletitle{Deep Neighborhood-aware Proxy Hashing with Uniform
  Distribution Constraint for Cross-modal Retrieval}.
\newblock \bibinfo{journal}{\emph{ACM Transactions on Multimedia Computing,
  Communications and Applications}} \bibinfo{volume}{20}, \bibinfo{number}{6}
  (\bibinfo{year}{2024}), \bibinfo{pages}{1--23}.
\newblock


\bibitem[Inje et~al\mbox{.}(2024)]%
        {inje2024document}
\bibfield{author}{\bibinfo{person}{Bhushan Inje}, \bibinfo{person}{Kapil~Kumar
  Nagwanshi}, {and} \bibinfo{person}{Radhakrishna Rambola}.}
  \bibinfo{year}{2024}\natexlab{}.
\newblock \showarticletitle{Document retrieval using clustering-based Aquila
  hash-Q optimization with query expansion based on pseudo relevance feedback}.
\newblock \bibinfo{journal}{\emph{International Journal of Computers and
  Applications}} \bibinfo{volume}{46}, \bibinfo{number}{7}
  (\bibinfo{year}{2024}), \bibinfo{pages}{496--507}.
\newblock


\bibitem[Karpukhin et~al\mbox{.}(2020)]%
        {karpukhin2020dense}
\bibfield{author}{\bibinfo{person}{Vladimir Karpukhin}, \bibinfo{person}{Barlas
  O{\u{g}}uz}, \bibinfo{person}{Sewon Min}, \bibinfo{person}{Patrick Lewis},
  \bibinfo{person}{Ledell Wu}, \bibinfo{person}{Sergey Edunov},
  \bibinfo{person}{Danqi Chen}, {and} \bibinfo{person}{Wen-tau Yih}.}
  \bibinfo{year}{2020}\natexlab{}.
\newblock \showarticletitle{Dense passage retrieval for open-domain question
  answering}.
\newblock \bibinfo{journal}{\emph{arXiv preprint arXiv:2004.04906}}
  (\bibinfo{year}{2020}).
\newblock


\bibitem[Kenton and Toutanova(2019)]%
        {kenton2019bert}
\bibfield{author}{\bibinfo{person}{Jacob Devlin Ming-Wei~Chang Kenton} {and}
  \bibinfo{person}{Lee~Kristina Toutanova}.} \bibinfo{year}{2019}\natexlab{}.
\newblock \showarticletitle{Bert: Pre-training of deep bidirectional
  transformers for language understanding}. In
  \bibinfo{booktitle}{\emph{Proceedings of naacL-HLT}},
  Vol.~\bibinfo{volume}{1}. Minneapolis, Minnesota, \bibinfo{pages}{2}.
\newblock


\bibitem[Kingma(2013)]%
        {kingma2013auto}
\bibfield{author}{\bibinfo{person}{Diederik~P Kingma}.}
  \bibinfo{year}{2013}\natexlab{}.
\newblock \showarticletitle{Auto-encoding variational bayes}.
\newblock \bibinfo{journal}{\emph{arXiv preprint arXiv:1312.6114}}
  (\bibinfo{year}{2013}).
\newblock


\bibitem[Kiros et~al\mbox{.}(2015)]%
        {kiros2015skip}
\bibfield{author}{\bibinfo{person}{Ryan Kiros}, \bibinfo{person}{Yukun Zhu},
  \bibinfo{person}{Russ~R Salakhutdinov}, \bibinfo{person}{Richard Zemel},
  \bibinfo{person}{Raquel Urtasun}, \bibinfo{person}{Antonio Torralba}, {and}
  \bibinfo{person}{Sanja Fidler}.} \bibinfo{year}{2015}\natexlab{}.
\newblock \showarticletitle{Skip-thought vectors}.
\newblock \bibinfo{journal}{\emph{Advances in neural information processing
  systems}}  \bibinfo{volume}{28} (\bibinfo{year}{2015}).
\newblock


\bibitem[Kratzwald and Feuerriegel(2018)]%
        {kratzwald2018adaptive}
\bibfield{author}{\bibinfo{person}{Bernhard Kratzwald} {and}
  \bibinfo{person}{Stefan Feuerriegel}.} \bibinfo{year}{2018}\natexlab{}.
\newblock \showarticletitle{Adaptive Document Retrieval for Deep Question
  Answering}. In \bibinfo{booktitle}{\emph{Proceedings of the 2018 Conference
  on Empirical Methods in Natural Language Processing}}.
  \bibinfo{pages}{576--581}.
\newblock


\bibitem[Kulis and Darrell(2009)]%
        {kulis2009learning}
\bibfield{author}{\bibinfo{person}{Brian Kulis} {and} \bibinfo{person}{Trevor
  Darrell}.} \bibinfo{year}{2009}\natexlab{}.
\newblock \showarticletitle{Learning to hash with binary reconstructive
  embeddings}.
\newblock \bibinfo{journal}{\emph{Advances in neural information processing
  systems}}  \bibinfo{volume}{22} (\bibinfo{year}{2009}).
\newblock


\bibitem[Kulis and Grauman(2011)]%
        {kulis2011kernelized}
\bibfield{author}{\bibinfo{person}{Brian Kulis} {and} \bibinfo{person}{Kristen
  Grauman}.} \bibinfo{year}{2011}\natexlab{}.
\newblock \showarticletitle{Kernelized locality-sensitive hashing}.
\newblock \bibinfo{journal}{\emph{IEEE Transactions on Pattern Analysis and
  Machine Intelligence}} \bibinfo{volume}{34}, \bibinfo{number}{6}
  (\bibinfo{year}{2011}), \bibinfo{pages}{1092--1104}.
\newblock


\bibitem[Lai et~al\mbox{.}(2018)]%
        {lai2018improved}
\bibfield{author}{\bibinfo{person}{Hanjiang Lai}, \bibinfo{person}{Yan Pan},
  \bibinfo{person}{Si Liu}, \bibinfo{person}{Zhenbin Weng}, {and}
  \bibinfo{person}{Jian Yin}.} \bibinfo{year}{2018}\natexlab{}.
\newblock \showarticletitle{Improved search in Hamming space using deep
  multi-index hashing}.
\newblock \bibinfo{journal}{\emph{IEEE Transactions on Circuits and Systems for
  Video Technology}} \bibinfo{volume}{29}, \bibinfo{number}{9}
  (\bibinfo{year}{2018}), \bibinfo{pages}{2844--2855}.
\newblock


\bibitem[Lan et~al\mbox{.}(2023)]%
        {lan2023towards}
\bibfield{author}{\bibinfo{person}{Tian Lan}, \bibinfo{person}{Xian-Ling Mao},
  \bibinfo{person}{Wei Wei}, \bibinfo{person}{Xiaoyan Gao}, {and}
  \bibinfo{person}{Heyan Huang}.} \bibinfo{year}{2023}\natexlab{}.
\newblock \showarticletitle{Towards Efficient Coarse-grained Dialogue Response
  Selection}.
\newblock \bibinfo{journal}{\emph{ACM Transactions on Information Systems}}
  \bibinfo{volume}{42}, \bibinfo{number}{2} (\bibinfo{year}{2023}),
  \bibinfo{pages}{1--32}.
\newblock


\bibitem[Le and Mikolov(2014)]%
        {le2014distributed}
\bibfield{author}{\bibinfo{person}{Quoc Le} {and} \bibinfo{person}{Tomas
  Mikolov}.} \bibinfo{year}{2014}\natexlab{}.
\newblock \showarticletitle{Distributed representations of sentences and
  documents}. In \bibinfo{booktitle}{\emph{International conference on machine
  learning}}. PMLR, \bibinfo{pages}{1188--1196}.
\newblock


\bibitem[LeCun et~al\mbox{.}(2015)]%
        {lecun2015deep}
\bibfield{author}{\bibinfo{person}{Yann LeCun}, \bibinfo{person}{Yoshua
  Bengio}, {and} \bibinfo{person}{Geoffrey Hinton}.}
  \bibinfo{year}{2015}\natexlab{}.
\newblock \showarticletitle{Deep learning}.
\newblock \bibinfo{journal}{\emph{nature}} \bibinfo{volume}{521},
  \bibinfo{number}{7553} (\bibinfo{year}{2015}), \bibinfo{pages}{436--444}.
\newblock


\bibitem[Lee et~al\mbox{.}(2024)]%
        {lee2024nv}
\bibfield{author}{\bibinfo{person}{Chankyu Lee}, \bibinfo{person}{Rajarshi
  Roy}, \bibinfo{person}{Mengyao Xu}, \bibinfo{person}{Jonathan Raiman},
  \bibinfo{person}{Mohammad Shoeybi}, \bibinfo{person}{Bryan Catanzaro}, {and}
  \bibinfo{person}{Wei Ping}.} \bibinfo{year}{2024}\natexlab{}.
\newblock \showarticletitle{NV-Embed: Improved Techniques for Training LLMs as
  Generalist Embedding Models}.
\newblock \bibinfo{journal}{\emph{arXiv preprint arXiv:2405.17428}}
  (\bibinfo{year}{2024}).
\newblock


\bibitem[Lehmann et~al\mbox{.}(2015)]%
        {lehmann2015dbpedia}
\bibfield{author}{\bibinfo{person}{Jens Lehmann}, \bibinfo{person}{Robert
  Isele}, \bibinfo{person}{Max Jakob}, \bibinfo{person}{Anja Jentzsch},
  \bibinfo{person}{Dimitris Kontokostas}, \bibinfo{person}{Pablo~N Mendes},
  \bibinfo{person}{Sebastian Hellmann}, \bibinfo{person}{Mohamed Morsey},
  \bibinfo{person}{Patrick Van~Kleef}, \bibinfo{person}{S{\"o}ren Auer},
  {et~al\mbox{.}}} \bibinfo{year}{2015}\natexlab{}.
\newblock \showarticletitle{Dbpedia--a large-scale, multilingual knowledge base
  extracted from wikipedia}.
\newblock \bibinfo{journal}{\emph{Semantic web}} \bibinfo{volume}{6},
  \bibinfo{number}{2} (\bibinfo{year}{2015}), \bibinfo{pages}{167--195}.
\newblock


\bibitem[Li et~al\mbox{.}(2021)]%
        {li2021dcspell}
\bibfield{author}{\bibinfo{person}{Jing Li}, \bibinfo{person}{Gaosheng Wu},
  \bibinfo{person}{Dafei Yin}, \bibinfo{person}{Haozhao Wang}, {and}
  \bibinfo{person}{Yonggang Wang}.} \bibinfo{year}{2021}\natexlab{}.
\newblock \showarticletitle{Dcspell: A detector-corrector framework for chinese
  spelling error correction}. In \bibinfo{booktitle}{\emph{Proceedings of the
  44th International ACM SIGIR Conference on Research and Development in
  Information Retrieval}}. \bibinfo{pages}{1870--1874}.
\newblock


\bibitem[Li et~al\mbox{.}(2022)]%
        {li2022deep}
\bibfield{author}{\bibinfo{person}{Mingjie Li}, \bibinfo{person}{Yuan-Gen
  Wang}, \bibinfo{person}{Peng Zhang}, \bibinfo{person}{Hanpin Wang},
  \bibinfo{person}{Lisheng Fan}, \bibinfo{person}{Enxia Li}, {and}
  \bibinfo{person}{Wei Wang}.} \bibinfo{year}{2022}\natexlab{}.
\newblock \showarticletitle{Deep learning for approximate nearest neighbour
  search: A survey and future directions}.
\newblock \bibinfo{journal}{\emph{IEEE Transactions on Knowledge and Data
  Engineering}} \bibinfo{volume}{35}, \bibinfo{number}{9}
  (\bibinfo{year}{2022}), \bibinfo{pages}{8997--9018}.
\newblock


\bibitem[Li et~al\mbox{.}(2024)]%
        {li2024consider}
\bibfield{author}{\bibinfo{person}{Rui Li}, \bibinfo{person}{Liyang He},
  \bibinfo{person}{Qi Liu}, \bibinfo{person}{Yuze Zhao}, \bibinfo{person}{Zheng
  Zhang}, \bibinfo{person}{Zhenya Huang}, \bibinfo{person}{Yu Su}, {and}
  \bibinfo{person}{Shijin Wang}.} \bibinfo{year}{2024}\natexlab{}.
\newblock \showarticletitle{CONSIDER: Commonalities and Specialties Driven
  Multilingual Code Retrieval Framework}. In
  \bibinfo{booktitle}{\emph{Proceedings of the AAAI Conference on Artificial
  Intelligence}}, Vol.~\bibinfo{volume}{38}. \bibinfo{pages}{8679--8687}.
\newblock


\bibitem[Li et~al\mbox{.}(2019a)]%
        {li2019problems}
\bibfield{author}{\bibinfo{person}{Shuqing Li}, \bibinfo{person}{Fusen Jiao},
  \bibinfo{person}{Yong Zhang}, {and} \bibinfo{person}{Xia Xu}.}
  \bibinfo{year}{2019}\natexlab{a}.
\newblock \showarticletitle{Problems and changes in digital libraries in the
  age of big data from the perspective of user services}.
\newblock \bibinfo{journal}{\emph{The Journal of Academic Librarianship}}
  \bibinfo{volume}{45}, \bibinfo{number}{1} (\bibinfo{year}{2019}),
  \bibinfo{pages}{22--30}.
\newblock


\bibitem[Li et~al\mbox{.}(2019b)]%
        {li2019approximate}
\bibfield{author}{\bibinfo{person}{Wen Li}, \bibinfo{person}{Ying Zhang},
  \bibinfo{person}{Yifang Sun}, \bibinfo{person}{Wei Wang},
  \bibinfo{person}{Mingjie Li}, \bibinfo{person}{Wenjie Zhang}, {and}
  \bibinfo{person}{Xuemin Lin}.} \bibinfo{year}{2019}\natexlab{b}.
\newblock \showarticletitle{Approximate nearest neighbor search on high
  dimensional data—experiments, analyses, and improvement}.
\newblock \bibinfo{journal}{\emph{IEEE Transactions on Knowledge and Data
  Engineering}} \bibinfo{volume}{32}, \bibinfo{number}{8}
  (\bibinfo{year}{2019}), \bibinfo{pages}{1475--1488}.
\newblock


\bibitem[Liu et~al\mbox{.}(2020)]%
        {liu2020online}
\bibfield{author}{\bibinfo{person}{Chong Liu}, \bibinfo{person}{Defu Lian},
  \bibinfo{person}{Min Nie}, {and} \bibinfo{person}{Xia Hu}.}
  \bibinfo{year}{2020}\natexlab{}.
\newblock \showarticletitle{Online optimized product quantization}. In
  \bibinfo{booktitle}{\emph{2020 IEEE International Conference on Data Mining
  (ICDM)}}. IEEE, \bibinfo{pages}{362--371}.
\newblock


\bibitem[Liu et~al\mbox{.}(2018)]%
        {liu2018finding}
\bibfield{author}{\bibinfo{person}{Qi Liu}, \bibinfo{person}{Zai Huang},
  \bibinfo{person}{Zhenya Huang}, \bibinfo{person}{Chuanren Liu},
  \bibinfo{person}{Enhong Chen}, \bibinfo{person}{Yu Su}, {and}
  \bibinfo{person}{Guoping Hu}.} \bibinfo{year}{2018}\natexlab{}.
\newblock \showarticletitle{Finding similar exercises in online education
  systems}. In \bibinfo{booktitle}{\emph{Proceedings of the 24th ACM SIGKDD
  International Conference on Knowledge Discovery \& Data Mining}}.
  \bibinfo{pages}{1821--1830}.
\newblock


\bibitem[Liu et~al\mbox{.}(2015)]%
        {liu2015data}
\bibfield{author}{\bibinfo{person}{Qingyun Liu}, \bibinfo{person}{Hongtao Xie},
  \bibinfo{person}{Yizhi Liu}, \bibinfo{person}{Chuang Zhang}, {and}
  \bibinfo{person}{Li Guo}.} \bibinfo{year}{2015}\natexlab{}.
\newblock \showarticletitle{Data-oriented multi-index hashing}. In
  \bibinfo{booktitle}{\emph{2015 IEEE International Conference on Multimedia
  and Expo (ICME)}}. IEEE, \bibinfo{pages}{1--6}.
\newblock


\bibitem[Liu et~al\mbox{.}(2014)]%
        {liu2014literature}
\bibfield{author}{\bibinfo{person}{Shengbo Liu}, \bibinfo{person}{Chaomei
  Chen}, \bibinfo{person}{Kun Ding}, \bibinfo{person}{Bo Wang},
  \bibinfo{person}{Kan Xu}, {and} \bibinfo{person}{Yuan Lin}.}
  \bibinfo{year}{2014}\natexlab{}.
\newblock \showarticletitle{Literature retrieval based on citation context}.
\newblock \bibinfo{journal}{\emph{Scientometrics}}  \bibinfo{volume}{101}
  (\bibinfo{year}{2014}), \bibinfo{pages}{1293--1307}.
\newblock


\bibitem[Liu et~al\mbox{.}(2024)]%
        {liu2024kat}
\bibfield{author}{\bibinfo{person}{Tianqi Liu}, \bibinfo{person}{Xinxin Zhang},
  \bibinfo{person}{Wenzheng Wang}, {and} \bibinfo{person}{Weisong Mu}.}
  \bibinfo{year}{2024}\natexlab{}.
\newblock \showarticletitle{KAT: knowledge-aware attentive recommendation model
  integrating two-terminal neighbor features}.
\newblock \bibinfo{journal}{\emph{International Journal of Machine Learning and
  Cybernetics}} (\bibinfo{year}{2024}), \bibinfo{pages}{1--18}.
\newblock


\bibitem[Liu et~al\mbox{.}(2012)]%
        {liu2012supervised}
\bibfield{author}{\bibinfo{person}{Wei Liu}, \bibinfo{person}{Jun Wang},
  \bibinfo{person}{Rongrong Ji}, \bibinfo{person}{Yu-Gang Jiang}, {and}
  \bibinfo{person}{Shih-Fu Chang}.} \bibinfo{year}{2012}\natexlab{}.
\newblock \showarticletitle{Supervised hashing with kernels}. In
  \bibinfo{booktitle}{\emph{2012 IEEE conference on computer vision and pattern
  recognition}}. IEEE, \bibinfo{pages}{2074--2081}.
\newblock


\bibitem[Liu et~al\mbox{.}(2011)]%
        {liu2011hashing}
\bibfield{author}{\bibinfo{person}{Wei Liu}, \bibinfo{person}{Jun Wang},
  \bibinfo{person}{Sanjiv Kumar}, {and} \bibinfo{person}{Shih-Fu Chang}.}
  \bibinfo{year}{2011}\natexlab{}.
\newblock \showarticletitle{Hashing with graphs}. In
  \bibinfo{booktitle}{\emph{Proceedings of the 28th international conference on
  machine learning (ICML-11)}}. \bibinfo{pages}{1--8}.
\newblock


\bibitem[Loubna et~al\mbox{.}({[n.\,d.]})]%
        {loubnaissn}
\bibfield{author}{\bibinfo{person}{ALI Loubna}, \bibinfo{person}{Turan~Can
  Gun}, {and} \bibinfo{person}{Waseem Alhasan}.}
  \bibinfo{year}{[n.\,d.]}\natexlab{}.
\newblock \showarticletitle{ISSN 2522-9400 European Modern Studies Journal Vol
  8 No 3}.
\newblock  (\bibinfo{year}{[n.\,d.]}).
\newblock


\bibitem[Lu et~al\mbox{.}(2019)]%
        {lu2019efficient}
\bibfield{author}{\bibinfo{person}{Xu Lu}, \bibinfo{person}{Lei Zhu},
  \bibinfo{person}{Jingjing Li}, \bibinfo{person}{Huaxiang Zhang}, {and}
  \bibinfo{person}{Heng~Tao Shen}.} \bibinfo{year}{2019}\natexlab{}.
\newblock \showarticletitle{Efficient supervised discrete multi-view hashing
  for large-scale multimedia search}.
\newblock \bibinfo{journal}{\emph{IEEE Transactions on Multimedia}}
  \bibinfo{volume}{22}, \bibinfo{number}{8} (\bibinfo{year}{2019}),
  \bibinfo{pages}{2048--2060}.
\newblock


\bibitem[Luo et~al\mbox{.}(2023)]%
        {luo2023survey}
\bibfield{author}{\bibinfo{person}{Xiao Luo}, \bibinfo{person}{Haixin Wang},
  \bibinfo{person}{Daqing Wu}, \bibinfo{person}{Chong Chen},
  \bibinfo{person}{Minghua Deng}, \bibinfo{person}{Jianqiang Huang}, {and}
  \bibinfo{person}{Xian-Sheng Hua}.} \bibinfo{year}{2023}\natexlab{}.
\newblock \showarticletitle{A survey on deep hashing methods}.
\newblock \bibinfo{journal}{\emph{ACM Transactions on Knowledge Discovery from
  Data}} \bibinfo{volume}{17}, \bibinfo{number}{1} (\bibinfo{year}{2023}),
  \bibinfo{pages}{1--50}.
\newblock


\bibitem[Luo et~al\mbox{.}(2018)]%
        {luo2018asymmetric}
\bibfield{author}{\bibinfo{person}{Xin Luo}, \bibinfo{person}{Peng-Fei Zhang},
  \bibinfo{person}{Ye Wu}, \bibinfo{person}{Zhen-Duo Chen},
  \bibinfo{person}{Hua-Junjie Huang}, {and} \bibinfo{person}{Xin-Shun Xu}.}
  \bibinfo{year}{2018}\natexlab{}.
\newblock \showarticletitle{Asymmetric discrete cross-modal hashing}. In
  \bibinfo{booktitle}{\emph{Proceedings of the 2018 ACM on International
  Conference on Multimedia Retrieval}}. \bibinfo{pages}{204--212}.
\newblock


\bibitem[Ma et~al\mbox{.}(2024)]%
        {ma2024volta}
\bibfield{author}{\bibinfo{person}{Yueen Ma}, \bibinfo{person}{Dafeng Chi},
  \bibinfo{person}{Jingjing Li}, \bibinfo{person}{Kai Song},
  \bibinfo{person}{Yuzheng Zhuang}, {and} \bibinfo{person}{Irwin King}.}
  \bibinfo{year}{2024}\natexlab{}.
\newblock \showarticletitle{VOLTA: Improving Generative Diversity by
  Variational Mutual Information Maximizing Autoencoder}. In
  \bibinfo{booktitle}{\emph{Findings of the Association for Computational
  Linguistics: NAACL 2024}}. \bibinfo{pages}{364--378}.
\newblock


\bibitem[Malakhov et~al\mbox{.}(2023)]%
        {malakhov2023developing}
\bibfield{author}{\bibinfo{person}{Kyrylo Malakhov}, \bibinfo{person}{Mykola
  Petrenko}, {and} \bibinfo{person}{Ellen Cohn}.}
  \bibinfo{year}{2023}\natexlab{}.
\newblock \showarticletitle{Developing an ontology-based system for semantic
  processing of scientific digital libraries}.
\newblock \bibinfo{journal}{\emph{South African Computer Journal}}
  \bibinfo{volume}{35}, \bibinfo{number}{1} (\bibinfo{year}{2023}),
  \bibinfo{pages}{19--36}.
\newblock


\bibitem[Malkov and Yashunin(2018)]%
        {malkov2018efficient}
\bibfield{author}{\bibinfo{person}{Yu~A Malkov} {and} \bibinfo{person}{Dmitry~A
  Yashunin}.} \bibinfo{year}{2018}\natexlab{}.
\newblock \showarticletitle{Efficient and robust approximate nearest neighbor
  search using hierarchical navigable small world graphs}.
\newblock \bibinfo{journal}{\emph{IEEE transactions on pattern analysis and
  machine intelligence}} \bibinfo{volume}{42}, \bibinfo{number}{4}
  (\bibinfo{year}{2018}), \bibinfo{pages}{824--836}.
\newblock


\bibitem[Mena and {\~N}anculef(2019)]%
        {mena2019binary}
\bibfield{author}{\bibinfo{person}{Francisco Mena} {and}
  \bibinfo{person}{Ricardo {\~N}anculef}.} \bibinfo{year}{2019}\natexlab{}.
\newblock \showarticletitle{A binary variational autoencoder for hashing}. In
  \bibinfo{booktitle}{\emph{Progress in Pattern Recognition, Image Analysis,
  Computer Vision, and Applications: 24th Iberoamerican Congress, CIARP 2019,
  Havana, Cuba, October 28-31, 2019, Proceedings 24}}. Springer,
  \bibinfo{pages}{131--141}.
\newblock


\bibitem[Meng et~al\mbox{.}({[n.\,d.]})]%
        {mengsfr}
\bibfield{author}{\bibinfo{person}{Rui Meng}, \bibinfo{person}{Ye Liu},
  \bibinfo{person}{Shafiq~Rayhan Joty}, \bibinfo{person}{Caiming Xiong},
  \bibinfo{person}{Yingbo Zhou}, {and} \bibinfo{person}{Semih Yavuz}.}
  \bibinfo{year}{[n.\,d.]}\natexlab{}.
\newblock \showarticletitle{Sfr-embedding-2: Advanced text embedding with
  multi-stage training, 2024}.
\newblock \bibinfo{journal}{\emph{URL https://huggingface.
  co/Salesforce/SFR-Embedding-2\_R}} (\bibinfo{year}{[n.\,d.]}).
\newblock


\bibitem[Miao et~al\mbox{.}(2018)]%
        {miao2018approximate}
\bibfield{author}{\bibinfo{person}{Jianhui Miao}, \bibinfo{person}{Zhiyang Li},
  \bibinfo{person}{Wenyu Qu}, \bibinfo{person}{Zeyan Zhou},
  \bibinfo{person}{Zhaobin Liu}, {and} \bibinfo{person}{Weijiang Liu}.}
  \bibinfo{year}{2018}\natexlab{}.
\newblock \showarticletitle{Approximate Nearest Neighbor Search Based on
  Hierarchical Multi-Index Hashing}. In \bibinfo{booktitle}{\emph{2018 IEEE
  SmartWorld, Ubiquitous Intelligence \& Computing, Advanced \& Trusted
  Computing, Scalable Computing \& Communications, Cloud \& Big Data Computing,
  Internet of People and Smart City Innovation
  (SmartWorld/SCALCOM/UIC/ATC/CBDCom/IOP/SCI)}}. IEEE,
  \bibinfo{pages}{1791--1796}.
\newblock


\bibitem[Mikolov(2013)]%
        {mikolov2013efficient}
\bibfield{author}{\bibinfo{person}{Tomas Mikolov}.}
  \bibinfo{year}{2013}\natexlab{}.
\newblock \showarticletitle{Efficient estimation of word representations in
  vector space}.
\newblock \bibinfo{journal}{\emph{arXiv preprint arXiv:1301.3781}}
  \bibinfo{volume}{3781} (\bibinfo{year}{2013}).
\newblock


\bibitem[Muangprathub et~al\mbox{.}(2021)]%
        {muangprathub2021document}
\bibfield{author}{\bibinfo{person}{Jirapond Muangprathub},
  \bibinfo{person}{Siriwan Kajornkasirat}, {and} \bibinfo{person}{Apirat
  Wanichsombat}.} \bibinfo{year}{2021}\natexlab{}.
\newblock \showarticletitle{Document plagiarism detection using a new concept
  similarity in formal concept analysis}.
\newblock \bibinfo{journal}{\emph{Journal of Applied Mathematics}}
  \bibinfo{volume}{2021}, \bibinfo{number}{1} (\bibinfo{year}{2021}),
  \bibinfo{pages}{6662984}.
\newblock


\bibitem[{\~N}anculef et~al\mbox{.}(2021)]%
        {nanculef2021self}
\bibfield{author}{\bibinfo{person}{Ricardo {\~N}anculef},
  \bibinfo{person}{Francisco Mena}, \bibinfo{person}{Antonio Macaluso},
  \bibinfo{person}{Stefano Lodi}, {and} \bibinfo{person}{Claudio Sartori}.}
  \bibinfo{year}{2021}\natexlab{}.
\newblock \showarticletitle{Self-supervised bernoulli autoencoders for
  semi-supervised hashing}. In \bibinfo{booktitle}{\emph{Progress in Pattern
  Recognition, Image Analysis, Computer Vision, and Applications: 25th
  Iberoamerican Congress, CIARP 2021, Porto, Portugal, May 10--13, 2021,
  Revised Selected Papers 25}}. Springer, \bibinfo{pages}{258--268}.
\newblock


\bibitem[Nascimento and De~Carvalho(2011)]%
        {nascimento2011spectral}
\bibfield{author}{\bibinfo{person}{Maria~CV Nascimento} {and}
  \bibinfo{person}{Andre~CPLF De~Carvalho}.} \bibinfo{year}{2011}\natexlab{}.
\newblock \showarticletitle{Spectral methods for graph clustering--a survey}.
\newblock \bibinfo{journal}{\emph{European Journal of Operational Research}}
  \bibinfo{volume}{211}, \bibinfo{number}{2} (\bibinfo{year}{2011}),
  \bibinfo{pages}{221--231}.
\newblock


\bibitem[Ng et~al\mbox{.}(2001)]%
        {ng2001spectral}
\bibfield{author}{\bibinfo{person}{Andrew Ng}, \bibinfo{person}{Michael
  Jordan}, {and} \bibinfo{person}{Yair Weiss}.}
  \bibinfo{year}{2001}\natexlab{}.
\newblock \showarticletitle{On spectral clustering: Analysis and an algorithm}.
\newblock \bibinfo{journal}{\emph{Advances in neural information processing
  systems}}  \bibinfo{volume}{14} (\bibinfo{year}{2001}).
\newblock


\bibitem[Nie et~al\mbox{.}(2020)]%
        {nie2020dc}
\bibfield{author}{\bibinfo{person}{Ping Nie}, \bibinfo{person}{Yuyu Zhang},
  \bibinfo{person}{Xiubo Geng}, \bibinfo{person}{Arun Ramamurthy},
  \bibinfo{person}{Le Song}, {and} \bibinfo{person}{Daxin Jiang}.}
  \bibinfo{year}{2020}\natexlab{}.
\newblock \showarticletitle{Dc-bert: Decoupling question and document for
  efficient contextual encoding}. In \bibinfo{booktitle}{\emph{Proceedings of
  the 43rd international ACM SIGIR conference on research and development in
  information retrieval}}. \bibinfo{pages}{1829--1832}.
\newblock


\bibitem[Norouzi et~al\mbox{.}(2012)]%
        {norouzi2012fast}
\bibfield{author}{\bibinfo{person}{Mohammad Norouzi}, \bibinfo{person}{Ali
  Punjani}, {and} \bibinfo{person}{David~J Fleet}.}
  \bibinfo{year}{2012}\natexlab{}.
\newblock \showarticletitle{Fast search in hamming space with multi-index
  hashing}. In \bibinfo{booktitle}{\emph{2012 IEEE conference on computer
  vision and pattern recognition}}. IEEE, \bibinfo{pages}{3108--3115}.
\newblock


\bibitem[Nowozin et~al\mbox{.}(2016)]%
        {nowozin2016f}
\bibfield{author}{\bibinfo{person}{Sebastian Nowozin}, \bibinfo{person}{Botond
  Cseke}, {and} \bibinfo{person}{Ryota Tomioka}.}
  \bibinfo{year}{2016}\natexlab{}.
\newblock \showarticletitle{f-gan: Training generative neural samplers using
  variational divergence minimization}.
\newblock \bibinfo{journal}{\emph{Advances in neural information processing
  systems}}  \bibinfo{volume}{29} (\bibinfo{year}{2016}).
\newblock


\bibitem[Ou et~al\mbox{.}(2021a)]%
        {ou2021integrating}
\bibfield{author}{\bibinfo{person}{Zijing Ou}, \bibinfo{person}{Qinliang Su},
  \bibinfo{person}{Jianxing Yu}, \bibinfo{person}{Bang Liu},
  \bibinfo{person}{Jingwen Wang}, \bibinfo{person}{Ruihui Zhao},
  \bibinfo{person}{Changyou Chen}, {and} \bibinfo{person}{Yefeng Zheng}.}
  \bibinfo{year}{2021}\natexlab{a}.
\newblock \showarticletitle{Integrating Semantics and Neighborhood Information
  with Graph-Driven Generative Models for Document Retrieval}. In
  \bibinfo{booktitle}{\emph{Proceedings of the 59th Annual Meeting of the
  Association for Computational Linguistics and the 11th International Joint
  Conference on Natural Language Processing (Volume 1: Long Papers)}}.
  \bibinfo{pages}{2238--2249}.
\newblock


\bibitem[Ou et~al\mbox{.}(2021b)]%
        {ou2021refining}
\bibfield{author}{\bibinfo{person}{Zijing Ou}, \bibinfo{person}{Qinliang Su},
  \bibinfo{person}{Jianxing Yu}, \bibinfo{person}{Ruihui Zhao},
  \bibinfo{person}{Yefeng Zheng}, {and} \bibinfo{person}{Bang Liu}.}
  \bibinfo{year}{2021}\natexlab{b}.
\newblock \showarticletitle{Refining BERT embeddings for document hashing via
  mutual information maximization}.
\newblock \bibinfo{journal}{\emph{arXiv preprint arXiv:2109.02867}}
  (\bibinfo{year}{2021}).
\newblock


\bibitem[Park et~al\mbox{.}(2002)]%
        {park2002web}
\bibfield{author}{\bibinfo{person}{Eui-Kyu Park}, \bibinfo{person}{Seong-In
  Moon}, \bibinfo{person}{Dong-Yul Ra}, {and} \bibinfo{person}{Myung-Gil
  Jang}.} \bibinfo{year}{2002}\natexlab{}.
\newblock \showarticletitle{Web Document Retrieval Using Sentence-Query
  Similarity.}. In \bibinfo{booktitle}{\emph{TREC}}.
\newblock


\bibitem[Peng et~al\mbox{.}(2023)]%
        {peng2023efficient}
\bibfield{author}{\bibinfo{person}{Yun Peng}, \bibinfo{person}{Byron Choi},
  \bibinfo{person}{Tsz~Nam Chan}, \bibinfo{person}{Jianye Yang}, {and}
  \bibinfo{person}{Jianliang Xu}.} \bibinfo{year}{2023}\natexlab{}.
\newblock \showarticletitle{Efficient approximate nearest neighbor search in
  multi-dimensional databases}.
\newblock \bibinfo{journal}{\emph{Proceedings of the ACM on Management of
  Data}} \bibinfo{volume}{1}, \bibinfo{number}{1} (\bibinfo{year}{2023}),
  \bibinfo{pages}{1--27}.
\newblock


\bibitem[Qian and Cheung(2022)]%
        {qian2022learning}
\bibfield{author}{\bibinfo{person}{Dong Qian} {and} \bibinfo{person}{William~K
  Cheung}.} \bibinfo{year}{2022}\natexlab{}.
\newblock \showarticletitle{Learning hierarchical variational autoencoders with
  mutual information maximization for autoregressive sequence modeling}.
\newblock \bibinfo{journal}{\emph{IEEE Transactions on Pattern Analysis and
  Machine Intelligence}} \bibinfo{volume}{45}, \bibinfo{number}{2}
  (\bibinfo{year}{2022}), \bibinfo{pages}{1949--1962}.
\newblock


\bibitem[Rastegari et~al\mbox{.}(2016)]%
        {rastegari2016xnor}
\bibfield{author}{\bibinfo{person}{Mohammad Rastegari},
  \bibinfo{person}{Vicente Ordonez}, \bibinfo{person}{Joseph Redmon}, {and}
  \bibinfo{person}{Ali Farhadi}.} \bibinfo{year}{2016}\natexlab{}.
\newblock \showarticletitle{Xnor-net: Imagenet classification using binary
  convolutional neural networks}. In \bibinfo{booktitle}{\emph{European
  conference on computer vision}}. Springer, \bibinfo{pages}{525--542}.
\newblock


\bibitem[Robertson et~al\mbox{.}(2009)]%
        {robertson2009probabilistic}
\bibfield{author}{\bibinfo{person}{Stephen Robertson}, \bibinfo{person}{Hugo
  Zaragoza}, {et~al\mbox{.}}} \bibinfo{year}{2009}\natexlab{}.
\newblock \showarticletitle{The probabilistic relevance framework: BM25 and
  beyond}.
\newblock \bibinfo{journal}{\emph{Foundations and Trends{\textregistered} in
  Information Retrieval}} \bibinfo{volume}{3}, \bibinfo{number}{4}
  (\bibinfo{year}{2009}), \bibinfo{pages}{333--389}.
\newblock


\bibitem[Rodrigues et~al\mbox{.}(2020)]%
        {rodrigues2020deep}
\bibfield{author}{\bibinfo{person}{Josiane Rodrigues}, \bibinfo{person}{Marco
  Cristo}, {and} \bibinfo{person}{Juan~G Colonna}.}
  \bibinfo{year}{2020}\natexlab{}.
\newblock \showarticletitle{Deep hashing for multi-label image retrieval: a
  survey}.
\newblock \bibinfo{journal}{\emph{Artificial Intelligence Review}}
  \bibinfo{volume}{53}, \bibinfo{number}{7} (\bibinfo{year}{2020}),
  \bibinfo{pages}{5261--5307}.
\newblock


\bibitem[Salakhutdinov and Hinton(2009)]%
        {salakhutdinov2009semantic}
\bibfield{author}{\bibinfo{person}{Ruslan Salakhutdinov} {and}
  \bibinfo{person}{Geoffrey Hinton}.} \bibinfo{year}{2009}\natexlab{}.
\newblock \showarticletitle{Semantic hashing}.
\newblock \bibinfo{journal}{\emph{International Journal of Approximate
  Reasoning}} \bibinfo{volume}{50}, \bibinfo{number}{7} (\bibinfo{year}{2009}),
  \bibinfo{pages}{969--978}.
\newblock


\bibitem[Shakir et~al\mbox{.}(2024)]%
        {shakir2024quantization}
\bibfield{author}{\bibinfo{person}{Aamir Shakir}, \bibinfo{person}{Tom Aarsen},
  {and} \bibinfo{person}{Sean Lee}.} \bibinfo{year}{2024}\natexlab{}.
\newblock \showarticletitle{Binary and Scalar Embedding Quantization for
  Significantly Faster \& Cheaper Retrieval}.
\newblock \bibinfo{journal}{\emph{Hugging Face Blog}} (\bibinfo{year}{2024}).
\newblock
\newblock
\shownote{https://huggingface.co/blog/embedding-quantization}.


\bibitem[Shen et~al\mbox{.}(2018)]%
        {shen2018nash}
\bibfield{author}{\bibinfo{person}{Dinghan Shen}, \bibinfo{person}{Qinliang
  Su}, \bibinfo{person}{Paidamoyo Chapfuwa}, \bibinfo{person}{Wenlin Wang},
  \bibinfo{person}{Guoyin Wang}, \bibinfo{person}{Lawrence Carin}, {and}
  \bibinfo{person}{Ricardo Henao}.} \bibinfo{year}{2018}\natexlab{}.
\newblock \showarticletitle{Nash: Toward end-to-end neural architecture for
  generative semantic hashing}.
\newblock \bibinfo{journal}{\emph{arXiv preprint arXiv:1805.05361}}
  (\bibinfo{year}{2018}).
\newblock


\bibitem[Shen et~al\mbox{.}(2019)]%
        {shen2019unsupervised}
\bibfield{author}{\bibinfo{person}{Yuming Shen}, \bibinfo{person}{Li Liu},
  {and} \bibinfo{person}{Ling Shao}.} \bibinfo{year}{2019}\natexlab{}.
\newblock \showarticletitle{Unsupervised binary representation learning with
  deep variational networks}.
\newblock \bibinfo{journal}{\emph{International Journal of Computer Vision}}
  \bibinfo{volume}{127}, \bibinfo{number}{11} (\bibinfo{year}{2019}),
  \bibinfo{pages}{1614--1628}.
\newblock


\bibitem[Shi et~al\mbox{.}(2020)]%
        {shi2020anchor}
\bibfield{author}{\bibinfo{person}{Xiaoshuang Shi}, \bibinfo{person}{Zhenhua
  Guo}, \bibinfo{person}{Fuyong Xing}, \bibinfo{person}{Yun Liang}, {and}
  \bibinfo{person}{Lin Yang}.} \bibinfo{year}{2020}\natexlab{}.
\newblock \showarticletitle{Anchor-based self-ensembling for semi-supervised
  deep pairwise hashing}.
\newblock \bibinfo{journal}{\emph{International Journal of Computer Vision}}
  \bibinfo{volume}{128}, \bibinfo{number}{8} (\bibinfo{year}{2020}),
  \bibinfo{pages}{2307--2324}.
\newblock


\bibitem[Stratos and Wiseman(2020)]%
        {stratos2020learning}
\bibfield{author}{\bibinfo{person}{Karl Stratos} {and} \bibinfo{person}{Sam
  Wiseman}.} \bibinfo{year}{2020}\natexlab{}.
\newblock \showarticletitle{Learning discrete structured representations by
  adversarially maximizing mutual information}. In
  \bibinfo{booktitle}{\emph{International Conference on Machine Learning}}.
  PMLR, \bibinfo{pages}{9144--9154}.
\newblock


\bibitem[Tan et~al\mbox{.}(2022a)]%
        {tan2022bit}
\bibfield{author}{\bibinfo{person}{Wentao Tan}, \bibinfo{person}{Lei Zhu},
  \bibinfo{person}{Weili Guan}, \bibinfo{person}{Jingjing Li}, {and}
  \bibinfo{person}{Zhiyong Cheng}.} \bibinfo{year}{2022}\natexlab{a}.
\newblock \showarticletitle{Bit-aware semantic transformer hashing for
  multi-modal retrieval}. In \bibinfo{booktitle}{\emph{Proceedings of the 45th
  international ACM SIGIR conference on research and development in information
  retrieval}}. \bibinfo{pages}{982--991}.
\newblock


\bibitem[Tan et~al\mbox{.}(2022b)]%
        {tan2022teacher}
\bibfield{author}{\bibinfo{person}{Wentao Tan}, \bibinfo{person}{Lei Zhu},
  \bibinfo{person}{Jingjing Li}, \bibinfo{person}{Huaxiang Zhang}, {and}
  \bibinfo{person}{Junwei Han}.} \bibinfo{year}{2022}\natexlab{b}.
\newblock \showarticletitle{Teacher-student learning: Efficient hierarchical
  message aggregation hashing for cross-modal retrieval}.
\newblock \bibinfo{journal}{\emph{IEEE Transactions on Multimedia}}
  \bibinfo{volume}{25} (\bibinfo{year}{2022}), \bibinfo{pages}{4520--4532}.
\newblock


\bibitem[Thakur et~al\mbox{.}(2021)]%
        {Thakur}
\bibfield{author}{\bibinfo{person}{Nandan Thakur}, \bibinfo{person}{Nils
  Reimers}, \bibinfo{person}{Andreas R{\"u}ckl{\'e}}, \bibinfo{person}{Abhishek
  Srivastava}, {and} \bibinfo{person}{Iryna Gurevych}.}
  \bibinfo{year}{2021}\natexlab{}.
\newblock \showarticletitle{Beir: A heterogenous benchmark for zero-shot
  evaluation of information retrieval models}.
\newblock \bibinfo{journal}{\emph{arXiv preprint arXiv:2104.08663}}
  (\bibinfo{year}{2021}).
\newblock


\bibitem[TONG et~al\mbox{.}(2024)]%
        {tong2024efficient}
\bibfield{author}{\bibinfo{person}{Wei TONG}, \bibinfo{person}{Liyang HE},
  \bibinfo{person}{Rui LI}, \bibinfo{person}{Wei HUANG},
  \bibinfo{person}{Zhenya HUANG}, {and} \bibinfo{person}{Qi LIU}.}
  \bibinfo{year}{2024}\natexlab{}.
\newblock \showarticletitle{Efficient similar exercise retrieval model based on
  unsupervised semantic hashing}.
\newblock \bibinfo{journal}{\emph{Journal of Computer Applications}}
  \bibinfo{volume}{44}, \bibinfo{number}{1} (\bibinfo{year}{2024}),
  \bibinfo{pages}{206}.
\newblock


\bibitem[Tong et~al\mbox{.}(2020)]%
        {tong2020exploiting}
\bibfield{author}{\bibinfo{person}{Wei Tong}, \bibinfo{person}{Shiwei Tong},
  \bibinfo{person}{Wei Hunag}, \bibinfo{person}{Liyang He},
  \bibinfo{person}{Jianhui Ma}, \bibinfo{person}{Qi Liu}, {and}
  \bibinfo{person}{Enhong Chen}.} \bibinfo{year}{2020}\natexlab{}.
\newblock \showarticletitle{Exploiting knowledge hierarchy for finding similar
  exercises in online education systems}. In \bibinfo{booktitle}{\emph{2020
  IEEE International Conference on Data Mining (ICDM)}}. IEEE,
  \bibinfo{pages}{1298--1303}.
\newblock


\bibitem[Villani(2021)]%
        {villani2021topics}
\bibfield{author}{\bibinfo{person}{C{\'e}dric Villani}.}
  \bibinfo{year}{2021}\natexlab{}.
\newblock \bibinfo{booktitle}{\emph{Topics in optimal transportation}}.
  Vol.~\bibinfo{volume}{58}.
\newblock \bibinfo{publisher}{American Mathematical Soc.}
\newblock


\bibitem[Von~Luxburg(2007)]%
        {von2007tutorial}
\bibfield{author}{\bibinfo{person}{Ulrike Von~Luxburg}.}
  \bibinfo{year}{2007}\natexlab{}.
\newblock \showarticletitle{A tutorial on spectral clustering}.
\newblock \bibinfo{journal}{\emph{Statistics and computing}}
  \bibinfo{volume}{17} (\bibinfo{year}{2007}), \bibinfo{pages}{395--416}.
\newblock


\bibitem[Wainwright et~al\mbox{.}(2008)]%
        {wainwright2008graphical}
\bibfield{author}{\bibinfo{person}{Martin~J Wainwright},
  \bibinfo{person}{Michael~I Jordan}, {et~al\mbox{.}}}
  \bibinfo{year}{2008}\natexlab{}.
\newblock \showarticletitle{Graphical models, exponential families, and
  variational inference}.
\newblock \bibinfo{journal}{\emph{Foundations and Trends{\textregistered} in
  Machine Learning}} \bibinfo{volume}{1}, \bibinfo{number}{1--2}
  (\bibinfo{year}{2008}), \bibinfo{pages}{1--305}.
\newblock


\bibitem[Wan et~al\mbox{.}(2013)]%
        {wan2013data}
\bibfield{author}{\bibinfo{person}{Ji Wan}, \bibinfo{person}{Sheng Tang},
  \bibinfo{person}{Yongdong Zhang}, \bibinfo{person}{Lei Huang}, {and}
  \bibinfo{person}{Jintao Li}.} \bibinfo{year}{2013}\natexlab{}.
\newblock \showarticletitle{Data driven multi-index hashing}. In
  \bibinfo{booktitle}{\emph{2013 IEEE International Conference on Image
  Processing}}. IEEE, \bibinfo{pages}{2670--2673}.
\newblock


\bibitem[Wang et~al\mbox{.}(2012)]%
        {wang2012semi}
\bibfield{author}{\bibinfo{person}{Jun Wang}, \bibinfo{person}{Sanjiv Kumar},
  {and} \bibinfo{person}{Shih-Fu Chang}.} \bibinfo{year}{2012}\natexlab{}.
\newblock \showarticletitle{Semi-supervised hashing for large-scale search}.
\newblock \bibinfo{journal}{\emph{IEEE transactions on pattern analysis and
  machine intelligence}} \bibinfo{volume}{34}, \bibinfo{number}{12}
  (\bibinfo{year}{2012}), \bibinfo{pages}{2393--2406}.
\newblock


\bibitem[Wang et~al\mbox{.}(2015b)]%
        {wang2015learning}
\bibfield{author}{\bibinfo{person}{Jun Wang}, \bibinfo{person}{Wei Liu},
  \bibinfo{person}{Sanjiv Kumar}, {and} \bibinfo{person}{Shih-Fu Chang}.}
  \bibinfo{year}{2015}\natexlab{b}.
\newblock \showarticletitle{Learning to hash for indexing big data—A survey}.
\newblock \bibinfo{journal}{\emph{Proc. IEEE}} \bibinfo{volume}{104},
  \bibinfo{number}{1} (\bibinfo{year}{2015}), \bibinfo{pages}{34--57}.
\newblock


\bibitem[Wang et~al\mbox{.}(2013a)]%
        {wang2013learning}
\bibfield{author}{\bibinfo{person}{Jun Wang}, \bibinfo{person}{Wei Liu},
  \bibinfo{person}{Andy~X Sun}, {and} \bibinfo{person}{Yu-Gang Jiang}.}
  \bibinfo{year}{2013}\natexlab{a}.
\newblock \showarticletitle{Learning hash codes with listwise supervision}. In
  \bibinfo{booktitle}{\emph{Proceedings of the IEEE international conference on
  computer vision}}. \bibinfo{pages}{3032--3039}.
\newblock


\bibitem[Wang et~al\mbox{.}(2017)]%
        {wang2017survey}
\bibfield{author}{\bibinfo{person}{Jingdong Wang}, \bibinfo{person}{Ting
  Zhang}, \bibinfo{person}{Nicu Sebe}, \bibinfo{person}{Heng~Tao Shen},
  {et~al\mbox{.}}} \bibinfo{year}{2017}\natexlab{}.
\newblock \showarticletitle{A survey on learning to hash}.
\newblock \bibinfo{journal}{\emph{IEEE transactions on pattern analysis and
  machine intelligence}} \bibinfo{volume}{40}, \bibinfo{number}{4}
  (\bibinfo{year}{2017}), \bibinfo{pages}{769--790}.
\newblock


\bibitem[Wang et~al\mbox{.}(2023c)]%
        {wang2023deep}
\bibfield{author}{\bibinfo{person}{Liangdao Wang}, \bibinfo{person}{Yan Pan},
  \bibinfo{person}{Cong Liu}, \bibinfo{person}{Hanjiang Lai},
  \bibinfo{person}{Jian Yin}, {and} \bibinfo{person}{Ye Liu}.}
  \bibinfo{year}{2023}\natexlab{c}.
\newblock \showarticletitle{Deep hashing with minimal-distance-separated hash
  centers}. In \bibinfo{booktitle}{\emph{Proceedings of the IEEE/CVF conference
  on computer vision and pattern recognition}}. \bibinfo{pages}{23455--23464}.
\newblock


\bibitem[Wang et~al\mbox{.}(2015a)]%
        {wang2015multi}
\bibfield{author}{\bibinfo{person}{Manlin Wang}, \bibinfo{person}{Xiaokang
  Feng}, {and} \bibinfo{person}{Jiangtao Cui}.}
  \bibinfo{year}{2015}\natexlab{a}.
\newblock \showarticletitle{Multi-index hashing with repeat-bits in Hamming
  space}. In \bibinfo{booktitle}{\emph{2015 12th International Conference on
  Fuzzy Systems and Knowledge Discovery (FSKD)}}. IEEE,
  \bibinfo{pages}{1307--1313}.
\newblock


\bibitem[Wang et~al\mbox{.}(2013b)]%
        {wang2013semantic}
\bibfield{author}{\bibinfo{person}{Qifan Wang}, \bibinfo{person}{Dan Zhang},
  {and} \bibinfo{person}{Luo Si}.} \bibinfo{year}{2013}\natexlab{b}.
\newblock \showarticletitle{Semantic hashing using tags and topic modeling}. In
  \bibinfo{booktitle}{\emph{Proceedings of the 36th international ACM SIGIR
  conference on Research and development in information retrieval}}.
  \bibinfo{pages}{213--222}.
\newblock


\bibitem[Wang et~al\mbox{.}(2023b)]%
        {wang2023colbert}
\bibfield{author}{\bibinfo{person}{Xiao Wang}, \bibinfo{person}{Craig
  Macdonald}, \bibinfo{person}{Nicola Tonellotto}, {and} \bibinfo{person}{Iadh
  Ounis}.} \bibinfo{year}{2023}\natexlab{b}.
\newblock \showarticletitle{ColBERT-PRF: Semantic pseudo-relevance feedback for
  dense passage and document retrieval}.
\newblock \bibinfo{journal}{\emph{ACM Transactions on the Web}}
  \bibinfo{volume}{17}, \bibinfo{number}{1} (\bibinfo{year}{2023}),
  \bibinfo{pages}{1--39}.
\newblock


\bibitem[Wang et~al\mbox{.}(2021)]%
        {wang2021high}
\bibfield{author}{\bibinfo{person}{Yongxin Wang}, \bibinfo{person}{Zhen-Duo
  Chen}, \bibinfo{person}{Xin Luo}, {and} \bibinfo{person}{Xin-Shun Xu}.}
  \bibinfo{year}{2021}\natexlab{}.
\newblock \showarticletitle{High-dimensional sparse cross-modal hashing with
  fine-grained similarity embedding}. In \bibinfo{booktitle}{\emph{Proceedings
  of the Web Conference 2021}}. \bibinfo{pages}{2900--2909}.
\newblock


\bibitem[Wang et~al\mbox{.}(2023a)]%
        {wang2023codet5}
\bibfield{author}{\bibinfo{person}{Yue Wang}, \bibinfo{person}{Hung Le},
  \bibinfo{person}{Akhilesh Gotmare}, \bibinfo{person}{Nghi Bui},
  \bibinfo{person}{Junnan Li}, {and} \bibinfo{person}{Steven Hoi}.}
  \bibinfo{year}{2023}\natexlab{a}.
\newblock \showarticletitle{CodeT5+: Open Code Large Language Models for Code
  Understanding and Generation}. In \bibinfo{booktitle}{\emph{Proceedings of
  the 2023 Conference on Empirical Methods in Natural Language Processing}}.
  \bibinfo{pages}{1069--1088}.
\newblock


\bibitem[Wang et~al\mbox{.}(2019)]%
        {wang2019multi}
\bibfield{author}{\bibinfo{person}{Zhiguo Wang}, \bibinfo{person}{Patrick Ng},
  \bibinfo{person}{Xiaofei Ma}, \bibinfo{person}{Ramesh Nallapati}, {and}
  \bibinfo{person}{Bing Xiang}.} \bibinfo{year}{2019}\natexlab{}.
\newblock \showarticletitle{Multi-passage BERT: A Globally Normalized BERT
  Model for Open-domain Question Answering}. In
  \bibinfo{booktitle}{\emph{Proceedings of the 2019 Conference on Empirical
  Methods in Natural Language Processing and the 9th International Joint
  Conference on Natural Language Processing (EMNLP-IJCNLP)}}.
  \bibinfo{pages}{5878--5882}.
\newblock


\bibitem[Weiss et~al\mbox{.}(2008)]%
        {weiss2008spectral}
\bibfield{author}{\bibinfo{person}{Yair Weiss}, \bibinfo{person}{Antonio
  Torralba}, {and} \bibinfo{person}{Rob Fergus}.}
  \bibinfo{year}{2008}\natexlab{}.
\newblock \showarticletitle{Spectral hashing}.
\newblock \bibinfo{journal}{\emph{Advances in neural information processing
  systems}}  \bibinfo{volume}{21} (\bibinfo{year}{2008}).
\newblock


\bibitem[Weng et~al\mbox{.}(2024)]%
        {weng2024fast}
\bibfield{author}{\bibinfo{person}{Shaoyuan Weng}, \bibinfo{person}{Zongwen
  Fan}, {and} \bibinfo{person}{Jin Gou}.} \bibinfo{year}{2024}\natexlab{}.
\newblock \showarticletitle{A fast DBSCAN algorithm using a bi-directional HNSW
  index structure for big data}.
\newblock \bibinfo{journal}{\emph{International Journal of Machine Learning and
  Cybernetics}} (\bibinfo{year}{2024}), \bibinfo{pages}{1--24}.
\newblock


\bibitem[Xie et~al\mbox{.}(2024)]%
        {xie2024improving}
\bibfield{author}{\bibinfo{person}{Xinran Xie}, \bibinfo{person}{Rui Chen},
  \bibinfo{person}{TaiLai Peng}, \bibinfo{person}{Dekun Lin}, {and}
  \bibinfo{person}{Zhe Cui}.} \bibinfo{year}{2024}\natexlab{}.
\newblock \showarticletitle{Improving Zero-Shot Information Retrieval with
  Mutual Validation of Generative and Pseudo-Relevance Feedback}. In
  \bibinfo{booktitle}{\emph{Asia-Pacific Web (APWeb) and Web-Age Information
  Management (WAIM) Joint International Conference on Web and Big Data}}.
  Springer, \bibinfo{pages}{113--129}.
\newblock


\bibitem[Xu and Li(2020)]%
        {xu2020hashing}
\bibfield{author}{\bibinfo{person}{Dong Xu} {and} \bibinfo{person}{Wu-Jun Li}.}
  \bibinfo{year}{2020}\natexlab{}.
\newblock \showarticletitle{Hashing based answer selection}. In
  \bibinfo{booktitle}{\emph{Proceedings of the AAAI Conference on Artificial
  Intelligence}}, Vol.~\bibinfo{volume}{34}. \bibinfo{pages}{9330--9337}.
\newblock


\bibitem[Xu et~al\mbox{.}(2018)]%
        {xu2018online}
\bibfield{author}{\bibinfo{person}{Donna Xu}, \bibinfo{person}{Ivor~W Tsang},
  {and} \bibinfo{person}{Ying Zhang}.} \bibinfo{year}{2018}\natexlab{}.
\newblock \showarticletitle{Online product quantization}.
\newblock \bibinfo{journal}{\emph{IEEE Transactions on Knowledge and Data
  Engineering}} \bibinfo{volume}{30}, \bibinfo{number}{11}
  (\bibinfo{year}{2018}), \bibinfo{pages}{2185--2198}.
\newblock


\bibitem[Xu et~al\mbox{.}(2015)]%
        {xu2015convolutional}
\bibfield{author}{\bibinfo{person}{Jiaming Xu}, \bibinfo{person}{Peng Wang},
  \bibinfo{person}{Guanhua Tian}, \bibinfo{person}{Bo Xu}, \bibinfo{person}{Jun
  Zhao}, \bibinfo{person}{Fangyuan Wang}, {and} \bibinfo{person}{Hongwei Hao}.}
  \bibinfo{year}{2015}\natexlab{}.
\newblock \showarticletitle{Convolutional neural networks for text hashing}. In
  \bibinfo{booktitle}{\emph{Twenty-Fourth International Joint Conference on
  Artificial Intelligence}}.
\newblock


\bibitem[Xuan et~al\mbox{.}(2019)]%
        {xuan2019variational}
\bibfield{author}{\bibinfo{person}{Richeng Xuan}, \bibinfo{person}{Junho Shim},
  {and} \bibinfo{person}{Sang-goo Lee}.} \bibinfo{year}{2019}\natexlab{}.
\newblock \showarticletitle{Variational deep semantic text hashing with
  pairwise labels}. In \bibinfo{booktitle}{\emph{Proceedings of the 13th
  International Conference on Ubiquitous Information Management and
  Communication (IMCOM) 2019 13}}. Springer, \bibinfo{pages}{1076--1091}.
\newblock


\bibitem[Xuan et~al\mbox{.}(2020)]%
        {xuan2020conditional}
\bibfield{author}{\bibinfo{person}{Richeng Xuan}, \bibinfo{person}{Junho Shim},
  {and} \bibinfo{person}{Sang-Goo Lee}.} \bibinfo{year}{2020}\natexlab{}.
\newblock \showarticletitle{Conditional text hashing utilizing pair-wise multi
  class labels}.
\newblock \bibinfo{journal}{\emph{ICIC Express Letters}} \bibinfo{volume}{14},
  \bibinfo{number}{4} (\bibinfo{year}{2020}), \bibinfo{pages}{417--422}.
\newblock


\bibitem[Xuan et~al\mbox{.}(2024)]%
        {xuan2024fast}
\bibfield{author}{\bibinfo{person}{Richeng Xuan}, \bibinfo{person}{Junho Shim},
  {and} \bibinfo{person}{Sang-goo Lee}.} \bibinfo{year}{2024}\natexlab{}.
\newblock \showarticletitle{FAST PASSAGE RETRIEVAL IN WEIGHTED HAMMING SPACE
  FOR OPEN-DOMAIN QUESTION ANSWERING}.
\newblock \bibinfo{journal}{\emph{ICIC Express Letters, Part B: Applications}}
  \bibinfo{volume}{15}, \bibinfo{number}{4} (\bibinfo{year}{2024}),
  \bibinfo{pages}{373--380}.
\newblock


\bibitem[Yamada et~al\mbox{.}(2021)]%
        {yamada2021efficient}
\bibfield{author}{\bibinfo{person}{Ikuya Yamada}, \bibinfo{person}{Akari Asai},
  {and} \bibinfo{person}{Hannaneh Hajishirzi}.}
  \bibinfo{year}{2021}\natexlab{}.
\newblock \showarticletitle{Efficient passage retrieval with hashing for
  open-domain question answering}.
\newblock \bibinfo{journal}{\emph{arXiv preprint arXiv:2106.00882}}
  (\bibinfo{year}{2021}).
\newblock


\bibitem[Yang et~al\mbox{.}(2022)]%
        {yang2022learning}
\bibfield{author}{\bibinfo{person}{Xiaojiang Yang}, \bibinfo{person}{Junchi
  Yan}, \bibinfo{person}{Yu Cheng}, {and} \bibinfo{person}{Yizhe Zhang}.}
  \bibinfo{year}{2022}\natexlab{}.
\newblock \showarticletitle{Learning deep generative clustering via mutual
  information maximization}.
\newblock \bibinfo{journal}{\emph{IEEE Transactions on Neural Networks and
  Learning Systems}} \bibinfo{volume}{34}, \bibinfo{number}{9}
  (\bibinfo{year}{2022}), \bibinfo{pages}{6263--6275}.
\newblock


\bibitem[Yang et~al\mbox{.}(2020)]%
        {yang2020nonlinear}
\bibfield{author}{\bibinfo{person}{Zhan Yang}, \bibinfo{person}{Jun Long},
  \bibinfo{person}{Lei Zhu}, {and} \bibinfo{person}{Wenti Huang}.}
  \bibinfo{year}{2020}\natexlab{}.
\newblock \showarticletitle{Nonlinear robust discrete hashing for cross-modal
  retrieval}. In \bibinfo{booktitle}{\emph{Proceedings of the 43rd
  international ACM SIGIR conference on research and development in information
  retrieval}}. \bibinfo{pages}{1349--1358}.
\newblock


\bibitem[Yaxue(2020)]%
        {yaxue2020convolutional}
\bibfield{author}{\bibinfo{person}{Qin Yaxue}.}
  \bibinfo{year}{2020}\natexlab{}.
\newblock \showarticletitle{Convolutional neural networks for literature
  retrieval}. In \bibinfo{booktitle}{\emph{2020 International Conference on
  Computer Vision, Image and Deep Learning (CVIDL)}}. IEEE,
  \bibinfo{pages}{393--397}.
\newblock


\bibitem[Ye et~al\mbox{.}(2020)]%
        {ye2020unsupervised}
\bibfield{author}{\bibinfo{person}{Fanghua Ye}, \bibinfo{person}{Jarana
  Manotumruksa}, {and} \bibinfo{person}{Emine Yilmaz}.}
  \bibinfo{year}{2020}\natexlab{}.
\newblock \showarticletitle{Unsupervised few-bits semantic hashing with
  implicit topics modeling}. In \bibinfo{booktitle}{\emph{EMNLP (Findings)}},
  Vol.~\bibinfo{volume}{20}. Association for Computational Linguistics (ACL),
  \bibinfo{pages}{2566--2575}.
\newblock


\bibitem[Yin et~al\mbox{.}(2019)]%
        {yin2019understanding}
\bibfield{author}{\bibinfo{person}{Penghang Yin}, \bibinfo{person}{Jiancheng
  Lyu}, \bibinfo{person}{Shuai Zhang}, \bibinfo{person}{Stanley Osher},
  \bibinfo{person}{Yingyong Qi}, {and} \bibinfo{person}{Jack Xin}.}
  \bibinfo{year}{2019}\natexlab{}.
\newblock \showarticletitle{Understanding straight-through estimator in
  training activation quantized neural nets}.
\newblock \bibinfo{journal}{\emph{arXiv preprint arXiv:1903.05662}}
  (\bibinfo{year}{2019}).
\newblock


\bibitem[Yu et~al\mbox{.}(2024)]%
        {yu2024tprf}
\bibfield{author}{\bibinfo{person}{Chuting Yu}, \bibinfo{person}{Hang Li},
  \bibinfo{person}{Ahmed Mourad}, \bibinfo{person}{Bevan Koopman}, {and}
  \bibinfo{person}{Guido Zuccon}.} \bibinfo{year}{2024}\natexlab{}.
\newblock \showarticletitle{TPRF: A Transformer-based Pseudo-Relevance Feedback
  Model for Efficient and Effective Retrieval}.
\newblock \bibinfo{journal}{\emph{arXiv preprint arXiv:2401.13509}}
  (\bibinfo{year}{2024}).
\newblock


\bibitem[Yu et~al\mbox{.}(2015)]%
        {yu2015understanding}
\bibfield{author}{\bibinfo{person}{Zheng Yu}, \bibinfo{person}{Haixun Wang},
  \bibinfo{person}{Xuemin Lin}, {and} \bibinfo{person}{Min Wang}.}
  \bibinfo{year}{2015}\natexlab{}.
\newblock \showarticletitle{Understanding short texts through semantic
  enrichment and hashing}.
\newblock \bibinfo{journal}{\emph{IEEE Transactions on Knowledge and Data
  Engineering}} \bibinfo{volume}{28}, \bibinfo{number}{2}
  (\bibinfo{year}{2015}), \bibinfo{pages}{566--579}.
\newblock


\bibitem[Yuan et~al\mbox{.}(2020)]%
        {yuan2020central}
\bibfield{author}{\bibinfo{person}{Li Yuan}, \bibinfo{person}{Tao Wang},
  \bibinfo{person}{Xiaopeng Zhang}, \bibinfo{person}{Francis~EH Tay},
  \bibinfo{person}{Zequn Jie}, \bibinfo{person}{Wei Liu}, {and}
  \bibinfo{person}{Jiashi Feng}.} \bibinfo{year}{2020}\natexlab{}.
\newblock \showarticletitle{Central similarity quantization for efficient image
  and video retrieval}. In \bibinfo{booktitle}{\emph{Proceedings of the
  IEEE/CVF conference on computer vision and pattern recognition}}.
  \bibinfo{pages}{3083--3092}.
\newblock


\bibitem[Zhang et~al\mbox{.}(2010)]%
        {zhang2010self}
\bibfield{author}{\bibinfo{person}{Dell Zhang}, \bibinfo{person}{Jun Wang},
  \bibinfo{person}{Deng Cai}, {and} \bibinfo{person}{Jinsong Lu}.}
  \bibinfo{year}{2010}\natexlab{}.
\newblock \showarticletitle{Self-taught hashing for fast similarity search}. In
  \bibinfo{booktitle}{\emph{Proceedings of the 33rd international ACM SIGIR
  conference on Research and development in information retrieval}}.
  \bibinfo{pages}{18--25}.
\newblock


\bibitem[Zhang et~al\mbox{.}(2017a)]%
        {zhang2017semi}
\bibfield{author}{\bibinfo{person}{Peng-Fei Zhang},
  \bibinfo{person}{Chuan-Xiang Li}, \bibinfo{person}{Meng-Yuan Liu},
  \bibinfo{person}{Liqiang Nie}, {and} \bibinfo{person}{Xin-Shun Xu}.}
  \bibinfo{year}{2017}\natexlab{a}.
\newblock \showarticletitle{Semi-relaxation supervised hashing for cross-modal
  retrieval}. In \bibinfo{booktitle}{\emph{Proceedings of the 25th ACM
  international conference on Multimedia}}. \bibinfo{pages}{1762--1770}.
\newblock


\bibitem[Zhang et~al\mbox{.}(2024)]%
        {zhang2024document}
\bibfield{author}{\bibinfo{person}{Qian Zhang}, \bibinfo{person}{Qinliang Su},
  \bibinfo{person}{Jiayang Chen}, {and} \bibinfo{person}{Zhenpeng Song}.}
  \bibinfo{year}{2024}\natexlab{}.
\newblock \showarticletitle{Document Hashing with Multi-Grained
  Prototype-Induced Hierarchical Generative Model}. In
  \bibinfo{booktitle}{\emph{Findings of the Association for Computational
  Linguistics: EMNLP 2024}}. \bibinfo{pages}{321--333}.
\newblock


\bibitem[Zhang et~al\mbox{.}(2022)]%
        {zhang2022two}
\bibfield{author}{\bibinfo{person}{Xu Zhang}, \bibinfo{person}{Xinzheng Niu},
  \bibinfo{person}{Philippe Fournier-Viger}, {and} \bibinfo{person}{Bing
  Wang}.} \bibinfo{year}{2022}\natexlab{}.
\newblock \showarticletitle{Two-stage traffic clustering based on HNSW}. In
  \bibinfo{booktitle}{\emph{International Conference on Industrial, Engineering
  and Other Applications of Applied Intelligent Systems}}. Springer,
  \bibinfo{pages}{609--620}.
\newblock


\bibitem[Zhang et~al\mbox{.}(2017b)]%
        {zhang2017deconvolutional}
\bibfield{author}{\bibinfo{person}{Yizhe Zhang}, \bibinfo{person}{Dinghan
  Shen}, \bibinfo{person}{Guoyin Wang}, \bibinfo{person}{Zhe Gan},
  \bibinfo{person}{Ricardo Henao}, {and} \bibinfo{person}{Lawrence Carin}.}
  \bibinfo{year}{2017}\natexlab{b}.
\newblock \showarticletitle{Deconvolutional paragraph representation learning}.
\newblock \bibinfo{journal}{\emph{Advances in Neural Information Processing
  Systems}}  \bibinfo{volume}{30} (\bibinfo{year}{2017}).
\newblock


\bibitem[Zhang and Zhu(2019)]%
        {zhang2019doc2hash}
\bibfield{author}{\bibinfo{person}{Yifei Zhang} {and} \bibinfo{person}{Hao
  Zhu}.} \bibinfo{year}{2019}\natexlab{}.
\newblock \showarticletitle{Doc2hash: Learning discrete latent variables for
  documents retrieval}. In \bibinfo{booktitle}{\emph{Proceedings of the 2019
  Conference of the North American Chapter of the Association for Computational
  Linguistics: Human Language Technologies, Volume 1 (Long and Short Papers)}}.
  \bibinfo{pages}{2235--2240}.
\newblock


\bibitem[Zhang and Zhu(2020)]%
        {zhang2020discrete}
\bibfield{author}{\bibinfo{person}{Yifei Zhang} {and} \bibinfo{person}{Hao
  Zhu}.} \bibinfo{year}{2020}\natexlab{}.
\newblock \showarticletitle{Discrete wasserstein autoencoders for document
  retrieval}. In \bibinfo{booktitle}{\emph{ICASSP 2020-2020 IEEE International
  Conference on Acoustics, Speech and Signal Processing (ICASSP)}}. IEEE,
  \bibinfo{pages}{8159--8163}.
\newblock


\bibitem[Zhao et~al\mbox{.}(2020)]%
        {zhao2020song}
\bibfield{author}{\bibinfo{person}{Weijie Zhao}, \bibinfo{person}{Shulong Tan},
  {and} \bibinfo{person}{Ping Li}.} \bibinfo{year}{2020}\natexlab{}.
\newblock \showarticletitle{Song: Approximate nearest neighbor search on gpu}.
  In \bibinfo{booktitle}{\emph{2020 IEEE 36th International Conference on Data
  Engineering (ICDE)}}. IEEE, \bibinfo{pages}{1033--1044}.
\newblock


\bibitem[Zhao et~al\mbox{.}(2024)]%
        {zhao2024dense}
\bibfield{author}{\bibinfo{person}{Wayne~Xin Zhao}, \bibinfo{person}{Jing Liu},
  \bibinfo{person}{Ruiyang Ren}, {and} \bibinfo{person}{Ji-Rong Wen}.}
  \bibinfo{year}{2024}\natexlab{}.
\newblock \showarticletitle{Dense text retrieval based on pretrained language
  models: A survey}.
\newblock \bibinfo{journal}{\emph{ACM Transactions on Information Systems}}
  \bibinfo{volume}{42}, \bibinfo{number}{4} (\bibinfo{year}{2024}),
  \bibinfo{pages}{1--60}.
\newblock


\bibitem[Zheng et~al\mbox{.}(2019)]%
        {zheng2019fast}
\bibfield{author}{\bibinfo{person}{Chaoqun Zheng}, \bibinfo{person}{Lei Zhu},
  \bibinfo{person}{Xu Lu}, \bibinfo{person}{Jingjing Li},
  \bibinfo{person}{Zhiyong Cheng}, {and} \bibinfo{person}{Hanwang Zhang}.}
  \bibinfo{year}{2019}\natexlab{}.
\newblock \showarticletitle{Fast discrete collaborative multi-modal hashing for
  large-scale multimedia retrieval}.
\newblock \bibinfo{journal}{\emph{IEEE Transactions on Knowledge and Data
  Engineering}} \bibinfo{volume}{32}, \bibinfo{number}{11}
  (\bibinfo{year}{2019}), \bibinfo{pages}{2171--2184}.
\newblock


\bibitem[Zheng et~al\mbox{.}(2020)]%
        {zheng2020generative}
\bibfield{author}{\bibinfo{person}{Lin Zheng}, \bibinfo{person}{Qinliang Su},
  \bibinfo{person}{Dinghan Shen}, {and} \bibinfo{person}{Changyou Chen}.}
  \bibinfo{year}{2020}\natexlab{}.
\newblock \showarticletitle{Generative semantic hashing enhanced via Boltzmann
  machines}.
\newblock \bibinfo{journal}{\emph{arXiv preprint arXiv:2006.08858}}
  (\bibinfo{year}{2020}).
\newblock


\bibitem[Zhong et~al\mbox{.}(2016)]%
        {zhong2016overview}
\bibfield{author}{\bibinfo{person}{Guoqiang Zhong}, \bibinfo{person}{Li-Na
  Wang}, \bibinfo{person}{Xiao Ling}, {and} \bibinfo{person}{Junyu Dong}.}
  \bibinfo{year}{2016}\natexlab{}.
\newblock \showarticletitle{An overview on data representation learning: From
  traditional feature learning to recent deep learning}.
\newblock \bibinfo{journal}{\emph{The Journal of Finance and Data Science}}
  \bibinfo{volume}{2}, \bibinfo{number}{4} (\bibinfo{year}{2016}),
  \bibinfo{pages}{265--278}.
\newblock


\bibitem[Zhou et~al\mbox{.}(2008)]%
        {zhou2008real}
\bibfield{author}{\bibinfo{person}{Kun Zhou}, \bibinfo{person}{Qiming Hou},
  \bibinfo{person}{Rui Wang}, {and} \bibinfo{person}{Baining Guo}.}
  \bibinfo{year}{2008}\natexlab{}.
\newblock \showarticletitle{Real-time kd-tree construction on graphics
  hardware}.
\newblock \bibinfo{journal}{\emph{ACM Transactions on Graphics (TOG)}}
  \bibinfo{volume}{27}, \bibinfo{number}{5} (\bibinfo{year}{2008}),
  \bibinfo{pages}{1--11}.
\newblock


\bibitem[Zhu et~al\mbox{.}(2021)]%
        {zhu2021retrieving}
\bibfield{author}{\bibinfo{person}{Fengbin Zhu}, \bibinfo{person}{Wenqiang
  Lei}, \bibinfo{person}{Chao Wang}, \bibinfo{person}{Jianming Zheng},
  \bibinfo{person}{Soujanya Poria}, {and} \bibinfo{person}{Tat-Seng Chua}.}
  \bibinfo{year}{2021}\natexlab{}.
\newblock \showarticletitle{Retrieving and reading: A comprehensive survey on
  open-domain question answering}.
\newblock \bibinfo{journal}{\emph{arXiv preprint arXiv:2101.00774}}
  (\bibinfo{year}{2021}).
\newblock


\bibitem[Zhu et~al\mbox{.}(2020)]%
        {zhu2020deep}
\bibfield{author}{\bibinfo{person}{Lei Zhu}, \bibinfo{person}{Xu Lu},
  \bibinfo{person}{Zhiyong Cheng}, \bibinfo{person}{Jingjing Li}, {and}
  \bibinfo{person}{Huaxiang Zhang}.} \bibinfo{year}{2020}\natexlab{}.
\newblock \showarticletitle{Deep collaborative multi-view hashing for
  large-scale image search}.
\newblock \bibinfo{journal}{\emph{IEEE Transactions on Image Processing}}
  \bibinfo{volume}{29} (\bibinfo{year}{2020}), \bibinfo{pages}{4643--4655}.
\newblock


\bibitem[Zhu et~al\mbox{.}(2023)]%
        {zhu2023multi}
\bibfield{author}{\bibinfo{person}{Lei Zhu}, \bibinfo{person}{Chaoqun Zheng},
  \bibinfo{person}{Weili Guan}, \bibinfo{person}{Jingjing Li},
  \bibinfo{person}{Yang Yang}, {and} \bibinfo{person}{Heng~Tao Shen}.}
  \bibinfo{year}{2023}\natexlab{}.
\newblock \showarticletitle{Multi-modal hashing for efficient multimedia
  retrieval: A survey}.
\newblock \bibinfo{journal}{\emph{IEEE Transactions on Knowledge and Data
  Engineering}} \bibinfo{volume}{36}, \bibinfo{number}{1}
  (\bibinfo{year}{2023}), \bibinfo{pages}{239--260}.
\newblock


\bibitem[Zhuang et~al\mbox{.}(2024)]%
        {zhuang2024setwise}
\bibfield{author}{\bibinfo{person}{Shengyao Zhuang}, \bibinfo{person}{Honglei
  Zhuang}, \bibinfo{person}{Bevan Koopman}, {and} \bibinfo{person}{Guido
  Zuccon}.} \bibinfo{year}{2024}\natexlab{}.
\newblock \showarticletitle{A setwise approach for effective and highly
  efficient zero-shot ranking with large language models}. In
  \bibinfo{booktitle}{\emph{Proceedings of the 47th International ACM SIGIR
  Conference on Research and Development in Information Retrieval}}.
  \bibinfo{pages}{38--47}.
\newblock


\bibitem[Zhuang et~al\mbox{.}(2023)]%
        {zhuang2023efficiently}
\bibfield{author}{\bibinfo{person}{Yan Zhuang}, \bibinfo{person}{Qi Liu},
  \bibinfo{person}{Yuting Ning}, \bibinfo{person}{Weizhe Huang},
  \bibinfo{person}{Rui Lv}, \bibinfo{person}{Zhenya Huang},
  \bibinfo{person}{Guanhao Zhao}, \bibinfo{person}{Zheng Zhang},
  \bibinfo{person}{Qingyang Mao}, \bibinfo{person}{Shijin Wang},
  {et~al\mbox{.}}} \bibinfo{year}{2023}\natexlab{}.
\newblock \showarticletitle{Efficiently measuring the cognitive ability of
  llms: An adaptive testing perspective}.
\newblock \bibinfo{journal}{\emph{arXiv preprint arXiv:2306.10512}}
  (\bibinfo{year}{2023}).
\newblock


\end{thebibliography}

%%
%% If your work has an appendix, this is the place to put it.
\end{document}